\newcommand\ignore[1]{}
\documentclass[12pt,citesort]{article}
\usepackage{amsmath,amssymb,graphicx,url}
\usepackage{color}
\usepackage{axodraw}
\newcommand\Black{}

\newcommand\cut[1]{}

\newcommand{\half}{ {\scriptstyle \frac{1}{2} } }
\newcommand{\Half}{ {\frac{1}{2} } }

\newcommand\be{\begin{equation}}
\newcommand\ee{\end{equation}}
\newcommand\bea{\begin{eqnarray}}
\newcommand\eea{\end{eqnarray}}
\newcommand{\bdm}{\begin{displaymath}}
\newcommand{\edm}{\end{displaymath}}
\newcommand\nn{ \nonumber\\}

\newcommand{\E}[1]{e^{\textstyle #1}}

\newcommand{\dd}[1]{\partial_{#1}}
\def\dd{\partial}
\makeatletter
\@addtoreset{equation}{section}
\makeatother

\title{Analyticity for Multi-Regge Limits of the Bern-Dixon-Smirnov Amplitudes}
  
\author{ Richard  C. Brower\footnote{Physics Department,
Boston University, Boston MA 02215},
Horatiu Nastase\footnote{Global Edge Institute, Tokyo Institute of
Technology, Tokyo 152-8550, Japan},  
Howard J. Schnitzer\footnote{Theoretical Physics Group, Martin Fischer School of 
Physics,
Brandeis Univ., Waltham, MA 02454} \\
and  \\ Chung-I Tan\footnote{Physics Department, Brown University,
Providence, RI 02912}
}

\begin{document}

\maketitle

\vspace{-10cm}

\begin{flushright}
BRX-TH-600, Brown-HET-1565, TIT/HEP-586
\end{flushright}
%\vspace{-5cm}

\vspace{8cm}

\begin{abstract}

  As a consequence of the AdS/CFT correspondence, planar ${\cal N} =4$   super Yang-Mills $SU(N)$ theory is expected to exhibit stringy   behavior and multi-Regge asymptotic.  In this paper we extend our   recent investigation to consider issues of analyticity, a central   feature of Regge asymptotics.  We contrast flat-space open string   theory in the planar limit with the ${\cal N}=4$ super Yang-Mills theory, as represented by the Bern, Dixon and Smirnov~\cite{Bern:2005iz} (BDS) conjecture for n-gluon   scattering, believed to be exact for $n=4,5$ and modified only by a   function of cross-ratios for $n\geq 6$.  It is emphasized that   multi-Regge factorization should be applied to trajectories with   definite signature.  A variety of analyticity and factorization   constraints realized in flat space string theory are not satisfied   by the BDS conjecture, at least when the exponential factors are   truncate in the infra-red regulator below $O(\epsilon)$.

\end{abstract}

\newpage
\setcounter{tocdepth}{3}
\tableofcontents

\newpage

\section{Introduction}
\label{sec:introduction}

Regge asymptotics, combined with analyticity and crossing symmetry is
a potentially powerful tool in understanding the planar limit of Yang
Mills theory.  Indeed Regge constraints played a major role in the
original S-matrix program that led to the discovery of string theory
in flat space and in the context of ${\cal N} = 4$ super Yang Mills
and ${\cal N}= 8$ SUGRA, the recent work of Arkani-Hamed, Cachazo and
Kaplan~\cite{ArkaniHamed:2008gz} has shown again the utilility of
strong asymptotic constraints on the S-matrix. The full power of
Regge asymptotics also includes factorization which imposes
self-consistency conditions as one considers amplitudes with
increasing number of external lines. The reason is familiar in the use
of Feynman diagrams. The 4-point amplitude defines the Regge exchange
``propagator'' and the Reggeon two particle vertex. Then through the use
of factorization and cutting rules (or unitarity), these same
propagators and vertex functions form building blocks for a variety of
multi-Regge limits of the n-point functions. In fact the process is
iterative.  In the 5-point funtion, one encounters a new double Regge
vertex, which occurs in higher point functions and in the 6-point
function a new Regge-particle scattering amplitude.  This hierarchy
places severe non-perturbative constraints on the theory. The
properties of these are well established in flat space string theory,
but as we will show have unexpected realization in the conjecture  by Bern, Dixon and
Smirnov (BDS)~\cite{Bern:2005iz} for the maximal helicity
violating (MHV) planar n-point ${\cal N} =4$ gluon amplitudes for all
coupling $\lambda$, at least when the exponential factors
are truncate below $O(\epsilon)$, the infra-red regulator.

In the modern context of gauge/string duality, the use of
Regge properties is only beginning to be exploited, however there are some
interesting results for the classic example of gauge/string duality which
maps ${\cal N} = 4$ super Yang Mills theory into gravity (or IIB
super strings) in $AdS_5 \times S_5$.  For example in the closed
string sector, Brower, Polchinski, Strassler and
Tan~\cite{Brower:2006ea} have shown that the weak coupling BFKL
Pomeron is mapped at strong coupling into a dual BFKL Pomeron with
very similar properties. Direct extrapolation from the weak coupling perturbative sum to the strong coupling limit has also been made by Kotikov, Lipatov,  Onishchenko,  and Velizhanin~\cite{Kotikov}. In both limits conformal symmetry~\footnote{
  For the weak coupling BFKL equation this is referred as M\"obius
  invariance which in strong coupling is realized~\cite{Brower:2007qh,Brower:2007xg, Cornalba:2007fs}
  as the $SL(2,C)$ isometries of Euclidean $AdS_3$ subspace of $AdS_5$.} requires that the leading Regge singularity is a fixed $J$-plane cut
at intercept $j_0(\lambda$). At weak coupling the intercept, $j_0 = 1 + O(\lambda)$  , is near 1 corresponding to BFKL   Reggized two gluon exchange while for strong coupling the intercept, $j_0 = 2 - O(1/\sqrt{\lambda})$, is near  2 for the $AdS_5$ graviton. Interesting interpolation between
these weak and strong coupling limits has also been made by Stasto~\cite{Stasto:2007uv}. 

In the open string (or gluon scattering) sector, there is a
new opportunity due to the BDS conjecture~\cite{Bern:2005iz,Anastasiou:2003kj}. (For a recent review, see
\cite{Alday:2008yw}; for older developments, see \cite{Bern:1997nh,Bern:2004kq}).
The BDS 4-point gluon amplitude exhibits a
remarkably simple Regge asymptotic
form~\cite{Korchemskaya:1996je,Naculich:2007ub}, without even taking
the high energy limit. Just as in the flat space super string theory,
the $J$-plane is meromorphic with simple $J$-plane poles.

Recent work supports the view that the BDS amplitudes may also be
formulated as a world sheet sigma model for strings propagating in
$AdS_5 \times S_5$.  Specifically, Alday and Maldacena
\cite{Alday:2007hr,Alday:2007he} (see also \cite{Alday:2007mf}) computed the wide angle scattering at
strong coupling from a minimal surface, in close analogy with earlier
calculations of flat space superstring amplitudes.  Subsequently
Berkovits and Maldacena \cite{Berkovits:2008ic} have demonstrated the
equivalence of the gluon MHV planar amplitudes with Wilson loops at
all values of the coupling using a fermionic T-duality, and noted that
the MHV planar n-gluon scattering amplitude (world sheet tree
amplitudes) are greatly simplified using a stringy generalization of
the spinor helicity formalism (see Appendix A, Ref. \cite{Berkovits:2008ic}).

On the other hand, the ${\cal N}=4$ SYM gluon amplitudes are IR divergent, which require a cut-off, 
usually treated in dimensional regularization with $D=4-2\epsilon$, both in SYM and in its gravity dual background. 
In particular, the Regge trajectory of gluons is both IR divergent as $1/\epsilon$ 
and divergent at $t=0$ as $\log \;(- t/\mu^2)$, which 
as we will see complicates the details of the Regge identification~\footnote{Gluon regge trajectory  in supersymmetric models was first calculated to two loops in~\cite{Kotikov:2000pm}.}. The gluon amplitudes are then to be treated as 
ingredients in IR safe quantities, where we can take the cut-off to zero and obtain physical results.

All of this makes a comparison of the Regge limit for 
n-gluon BDS and the planar approximation to flat space superstring intriguing. (Throughout this paper, flat space string theory and  gluon scattering are compared for the leading planar and large $N_c$ approximation.)  It is interesting to understand how
the two theories realize Regge asymptotics and how these expressions
differ, particularly as many generic features of Regge amplitudes
reflect very general analyticity constraints on the planar amplitudes
for any renormalizable field theory with leading Regge asymptotics.
Ultimately these differences should be traceable to consequences of
string scattering in the $AdS_5\times S_5$ background dual to ${\cal N}=4$ SYM 
versus the usual flat space background, although
aside from a few comments, this  will be postponed to future investigations.

Since it is widely believed that the BDS amplitude is exact up to
 ${\cal O}(\epsilon)$ terms for n = 4 and
n =5 and can only be corrected by a function of cross-ratios for $n >
5$, departures from conventional Regge expectation deserve careful
scrutiny.  Dual conformal invariance  for the Wilson loop, 
was orginally found in perturbation theory~\cite{Drummond:2007au,Brandhuber:2007yx,Drummond:2007aua,Drummond:2007cf}).
Subsequently it has also been observed at strong coupling~\cite{Komargodski:2008wa,Beisert:2008iq} 
and extended to dual superconformal invariance in Ref.~\cite{Berkovits:2008ic,Drummond:2008vq,Brandhuber:2008pf}.
The need for a function
of cross-ratios at $n>5$ is seen at strong coupling in
\cite{Alday:2007he,Astefanesei:2007bk,Itoyama:2008je}, and at low
order perturbation theory for $n=6$~\cite{Drummond:2007bm,Bern:2008ap,Drummond:2008aq,Cachazo:2008hp}. Consequently this
comparison with flat space string theory casts light on the special
properties inherent in planar ${\cal N} =4$ gluon scattering
amplitudes.  Differences should be traceable to the world
sheet formulations of the flat space and gravity dual string.

In a recent paper \cite{Brower:2008nm} the authors investigated a limited number of
constraints in the single Regge and linear multi-Regge
behavior in the Euclidean region.   Here we take up the
issue of analytic continuation to regions describing physical
processes, which has also been considered by  Bartels et al.~\cite{Bartels:2008ce}. The combined
constraints of Regge asymptotics, factorization and analytic
continuation are subtle and very powerful. For an extensive analysis
of this subject for the open string amplitude in flat space, one may
see the review~\cite{Brower:1974yv}.  The reader is referred to 
this and Ref.~\cite{Brower:2008nm} for crucial results which are not
repeated here.

\cut{ TOO DETAILED FOR INTRODUCTION
In this paper analytic continuation to physical regions is discussed, where it is emphasized that factorization of multi-Regge amplitudes need only be expected for signature amplitudes.  Therefore discussion of these two issues plays a prominent role in this paper.  In this context, Bartels, et. al. \cite{Bartels:2008ce}  considered the multi-Regge behavior of BDS for one particular color ordering of the 6-point function, i.e. where the two incoming gluous were diagonally apposite in color ordering.  In this case, a violation of ``naive'' Regge factorization was found.  it is important to recognize that  this particular amplitude is just one of 8 inequivalent color combinations relevant for the linear multi-Regge limit in question; 6 of which do satisfy naive factorization, and two do not.  
 It has been  suggested  that  the BDS amplitudes can be ``repaired''  to restore factorization. However,  whether signatured factorization or naive factorization,   it appears that the required modification cannot be written as the analytic continuation of a function of cross-ratios. } 

It is important to stress the focus of this paper. The BDS amplitudes are writen as a product of two terms,
\be
{\cal A}_{BDS} = {\cal A}_{tree} M(\epsilon) \; ,
\ee
the tree amplitudes with all polarization dependence and
a scalar amplitude, $M(\epsilon)$, that is factorized into an IR divergent part
and a finite part as $\epsilon \rightarrow 0$. Our analysis is done
for the Regge limits {\bf after} truancating $\log({\cal M})$ below  $O(\epsilon)$. 

The  organization and main conclusions of the paper are as follows. We begin in Sec. 3 by reviewing the Regge form for the 4-point function, remarking on the singular structure of the Regge trajectory at the gluon pole, the continuation to the physical region and the definition of Regge exchanges of definite signature.  In Sec. 4, we consider the 5-point function and contrast the analytic properties of the double Regge vertex relative to flat space string theory. We note that the double Regge vertex does not obey the constraint needed for the absence of overlapping physical region discontinuities of the Steinman relation.  In Sec. 5, we explain how multi-Regge factorization of the 6-point function is realized in the signatured amplitudes for flat-space string theory by properly taking into account singularity in ``cross-ratio'' variables and the failure
of the BDS amplitudes to satisfy this property. In Sec. 6, we incorporate  color traces as well as gluon polarizations and show how factorization is realized for a general n-point amplitudes in the multi-Regge limit for flat space open string theory. Finally in Sec. 7, we discuss discontinuities in crossed-channel invariants, and find, paradoxically, the absence of Regge contribution in the ``triple-Regge'' limit for n-point amplitudes, $n \geq 6$. In Sec. 8 we conclude with some general discussions, including the possibility that some or most of these unconventional analyticity properties of multi-Regge amplitudes may be a result of the truncation of the log of  BDS amplitude below  $O(\epsilon)$.

Before proceeding to a detailed discussion of the BDS amplitudes, in
Sec.~\ref{sec:analytic}, we present a general method for continuing
individual planar multi-particle amplitudes.  Following the kinematical
approach of Alday and Maldacena~\cite{Alday:2007hr,Alday:2007he}, one
can define for planar amplitudes the Regge limit away from all the physical singularites
where amplitudes  are
real. {\em We refer to this limit as the Euclidean Regge limit.}
Subsequently one can analytically continue each planar amplitude in
the complex plane of invariants to a
particular physical region above all  unitarity
thresholds to properly define the complex phases.  Readers
familiar with this subject may wish to proceed to
Secs.~\ref{sec:constraints}-\ref{sec:5plus} where these methods are
applied to higher point functions.

\newpage
\section{Analytic Continuation of Planar Amplitudes}
\label{sec:analytic}

Both flat space open string theory amplitudes and the BDS n-gluon
amplitudes are defined as on-shell scattering amplitudes with a
restricted set of physical singularities due to the planar
structure. This allows one to define the phases of multi-Regge limits
by a systematic procedure. As we explain shortly, this procedure
takes two steps. First, for each planar amplitude, the Regge limit is
taken in the deep ``Euclidean'' region where it is real and analytic.
Second, this amplitude can be defined in the physical
scattering region by an analytic continuation in the ``upper half
plane''.  While in prinicple the analytic continuation can be
peformed on an independent set of $3n-10$ Mandelstam invariances that respect the constraints for on-shell scattering amplitudes, there is
a subtlety involving cross ratios that approach unity in the extreme
Regge limit, which we will explain briefly here for the 6-point function and more fully  in Sec. \ref{sec:FactorizationConstraints}.

The leading term in the large N limit for gauge theories, as
emphasized first by 't Hooft, restricts the perturbative expansion to
planar diagrams. This topological feature is shared by open
superstring scattering amplitudes in flat space and was one of the
first indications that Yang Mills theory, and even QCD, might be
equivalent to a string theory. The BDS conjecture also refers to the
planar approximation for ${\cal N} = 4$ super Yang Mills theory and
therefore may well share some properties with open string theory.  One
consequence of the planar approximation is that the n-point gluon
amplitude, ${\cal A}_n(k_i, \epsilon_i, a_i)$, is a sum over single color traces
for each permutation $\pi(i)$ modulo pure cyclic ordering:
\be
{\cal A}_n(k_i, \epsilon_i, a_i) = \sum_{\pi} Tr[T^{a_{\pi(1)}}  T^{a_{\pi(2)}}
\cdots T^{a_{\pi(n)}} ] A_n(k_{\pi(1)}, \epsilon_{\pi(1)}, k_{\pi(2)},  \epsilon_{\pi(2)}, \cdots, k_{\pi(n)},\epsilon_{\pi(n)})
\label{colorord}
\ee
In (\ref{colorord}), the $T^{a}$ are generators of $SU(N)$ in the
fundamental representation. For convenience, we will
in what follows extend the analysis to $U(N)$, with normalization: $\sum_a
T^a_{ij}T^a_{lm}=2 \delta_{im}\delta_{jl}$.

The MHV n-gluon scattering amplitudes for ${\cal N}=4$ SYM  and
for the open superstring theory~\cite{Stieberger:2006bh,Berkovits:2008ic} may be factored
into a product of the Born term, the planar
n-gluon tree amplitude   and a ``reduced amplitude'', $M_n(1,2,\cdots n)$:
\be
 A_n(k_{1}, \epsilon_{1},  \cdots, k_{n},\epsilon_{n}) = A_{n,tree}(k_{1}, \epsilon_{1},  \cdots, k_{n},\epsilon_{n}) \; 
M_n(k_{1}, \cdots, k_{n}) 
\ee
All the polarization dependence~\footnote{We shall adopt
  ``all-incoming'' momentum convention. However, occasionally, for
  convenience, we will switch to ``all-outgoing'' convention. For
  convenience we shall often use the shorthand notations such as
  $A_{n} (\pi(1),\pi(2), \cdots,\pi(n))$ for $A_{n}(k_{\pi(1)},
  \epsilon_{\pi(1)}, k_{\pi(2)}, \epsilon_{\pi(2)}, \cdots,
  k_{\pi(n)},\epsilon_{\pi(n)})$ or similarly for $A_{n,tree}$ and
  $M_n$ when the full set of cyclically ordered arguments is obvious.}
is in the conformal invariant MHV tree amplitude. The reduced
amplitude, $M_n = M_{n,BDS}[t^{[r]}_i/\mu^2]$ or $M_n =
M_{n,string}[\alpha' t^{[r]}_i]$, is a Lorentz scalar function of
invariants for adjacent momenta (i.e.  $t^{[r]}_i = (k_i + \cdots+
k_{i+r-1})^2 )$), with an intrinsic scale (the string tension
$1/\alpha'$ for flat space string theory and the IR cut-off $\mu^2$
for BDS). Since it is well known that the zero slope limit ($\alpha'
\rightarrow 0$) for flat space string scattering gives tree-level
Yang-Mills theory, comparision with BDS can be restricted to the
reduced amplitude.

The planar amplitude $A(1,2,\cdots n)$  is real and analytic
(no cuts or poles) for Euclidean or space-like invariants, $t^{[r]}_i <
0$.  Consequently it is convenient, when studying the Regge limits, to
first take the limit in the Euclidean region followed by analytic
continuation to the physical region to establish the complex
phase. The subtlety is that one must do this continuation in an {\bf
  independent set} of Mandelstam invariants staying on the mass and
energy-momentum shell. For an n-particle planar amplitude, there are
$n(n-3)/2$ BDS invariants but only $3n-10$ are independent.

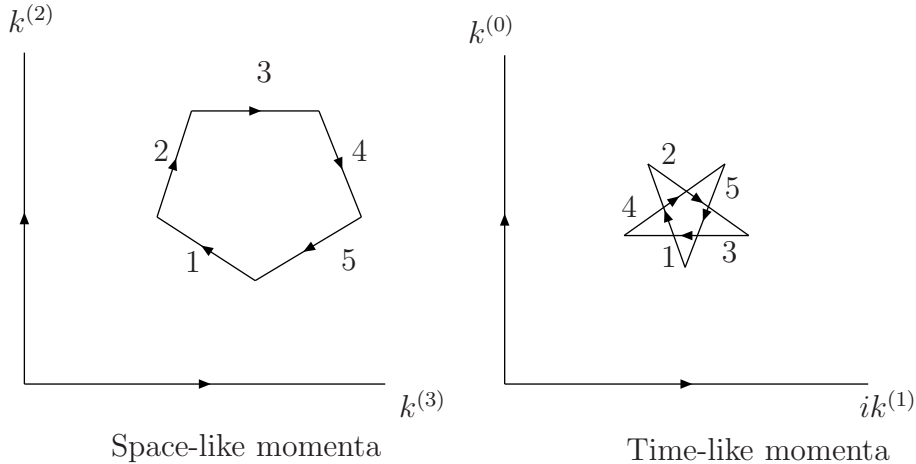
\begin{figure}[th!]
\begin{center}
  \begin{picture}(352,179) (38,-105)
    \SetWidth{0.5}
    \ArrowLine(44,-75)(180,-75)
       \ArrowLine(44,-75)(44,50)
    \ArrowLine(225,-75)(225,49)
    \ArrowLine(225,-75)(362,-75)
    \Text(77,-105)[lb]{{\Black{Space-like momenta}}}
    \Text(272,-104)[lb]{{\Black{Time-like momenta}}}
     \ArrowLine(131,-36)(94,-12)
      \ArrowLine(94,-12)(107,28)
     \ArrowLine(107,28)(155,28)
        \ArrowLine(155,28)(171,-12)
           \ArrowLine(171,-12)(131,-36)
         \ArrowLine(293,-31)(279,8)
          \ArrowLine(279,8)(317,-19)
             \ArrowLine(317,-19)(270,-19)
               \ArrowLine(270,-19)(308,8)
    \ArrowLine(308,8)(293,-31)
 \Text(186,-88)[lb]{{\Black{$k^{(3)}$}}}
    \Text(360,-88)[lb]{{\Black{$i k^{(1)}$}}}
    \Text(212,54)[lb]{{\Black{$ k^{(0)}$}}}
    \Text(132,39)[lb]{{\Black{$3$}}}
    \Text(168,9)[lb]{{\Black{$4$}}}
    \Text(164,-32)[lb]{{\Black{$5$}}}
    \Text(285,-31)[lb]{{\Black{$1$}}}
    \Text(270,-12)[lb]{{\Black{$4$}}}
    \Text(309,-5)[lb]{{\Black{$5$}}}
    \Text(308,-29)[lb]{{\Black{$3$}}}
    \Text(285,8)[lb]{{\Black{$2$}}}
    \Text(105,-33)[lb]{{\Black{$1$}}}
    \Text(93,9)[lb]{{\Black{$2$}}}
    \Text(38,58)[lb]{{\Black{$k^{(2)}$}}}
 
  \end{picture}
\end{center}
\caption{The 5-point gluonic amplitude, ${\cal A}_5(k_1,k_2,k_3,k_4,k_5)$ evaluated on 
shell ($k^2_i =0$, $\sum_i k_i =0$) with all BDS invariants ($t^{[r]}_i <0$) space-like.}
\label{fig:star}
\end{figure}

The procedure we choose to use closely follows the approach introduced
by Alday and Maldacena~\cite{Alday:2007hr,Alday:2007he} for the deep
Euclidean region (or wide angle scattering) for the planar n-point
amplitude. We will extend this to allow us to approach the
Regge limit, while avoiding all unitarity thresholds as needed
in the subsequent discussion. Following
Refs.~\cite{Brower:2008nm,Alday:2007hr,Alday:2007he} we first introduce
light-cone variables on the external legs: $k^\pm_i = k^{(0)}_i \pm
k^{(3)}$ and $\vec k^\perp_i = (k^{(1)}_i,k^{(2)}_i)$.  This
represents $4n$ variables that must be constrained to give $3n-10$
invariants by enforcing (i) the mass shell $k^2_i =- k^+_i k^- + \vec
k^\perp_i\cdot  \vec k^\perp_i=0$, (ii) energy momentum conservation $\sum_i
k_i =0$ and (iii) Lorentz invariance. The last is guaranteed for 
the BDS amplitudes because they are explicit functions of Lorentz
scalars. Consequently in light-cone coordinates, only the mass shell
and energy momentum constraints need to be explicitly respected to
satisfy all the non-linear constraints. The general solution for
arbitrary n, as realized by both
Alday-Maldacena~\cite{Alday:2007hr,Alday:2007he} and
Arkani-Hamed-Kaplan~\cite{ArkaniHamed:2008yf} requires analytically
continuing to a $(2,2)$ metric by taking $k^{(1)} \rightarrow i
k^{(1)}$ pure imaginary. Thus there is a space-like geometry in the 2-3
plane and a time-like geometry in the 0-1 plane.  To satisfy
energy-momentum conditions we construct closed ``polygons'' in each
plane. Next to satisfy the on-shell condition, $(k^{(0)}_i)^2 + (i
k^{(1)}_i)^2 = (k^{(2)}_i)^2 + (k^{(3)}_i)^2 $ we must have the sides
of equal length for the i-th gluon in the two planes. As an
illustration consider the 5-point function in Fig.~\ref{fig:star}.

\begin{figure}[h!]
\begin{center}
  \begin{picture}(356,182) (42,-98)
    \SetWidth{0.5}
   \ArrowLine(47,-69)(182,-69)
    \ArrowLine(47,-69)(47,56)
    \ArrowLine(227,-69)(227,55)
    \ArrowLine(227,-69)(364,-69)
    \Text(79,-98)[lb]{{\Black{Space-like momenta}}}
    \Text(274,-97)[lb]{{\Black{Time-like momenta}}}
     \ArrowLine(135,-30)(95,-5)
        \ArrowLine(95,-5)(95,35)
            \ArrowLine(95,35)(135,55)
                \ArrowLine(135,55)(175,35)
                    \ArrowLine(175,35)(175,-5)
    \ArrowLine(175,-5)(135,-30)

     \ArrowLine(300,-30)(300,16)
     \ArrowLine(302,16)(302,-30)
     \ArrowLine(299,-30)(299,16)
     \ArrowLine(301,16)(301,-30)
    \Text(368,-84)[lb]{{\Black{$i k^{(1)}$}}}
    \Text(216,68)[lb]{{\Black{$ k^{(0)}$}}}
    \Text(183,-83)[lb]{{\Black{$k^{(3)}$}}}
    \Text(42,66)[lb]{{\Black{$k^{(2)}$}}}
     \Text(108,-27)[lb]{{\Black{$1$}}}
    \Text(80,15)[lb]{{\Black{$2$}}}
    \Text(110,53)[lb]{{\Black{$3$}}}
    \Text(160,53)[lb]{{\Black{$4$}}}
    \Text(187,15)[lb]{{\Black{$5$}}}
   \Text(160,-27)[lb]{{\Black{$6$}}}
\end{picture}
\end{center}
\caption{The 6-point gluonic amplitude, ${
    A}_6(k_1,k_2,k_3,k_4,k_5,k_6)$, evaluated on shell ($k^2_i =0$,
  $\sum_i k_i =0$) with all BDS invariants ($t^{[r]}_i <0$)
  space-like. Note here it was possible to set  $i k^{(1)}_i$ to a constant.}
\label{fig:hex}
\end{figure}
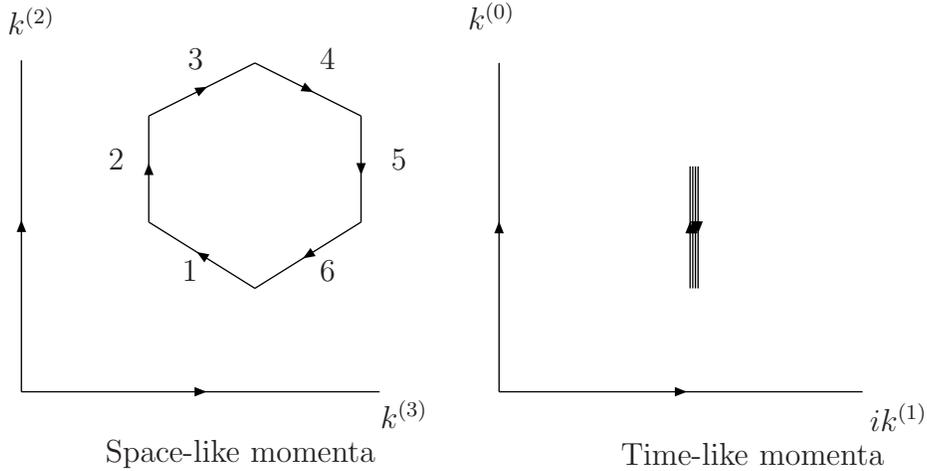

We choose a pentagon in the space-like 2-3 plane and a star in the time
like 0-1 plane. This ensures that all BDS invariants, which in this
case are two body invariants $(k_i + k_{i+1})^2$, are Euclidean!
In fact this construction works for any odd n by choosing an n-gon
in the space like plane and a ``star'' where cyclic order takes
you to the ``opposite'' side. The case of even n is simpler because
the ``star'' can now be taken to be a line back and forth in $k^{(0)}$ at $k^{(1)} =0$ so one does not need the 
additional ``energy-like'' component (see Fig.~\ref{fig:hex}).

\begin{figure}
\begin{center}
%\fcolorbox{white}{white}{
  \begin{picture}(256,131) (170,-80)
    \SetWidth{0.5}
%    \SetColor{Black}
    \Vertex(322,3){1.41}
    \Vertex(338,4){1.41}
    \Vertex(353,4){1.41}
    \Vertex(342,-33){1.41}
    \Vertex(354,-33){1.41}
    \Text(312,26)[lb]{{\Black{$s_3$}}}
    \SetWidth{1.0}
    \LongArrow(176,-33)(176,-30)
    \SetWidth{0.8}
    \Photon(176,-70)(176,-33){1.5}{5.5}
    \SetWidth{1.0}
    \LongArrow(289,13)(289,16)
    \SetWidth{0.8}
    \Photon(289,-25)(289,12){1.5}{5.5}
    \Photon(281,-26)(233,-25){1.5}{7.5}
    \SetWidth{1.0}
    \LongArrow(280,-25)(284,-26)
    \SetWidth{0.8}
    \Photon(338,-26)(290,-25){1.5}{7.5}
    \SetWidth{1.0}
    \LongArrow(337,-25)(341,-26)
    \SetWidth{0.8}
    \Photon(394,-26)(346,-25){1.5}{7.5}
    \SetWidth{1.0}
    \LongArrow(394,-25)(398,-26)
    \LongArrow(400,15)(400,18)
    \SetWidth{0.8}
    \Photon(401,-23)(401,14){1.5}{5.5}
    \Text(283,32)[lb]{{\Black{$k_4$}}}
    \Text(395,35)[lb]{{\Black{$k_{n-1}$}}}
    \Text(172,-80)[lb]{{\Black{$-k_1$}}}
    \Text(397,-80)[lb]{{\Black{$-k_n$}}}
    \Text(316,-37)[lb]{{\Black{$q_3$}}}
    \Text(370,-39)[lb]{{\Black{$q_{n-3}$}}}
    \SetWidth{1.0}
    \LongArrow(402,-30)(402,-27)
    \SetWidth{0.8}
    \Photon(402,-68)(402,-31){1.5}{5.5}
    \Text(225,-41)[lb]{{\Black{$\kappa_{12}$}}}
    \Text(283,-38)[lb]{{\Black{$\kappa_{23}$}}}
    \SetWidth{1.0}
    \LongArrow(174,13)(174,16)
    \SetWidth{0.8}
    \Photon(175,-25)(175,12){1.5}{5.5}
    \Photon(223,-26)(175,-25){1.5}{7.5}
    \SetWidth{1.0}
    \LongArrow(222,-25)(226,-26)
    \LongArrow(229,13)(229,16)
    \SetWidth{0.8}
    \Photon(230,-26)(230,11){1.5}{5.5}
    \Text(171,27)[lb]{{\Black{$k_2$}}}
    \Text(228,28)[lb]{{\Black{$k_3$}}}
    \Text(254,-36)[lb]{{\Black{$q_2$}}}
    \Text(195,-36)[lb]{{\Black{$q_1$}}}
    \Text(195,27)[lb]{{\Black{$s_1$}}}
    \Text(255,25)[lb]{{\Black{$s_2$}}}
  \end{picture}
%}
\end{center}
\caption{Multiperipheral limit for the 2 to n-2 gluon scattering amplitude
in the tree approximation.}
\label{fig:multi_gluon.a}
\end{figure}
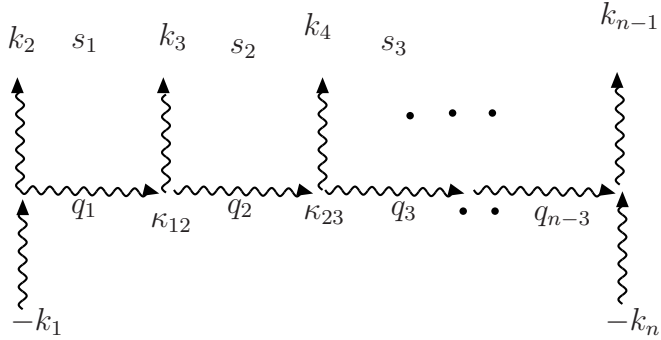

In the multi-Regge limit, particles are either right and left movers, with large $k^\pm$ components respectively. It is possible to approach the limit, $k^\pm \rightarrow  \infty$,  while staying  in the Euclidean region as depicted,  for example for the 5-point function, in Fig.~\ref{fig:5Regge.a}. This can clearly be generalized to any n.  (For n even, the necessary deformation involves primarily the left-hand side of Fig. \ref{fig:hex} with a corresponding elongation for the figure on the right.) Equivalently, for a general n-point amplitude in  the multi-Regge limit,
 instead of using independent on-shell momenta, one can use a set of  $3n-10$ independent invariants.  A natural set,  $s_1,
s_2, \cdots$, $t_1, t_2, \cdots$, $\kappa_{12}, \kappa_{23}, \cdots$, appropriate for a given multi-Regge region, 
has been discussed  in Ref. \cite{Brower:2008nm}, and this is  also
illustrated in Fig.  \ref{fig:multi_gluon.a}, (with $t_i=-q^2_i$). This set of independent variables was first introduced by N. Bali, G. F. Chew and A. Pignotti (BCP), \cite{Bali:1967zz}, and we shall refer to these as the BCP variables. The BCP set  is equivalent to the usual set  of BDS variables for $n=4$ and $5$. For $n\geq 6$, a BDS variable is either already a BCP variable or   can be expressed  in terms of the BCP variables through  a set of cross ratios, with  an accompanying set of 4-dimensional Gram-determinant constraints. 

Multi-Regge limit of planar amplitudes in the Euclidean region has been shown to factorize for both the flat-space string theory \cite{Brower:1974yv} and for the BDS n-gluon amplitudes \cite{Brower:2008nm} when the limit is taken in terms of an independent set of BCP invarints.  In Ref.  \cite{Brower:2008nm}, we have focussed on this Euclidean limit appropriate for a particular color ordering as for example depicted in Fig.~\ref{fig:multi_gluon.a}. Moreover
the factorization of the multi-Regge limit was achieved
precisely because all ``cross ratios'' either vanish or approach 1 in the Euclidean multi-Regge region. For example, there are three cross ratios for a 6-point BDS amplitude, $u_1$, $u_2$ and $u_3$.  In the multi-Regge limit, \be \Phi\equiv u_3 = \frac{s   s_{2}}{\Sigma_1\Sigma_2}\rightarrow 1, \ee with $u_1, u_2\rightarrow 0$. (See Eq. (5.13) and Appendix B of Ref.  \cite{Brower:2008nm}.)  The fact that $u_3\rightarrow 1$ follows from a non-linear Gram-determinant constraint in the multi-Regge limit.

\begin{figure}[th!]
\begin{center}
%\fcolorbox{white}{white}{
  \begin{picture}(380,197) (49,-97)
    \SetWidth{0.5}
%    \SetColor{Black}
 \ArrowLine(50,-58)(50,66)
   \ArrowLine(50,-58)(221,-58)
    \ArrowLine(255,-58)(255,66)
    \ArrowLine(255,-58)(391,-58)
    
     \ArrowLine(143,-11)(76,7)
     \ArrowLine(76,7)(132,20)
 \ArrowLine(132,20)(151,20)
  \ArrowLine(151,20)(210,7)
    \ArrowLine(210,7)(143,-11)
    \SetWidth{0.8}
   
    \Text(100,-14)[lb]{{\Black{$1$}}}
    \Text(95,22)[lb]{{\Black{$2$}}}
    \Text(138,32)[lb]{{\Black{$3$}}}
    \Text(180,21)[lb]{{\Black{$4$}}}
    \Text(171,-15)[lb]{{\Black{$5$}}}
    \Text(84,-96)[lb]{{\Black{Space-like momenta}}}
    \Text(280,-97)[lb]{{\Black{Time-like momenta}}}
    \SetWidth{0.5}
    
 \ArrowLine(332,-35)(309,32)
    \ArrowLine(309,32)(342,-20)
      \ArrowLine(342,-20)(323,-20)
        \ArrowLine(323,-20)(355,32)
   \ArrowLine(355,32)(332,-35)
   
    \SetWidth{0.8}

    \Text(306,11)[lb]{{\Black{$1$}}}
    \Text(321,20)[lb]{{\Black{$2$}}}
    \Text(337,20)[lb]{{\Black{$4$}}}
    \Text(354,10)[lb]{{\Black{$5$}}}
    \Text(331,-17)[lb]{{\Black{$3$}}}
    \Text(217,-76)[lb]{{\Black{$k^{(3)}$}}}
    \Text(400,-76)[lb]{{\Black{$ik^{(1)}$}}}
    \Text(238,84)[lb]{{\Black{$k^{(0)}$}}}
       \Text(57,75)[lb]{{\Black{$k^{(2)}$}}}
  \end{picture}
%}
\end{center}
\caption{The 5-point gluonic amplitude, ${\cal
    A}_5(k_1,k_2,k_3,k_4,k_5)$ in the double-Regge region, evaluated on
  shell ($k^2_i =0$, $\sum_i k_i =0$) with all BDS invariants
  ($t^{[r]}_i <0$) space-like.}
\label{fig:5Regge.a}
\end{figure}
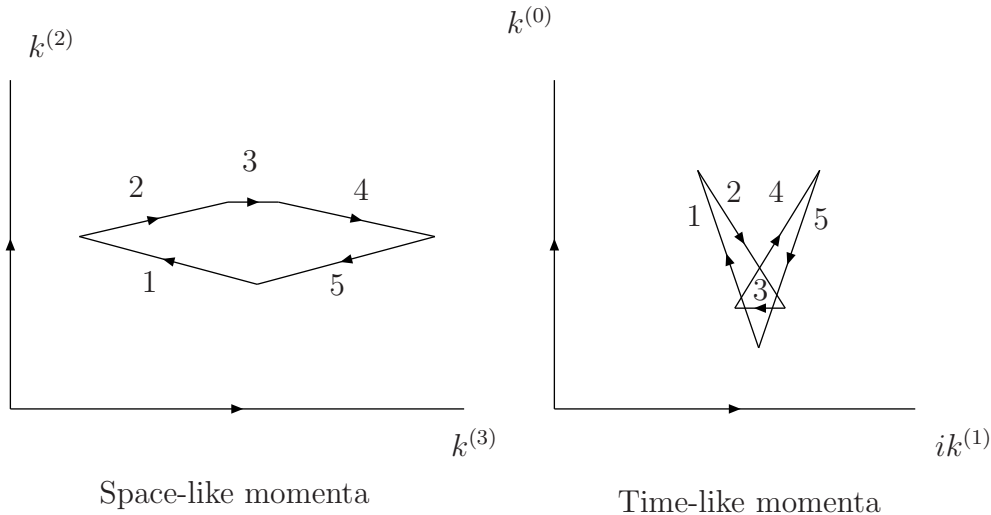

In this paper we focus on how to how to 
analytically continue these planar amplitudes  back to the physical region with momenta depicted in Fig. \ref{fig:multi_gluon.a}. This will be done in some details for the 4-point and 5-point amplitudes  in Sec. \ref{sec:4point} and Sec.~\ref{sec:sig_double} respectively. The continuation for general n-point amplitudes can in principle be done   by a suitable
generalization of the procedure described in Sec.~\ref{sec:sig_double}
for the 5-point amplitude.   Although this procedure provides a general construction to handle 6-point and
higher amplitudes where the non-linear constraints of the BDS invariants 
would otherwise be formidable, it has long been recognized that the simplicity of the multi-Regge amplitudes often masks the underlying analytic structure of the original  planar amplitudes \cite{Weis:1972ir}. While 
the full amplitude can be correctly continued to the physical region obeying all non-linear constraints and the $+  i \epsilon$  prescription for normal thresholds, this can not be done for the leading  Regge asymptotic term. 
In the Regge limit an alternate prescription requires relaxing the nonlinear Gram-determinant constraints during the course of the analytic continuation. For instance, for the a 6-point amplitude, each planar amplitude should be considered a function of $\Phi$, with a branch point singularity at $\Phi=0$. Depending on the color-ordering involved, the variable $\Phi$  can be continued to different points,
\be
\Phi \rightarrow 1, e^{-2\pi i }, e^{2\pi i}
\ee
by circling around the branch point at $\Phi=0$. Of the  8 independent color orderings,  the constraint $\Phi=1$ can be maintained for 6, but not for 2, in the multi-Regge region  \cite{Weis:1972ir}. In short the   ``on-shell'' constraints  and the ``multi-Regge limit''  do not  commute and one cannot make use of the simplified nonlinear Gram-determinant constraints in the course of analytic continuation for some set of BDS amplitudes.   One possible approach is to follow strictly using independent momentum components under $O(2,2)$ continuation. The conceptual advantage of this approach has recently been stressed by Nima Arkani-Hamed. In practice, this problem can be evade if we simply treat $\Phi$'s as independent variables until the physical region is reached. This procedure has been adopted  consistently in past in establishing analyticity and factorization for the total amplitude, i.e., after summing over all planar orderings, for flat-space string theory \cite{Weis:1972ir}.   We shall return to this point in Sec. 5.

\newpage
\section{Regge Behavior of 4-Point Function}
\label{sec:4point}

Let us review and contrast Regge properties for the
4-point amplitude in flat space string theory vs the BDS amplitude
for gluons. Our motivation is to strike  a cautionary note on comparing
traditional Regge behavior with the BDS amplitude. 

The 4-point amplitude has 3 Mandelstam invariants $s = - (k_1 + k_4)^2$,
$t =- (k_1 + k_2)^2$ and $u =-(k_1 + k_3)^2$, with $s+t+u=0$.  In the planar limit for
open strings in flat space, there are 3 planar amplitudes with
singularities for positive Mandelstam invariants s-t, t-u and s-u
corresponding to 3 independent permutations of the trace $Tr[1234]$,
$Tr[1243]$ and $Tr[1423]$ . To take the Regge limit while avoiding these singularities, in
this case discrete poles, we consider $s \rightarrow - \infty$,
($u\simeq -s $), with $t < 0$ for the s-t permutation.  The
``gluonic'' open string amplitude~\footnote{We suppress here
  the color and helicity dependence of the external gluon lines. Also the
6 extra dimensions of the super string are assumed to be compactified on a
6-d torus.} 
gives~\cite{Stieberger:2006bh}
\be
A(s,t)/{A}_{tree}(s,t) = \frac{\Gamma[2-\alpha(t)]\Gamma[2-\alpha(s)]}{\Gamma[3-\alpha(s) - \alpha(t)]}  = [1 - \alpha't - 
\alpha's] \int^1_0 dz z^{-\alpha(t) + 1} (1-z)^{-\alpha(s) +1}
\ee
with trajectory $\alpha(t) = 1+ \alpha' t $.
The tree-amplitude contains both $s$-channel and $t$-channel gluon
poles,  
\be
A_{tree}\sim \frac{s}{t} + \frac{t}{s}\;.
\ee
In the Regge limit $s\rightarrow -\infty$, the integral is dominated by
the region $z = O(-1/\alpha' s)$ and is easily computed 
\be
M_4(s,t)=A(s,t) /{ A}_{tree}  \simeq (-\alpha' s )^{\alpha(t)-1} \int^\infty_0 dy y^{-\alpha(t) + 1 } e^{- y} =  \Gamma[2- 
\alpha(t)] (-\alpha' s )^{\alpha(t)-1} 
\ee
No purely conformal
theory can have a leading Regge pole, because the trajectory function
requires a mass scale.  In the limit $\alpha'\rightarrow 0$, $M_4(s,t) \rightarrow 1$ as it must. Comparing this with 
the Regge limit of the
BDS amplitude, this scale must be provided by the IR cut-off.  In what follows, we will often introduce a notational 
simplification,
\be
\omega(t) = \alpha(t)-1\;, 
\ee
and this will be used for both flat-space string theory and for BDS.

The Reggeization of the gluon in non-supersymmetric Yang Mills
theories~\cite{Grisaru:1973vw,Grisaru:1974cf,Fadin:1975cb,Fadin:1996tb,Korchemskaya:1996je}, as well as 
supersymmetric Yang Mills
~\cite{Grisaru:1981ra,Grisaru:1982bi,Schnitzer:2007kh,Schnitzer:2007rn,Kotikov:2000pm},
has a long history. Consider the ${\cal N} =4$ SYM together with the BDS
conjecture for the corresponding planar on-shell 2-to-2 gluon scattering
amplitude, ${ A}_4(k_1 + k_2 \rightarrow - k_3 - k_4)$,
\be \label{eq:BDSfour}
M_4= A_4/{ A}_{tree} = { A}^2_{div}(s) { A}^2_{div}(t)
\; \E{\textstyle \frac{f(\lambda)}{8} \log^2(s/t) + \widetilde c(\lambda)}
\ee
where $\lambda = g^2 N$, $f(\lambda)$ is proportional to the cusp
anomalous dimension~\cite{Bern:2005iz,Alday:2007hr,Korchemsky:1985xj,
  Ivanov:1985np,Collins:1989bt,Magnea:1990zb,Catani:1998bh,Sterman:2002qn},
and $\tilde c(\lambda)$ is a constant. The Sudakov
form factor is 
\be
A_{div}(s) = \exp\left\{  -\frac{1}{16 }  f\left(\lambda\right) \log^2(-s/\mu^2) 
 +\left[\frac{1}{8 \epsilon}  f^{(-1)}\left(\lambda\right) + \frac{1}{4 }  g\left(\lambda\right)\right] \log(-s/\mu^2)  \right\}
\ee
up to an $s$- and $t$-independent divergent factor. Remarkably the cancellation
of the $\log^2(-s/\mu^2)$ and $\log^2(-t/\mu^2)$ in ${
  A}_{div}^2(s)$ and ${ A}_{div}^2(t)$ respectively with the
$\log^2(s/t)$ in Eq.~(\ref{eq:BDSfour}) immediately gives the Regge
amplitude \cite{Drummond:2007aua,Naculich:2007ub},
\be \label{eq:regge4}
M_4(s,t) =   \beta(t)   \left(\frac{- s}{\mu^2} \right)^{\alpha(t)-1} = \beta(t)   \left(\frac{- s}{\mu^2} \right)^{\omega(t)}
\ee
{\bf without} taking the Regge limit: $-s$ large at fixed $t$. The gluon trajectory function 
\be \label{eq:alpha}
\alpha(t) =  1+\omega(t)= 1 + \frac{1}{4 \epsilon} f^{(-1)}(\lambda) - \frac{1}{4} f(\lambda) \log(-t/
\mu^2) +\Half g(\lambda) +O(\epsilon)
\ee
depends on  $\mu$, the IR cut-off~\footnote{For the rest of the paper, we will mostly put $\mu^2=1$.}. 
With Regge residue 
\be
\beta(t) \equiv \gamma^2(t)=   {\rm constant}   \left(\frac{-t} {\mu^2  } \right)^{\omega(-\mu^2)} ,
 \label{eq:beta}
\ee
$M_4(s,t)$ is manifestly symmetric in $s\leftrightarrow t$. (See Fig. \ref{fig:4regge}.)

A moving Regge trajectory requires the scale breaking introduced by the IR cut-off playing the role of the Regge slope parameter $\alpha'$ in flat space string theory.  In this sense the Regge behavior of ${\cal N} = 4$ gluonic scattering is clearly a subtle affair.  Note that the trajectory (\ref{eq:alpha}), at $t$ large and positive goes as $ - \log(-t/\mu^2)$, turning complex rather than rising linearly. Unlike the flat-space string theory, there are no Regge recurrences at positive real $t$ and no scale for the slope $\alpha'$, consistent with a ${\cal N} =4$ conformal theory with no massive states. Conversely, in the deep Euclidean region where $t<0$, the trajectory is real and unbounded from below.

However due to the IR cut-off of a conformal theory there are some
unusual features to this Regge amplitude.  First the physical $J=1$
gluon pole at $t=0$ {\bf does not} lie on the trajectory. In fact the
trajectory is singular at $t=0$, presumably due to the multi-gluon
channel.  A more appropriate cut-off 
(as  explained in \cite{Naculich:2007ub,Grisaru:1981ra,Grisaru:1982bi}) 
would be to fix
the trajectory at $J=1$  for $t=0$, or use a ``Higgsing'' regularization scheme \cite{Grisaru:1973vw,Grisaru:1974cf} 
which 
consistently gives mass to the gluonic state, keeping it on the
trajectory and separated from the multi-gluon channel. All this is
just to remind ourselves that the conventional Reggeization of a field
should be considered in the context of a proper renormalizable field theory. To avoid
dependence on the IR cut-off it is better to restrict the Regge
hypothesis here to the combined limit with $s << t <<
0$. Alternatively we may take $d > 4-2 \epsilon$ with negative
$\epsilon$ which is IR finite, \cite{DelDuca:2008pj} and then the coupling is driven to zero
in the IR, there is no Regge behavior, but the gluon at $t=0$ does have
$J=1$ as it should.

Finally if we compare the flat space Regge limit with the Alday-Maldacena approach at strong coupling we see 
another important
contrast.  As demonstrated in Ref.~\cite{Brower:2006ea}, in flat space (and
strong coupling ${\cal N} =4$ closed strings) the Regge limit is a
result of the world sheet operator product expansion as the particle vertices, $V(k_i,z_i) = e^{ik X(z_i)}$, approach 
one another, which allows one to introduce a new on-shell vertex operator
\be
{\cal V}^{\pm}(k,z)  = (\partial_z X^\pm(z))^{1 + \alpha' t} e^{\mp k X(z)} 
\ee
for the emission and absorption of a Reggeon. This world sheet
approach naturally leads to the standard properties of flat space
string multi-Regge amplitudes. In the case of the 4-point BDS
amplitude, at least at strong coupling following the Alday-Maldacena
world sheet approach, there does not appear to be a similar result
because the Regge limit does not require such a limiting procedure for
the external operators.  This already underlines differences between
flat space string theory and MSYM.

\subsection{Analytic Continuation and 4-Point Signatured  Amplitude}

As an  illustration, let us consider the continuation of  4-point gluon Regge amplitude: $\gamma(t) (-s)^{\alpha(t)} \gamma(t)$.  We know the answer to analytic continuation so this is a practice problem doing it the hard way! The answer is to continue $ - s = e^{ - i(\pi-   \theta)}\; |s|$ on an arc in the upper-half s-plane (UHP), ($\theta$ from $ \pi$ to $0$), to get,
\be
\gamma(t) (-s)^{\alpha(t)} \gamma(t) \rightarrow \gamma(t) (s)^{\alpha(t)} e^{-i \pi \alpha(t)} \gamma(t) 
\ee
Now let's do it in light-cone co-ordinates:

\begin{figure}[h] 
\begin{center}
%\fcolorbox{white}{white}{
  \begin{picture}(203,194) (77,-18)
    \SetWidth{0.5}
    %\SetColor{Black}
    \Photon(125,74)(216,74){5.5}{6}
    \SetWidth{1.0}
    \Line(122,79)(122,140)
    \Line(217,81)(217,142)
    \Line(123,7)(123,68)
    \Line(217,5)(217,66)
    \Text(250,69)[lb]{{\Black{$\gamma(t)$}}}
    \SetWidth{0.5}
    \Vertex(217,73){6.4}
    \Vertex(123,75){6.4}
    \Text(77,68)[lb]{{\Black{$\gamma(t)$}}}
    \Text(216,160)[lb]{{\Black{$k_3$}}}
    \Text(119,159)[lb]{{\Black{$k_2$}}}
    \Text(168,10)[lb]{{\Black{$s$}}}
    \Text(123,-18)[lb]{{\Black{$k_1$}}}
    \Text(216,-14)[lb]{{\Black{$k_4$}}}
    \Text(169,92)[lb]{{\Black{$t$}}}
  \end{picture}
%}
\end{center}
\caption{The Regge limit for the  elastic {\bf planar} 4-point amplitude ${A}_4(s,t)$ with thresholds for $s \ge 0, t \ge 
0$ with a ``twisted'
Regge limit'' is $s \simeq -u \rightarrow - \infty$ and
$t < 0$.} 
\label{fig:4regge}
\end{figure}
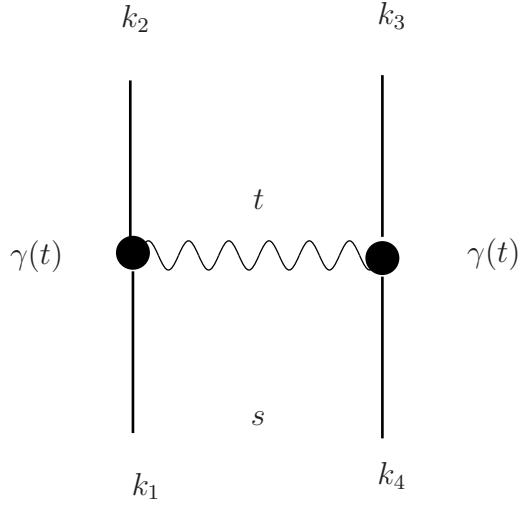

Start with $k_i = (k^+_i,k^-,k_\perp)$. The $A_4(s,t)$ BDS
amplitude has an Euclidean Regge limit for $u = -s -t \rightarrow
\infty, t<0$  or $u= - (k_2 + k_3)^2 \rightarrow \infty$, $t= -(k_1 +k_2)^2 < 0$. This  is satisfied by
\be
-k^+_1 =  k^+_2 \rightarrow  \infty \quad k^-_3 = - k^-_4 \rightarrow \infty 
\ee
To have these strict equalities (which is nice but not absolutely necessary)
we work in the brick-wall frame with $q^\perp = k^\perp_1 = k^\perp_2$ (and therefore $q^\perp = - k^\perp_3 = - k^
\perp_4$). So the 
conservation on E-p is exactly,
\be
k^\pm_1 + k^\pm_2 = - k^\pm_3 - k^\pm_4 = 0
\ee
so we can stay on the E-p and mass-shell $k^2_i =0$ with the continuation,
\be
k^\pm_1 \rightarrow e^{\pm i\theta} k^\pm_1 \quad,\quad k^\pm_2 \rightarrow e^{\pm i\theta} k^\pm_2 \quad,\quad
\ee
and $k^\pm_3,k^\pm_4$ unchanged,
and we see that  $t =- 2 k_1 k_2$ is not changed in the continuation, but
$s = - 2 k_1 k_4 \simeq  2 e^{i \theta} k^+_1 k^-_4$ is continued around the UHP as it should be!

Finally it is important to realize that more than one planar amplitude contributes to the same Regge limit. For example for the 4-point amplitude there are 3 distinct planar amplitude corresponding different cyclic orders of the external lines: $A(1,2,3,4)$, $A(2,1,3,4)$ and $A(1,3,2,4)$ with singularities for positive Mandelstam variables $s,t$;
$u,t$ and $s,u$ respectively.  The first two (with t-channel exchanges) contribute to the Regge limit $s \rightarrow \infty$ at fixed $t$. It is useful to introduce a variable $\eta = \pm 1$ to distinguish between $A_{\eta =1} = A(1,2,3,4)$ with singularities for possitive $s$ and $A_{\eta =- 1} = A(2,1,3,4)$ with singularities for negative $s \sim -u$. In the physical region, the leading Regge term is
\be
A_{\eta}(s,t) \sim (- \eta s)^{\alpha(t)} \;, 
\ee
a complex phase, $(-)^{\alpha(t)} = e^{\textstyle -i \pi \alpha(t)}$, for $\eta = 1$ and real for $\eta = -1$.  In general a non-degenerate Regge singularity in the $J$-plane contributes only to even or odd linear combination of
these two amplitudes,
\be
\widetilde A_{\sigma}(s,t) = A_{\eta =1}(s,t)  + \sigma A_{\eta = -1}(s,t) \sim (e^{-i \pi \alpha(t)}+\sigma ) s^{\alpha(t)}
\ee
distinguished by a quantum number referred to as ``signature'' ($\sigma = \pm$). Factorization only applies to Regge exchanges with definite signature. In open string theory, signature corresponds to state  even and odd  under
world sheet parity -- the eigenvalues of the twist operator $\Omega$. In the present context this operator, $\Omega$,  defines the signature factor,
\be
\xi_\sigma = 
 e^{-i\pi \alpha(t)} +\sigma \equiv \sqrt 2\; \Omega_{\sigma,\eta}\;  ( - \eta)^{\alpha(t)}
\ee
or applied to the amplitude, $\widetilde A_\sigma = \sqrt 2 \; \Omega_{\sigma\eta} A_\eta$,
where  $\Omega_{\sigma\eta} = [ (1 + \eta) + (1 - \eta) \sigma]/ 2\sqrt{2}$ or 
\be
\Omega_{\sigma\eta} = \frac{1}{\sqrt{2}} 
\left[\begin{array}{cc}
1 & \;\;\; 1\\
1 & -1\\
\end{array} \right]
\ee
with $(+,-)$ ordering of vectors indices of both $\sigma$ and $\eta$. Various  $\sqrt 2$ factors  have been inserted  so that  $\Omega^2 = 1$.

For definiteness, we shall express the signatured amplitudes  as 
\be
 \widetilde A_\sigma(s,t)= \widetilde \Pi_{\sigma}(t) \; s^{\alpha(t)}
\ee
where
\be
\widetilde \Pi_{\sigma}(t)= \xi_{\sigma}(t) \; \Gamma (t) \;.  \label{eq:pi}
\ee
 For flat-space string theory, we have 
 \be
 \Gamma(t)=\Gamma(1 -\alpha(t))  \label{eq:StringPropagator}
 \ee
  and, for BDS, 
  \be
  \Gamma(t)=\gamma^2(t) \left(\frac{ \mu^2}{-t}\right) \;. \label{eq:BDSPropagator}
 \ee
   Also, for future reference, note $\Omega^\dagger = \Omega$. It follows that one can  easily invert the process so that, in the physical region,
\be
A_\eta (s,t)=\frac{1}{\sqrt{2}} \; \Omega_{\eta \sigma}\; \widetilde A_\sigma(s,t)= \Gamma(t) \; (-\eta)^{\alpha(t)} \;s^{\alpha(t)} \;.
\ee

\newpage
\section{Analyticity of  the Double Regge Vertex}
\label{sec:constraints}

It has long been recognized that analyticity, when combined with other general principles such as unitarity, 
provides a powerful tool for constraining the allowed behavior for scattering amplitudes. For instance, one-loop  
MHV gluon amplitudes can be obtained using  unitarity techniques 
starting from tree-graphs. Conversely, given the BDS conjecture, analyticity and unitarity can in 
principle be used to test its validity.
 In this section, we discuss some aspects of analyticity constraints on
5-point amplitudes in various Regge limits. We contrast the properties
of the BDS amplitudes with the stringy expectations based on
flat-space open-string amplitudes. We first  focus on the BDS Reggeon-particle-Reggeon vertex, $G_2$, which,  
unlike the case of  flat-space string theory, has a rather special  analyticity structure in the $\kappa$ variable. We 
demonstrate that the BDS vertex, $G_{2,BDS}(t_1, t_2,\kappa_{12})$,  (see Fig. \ref{fig:regge5both}b, with $\kappa_{12}=s_1s_2/s$), obtained from 5-point  amplitude computed to 
$O(\epsilon^0)$, 
does not satisfy the Steinmann rules~\cite{Stapp:1971hh,Steinmann,Detar:1972nd,Tan:1972kr,Cahill:1973px}.  
This structure is shown to lead to an unusual behavior for the 5-point BDS amplitude in 
certain  limiting regions,\footnote{The  limit where $|s|>>|s_1|\rightarrow \infty$ and $|s_1|>>|s_2|$, with $s_2$ 
either large or fixed,   is referred to historically as the ``helicity-pole'' limit
~\cite{Detar:1971gn,Detar:1971dj,Detar:1972nd,Brower:1974yv}.   
For 5-point function, this lies outside the physical region. However, this is no longer the 
case for $n\geq 6$.} e.g., $|s|>>|s_1|>>|s_2|$, or $|s|>>|s_2|>>|s_1|$.  
The implication of this discussion for $n \geq 6 $ will be addressed   in Sec. \ref{sec:inclusive}.  In particular, we find the 
absence of a Regge contribution in the
  inclusive ``triple Regge'' limit for the 6-point function.

\subsection{Single- and Double-Regge Limits for 5-Gluon Amplitude}
\label{sec:sig_double}
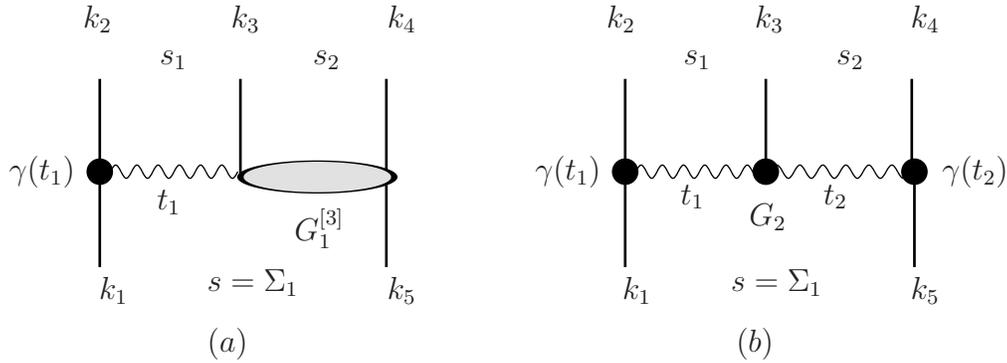
\begin{figure}[h!] 
\begin{center}
%\fcolorbox{white}{white}{
  \begin{picture}(380,131) (23,-53)
    \SetWidth{1.0}
 %   \SetColor{Black}
 \Line(305,20)(305,53)
    \SetWidth{0.5}
    \Vertex(305,18){5}
    \SetWidth{1.0}
   \Line(360,20)(360,53)
    \SetWidth{0.5}
    \Vertex(361,18){5}
    \Photon(304,18)(256,18){2.5}{6}
    \SetWidth{1.0}
    \Line(361,-18)(361,15)
    \SetWidth{0.5}
    \Photon(307,18)(359,18){2.5}{6}
    \Text(252,-32)[lb]{{\Black{$k_1$}}}
    \Text(361,-32)[lb]{{\Black{$k_5$}}}
    \Text(302,71)[lb]{{\Black{$k_3$}}} 
    \Text(361,71)[lb]{{\Black{$k_4$}}}
    \Text(333,57)[lb]{{\Black{$s_2$}}}
    \Text(373,12)[lb]{{\Black{$\gamma(t_2)$}}}
    \SetWidth{1.0}
    \Line(252,-18)(252,19)
   \Line(252,53)(252,20)
    \Text(246,71)[lb]{{\Black{$k_2$}}}
    \Text(275,57)[lb]{{\Black{$s_1$}}}
    \Text(219,12)[lb]{{\Black{$\gamma(t_1)$}}}
    \Text(300,-4)[lb]{{\Black{$G_2$}}}
    \Text(274,4)[lb]{{\Black{$t_1$}}}
    \Text(328,4)[lb]{{\Black{$t_2$}}}
    \SetWidth{0.5}
    \Vertex(252,18){5}
    \Text(293,-29)[lb]{{\Black{$s = \Sigma_1$}}}
    \SetWidth{1.0}

    \Line(54,53)(54,20)
 \Line(107,17)(107,53) 
   \Line(162,17)(162,53)
    \SetWidth{0.5}
    \Photon(106,18)(58,18){2.5}{6}
    \SetWidth{1.0}
        \Line(162,-18)(162,13)
            \Line(54,-18)(54,15)
    \Text(54,-32)[lb]{{\Black{$k_1$}}}
    \Text(163,-32)[lb]{{\Black{$k_5$}}}
    \Text(104,71)[lb]{{\Black{$k_3$}}} 
    \Text(163,71)[lb]{{\Black{$k_4$}}}
    \Text(135,57)[lb]{{\Black{$s_2$}}}
   
    \Text(48,71)[lb]{{\Black{$k_2$}}}
    \Text(77,57)[lb]{{\Black{$s_1$}}}
    \Text(20,12)[lb]{{\Black{$\gamma(t_1)$}}}
    \Text(128,-11)[lb]{{\Black{$G_1^{[3]}$}}}
    \Text(76,2)[lb]{{\Black{$t_1$}}}
    \SetWidth{0.5}
    \Vertex(54,18){5}
    \Text(95,-29)[lb]{{\Black{$s = \Sigma_1$}}}
    \GOval(136,16)(6,29)(0){0.882}
    
        \Text(95,-53)[lb]{{\Black{$(a)$}}}
            \Text(295,-53)[lb]{{\Black{$(b)$}}}
  \end{picture}
\end{center}
\caption{Regge limits for 5-point amplitude. On the left, the single Regge limit
factorizes defining a new single Regge 3-particle vertex, $G_1^{[3]}(s_2,t_1,t_2,\kappa_{12})$ and on the right, the  double Regge limit
defines a new two-Reggeon vertex, $G_2(t_1,t_2,\kappa_{12})$.}
\label{fig:regge5both}
\end{figure}

In Ref. \cite{Brower:2008nm}, it was demonstrated how
BDS n-point amplitudes, $n\geq 5$, led to Regge behavior in various
Regge limits taken in the Euclidean region. For
example, a color-ordered 5-point amplitude, aside from lower-order
terms coming from the asymptotic expansion of the tree-amplitude,
again has the remarkable property observed in the 4-point function
of an {\tt exact single- and double-Regge form},
\bea
M_5(s_1, s_2, s, t_1, t_2)&=&A_5/A_{tree} = \gamma(t_1) (-s_1/\mu^2)^{\omega_1} G_1^{[3]}(s_2,t_1,t_2,\kappa_{12})
\nonumber\\
&=& \gamma(t_1) (-s_1/\mu^2)^{\omega_1} G_2(t_1, t_2,\kappa_{12}) (-s_2/\mu^2)^{\omega_2} \gamma(t_2) \; ,
\label{eq;exact5}
\eea
where $ \alpha(t_1)=\alpha_1=1+\omega_1$, $\alpha(t_2)=\alpha_2=1+\omega_2$ and
$\kappa=s/s_1s_2$.  The two expressions can be interpreted as 
either the single- or double-Regge limit as illustrated in
Fig. \ref{fig:regge5both}. The expression is in fact symmetric under
the interchange $s_1 \leftrightarrow s_2$. In addition to the Regge
trajectory, $\alpha(t)$, two particle vertex, $\gamma(t)$, one
encounters for the first time in the 5-point function the
Reggeon-particle--Reggeon vertex, $G_2(t_1,  t_2,\kappa_{12})$, and the
three particle Reggeon vertex, $G_1^{[3]}(s_2,t_1,t_2,\kappa_{12})$),
which must re-occur in higher point function if factorization is valid. We
discuss below   the analytic structure of vertices, in particularly
the Reggeon-particle--Reggeon vertex, $G_2$, in the $\kappa$ variable, and the continuation to  the
physical region for the  color-ordered $(12345)$, Fig. \ref{fig:regge5both}.  The  issue of physical region  {\bf 
factorization} in the linear
multi-Regge limits for $n\geq 5$ in summing over all  color orderings will be discussed in Sec. \ref{sec:5plus}.

The particular color-ordered 5-point amplitude considered in
Eq.~\ref{eq;exact5} is real in the Euclidean region, $s_1,s_2, s, t_1,
t_2<0$. The physical region where $s_1,s_2,s>0$ and $t_1,t_2<0$ is
reached by analytic continuation of the {\bf on-shell} scattering
amplitude.  The procedure follows the strategy outlined in
Sec.~\ref{sec:analytic}.

On shell, there are 5 independent BDS invariants, $s, s_1, s_2, t_1, t_2$.  As
we mentioned in Sec. \ref{sec:analytic}, the analytic continuation from the deep Euclidean
region to the physical region can best be understood in terms of
light-cone momentum components a la Alday and Maldacena~\cite{Alday:2007he}, and 
Arkani-Hamed and Kaplan~\cite{ArkaniHamed:2008yf}. Let us first provide an estimate. For $s_1, s_2, s
\rightarrow - \infty$, consider the Euclidean region where
\be
-  k_1^+\sim k^+_2 \rightarrow  \infty\; ,\quad  - k_5^-\sim  k^-_4 \rightarrow - \infty\;, 
\quad  {\rm with} \quad  k^-_3  < 0\; , \quad k^+_3 > 0 
\ee
However, due to  the on-shell condition, $k_3^2=0$,  we must have
\be
k^+_3 k^-_3 = (k^\perp_3)^2 < 0  \quad, \quad \kappa = s_1 s_2/s  \simeq (k^\perp_3)^2 < 0
\ee
We can follow  what \cite{Alday:2007he}
 and \cite{ArkaniHamed:2008yf} did by continuing to a $(2,2)$ metric
with $k^1_3 = i |k^1_3|$ to allow for this. We can be in a frame with all components of $k^2_3 = O(\mu)$.   Now the 
trick is almost the same. Continue
\be
k^\pm_1 \rightarrow e^{\pm i\theta} k^\pm_1 \quad,\quad k^\pm_2 \rightarrow e^{\pm i\theta} k^\pm_2 \quad,\quad
\ee
to get to $s>0, s_1 >0$ and $\kappa < 0$. Next continue,
\be
k^1_3 = e^{-i \theta/2} i |k^1_3|
\ee 
so that $\kappa$ goes to $\kappa > 0$ through the upper-half plane 
(UHP).\footnote{However, there are some details: we need to guarantee that $t_1<0 $ and $t_2<0$. 
To do this properly, we should follow the procedure outlined in Sec.~\ref{sec:analytic}. We will not go 
into those details here. We also note that we have reversed our convention, from $\kappa=s/s_1s_2$ used in Ref.  
\cite{Brower:2008nm}, to $\kappa=s_1s_2/s$. }

From Ref. \cite{Brower:2008nm}, we have found that the
Reggeon-Reggeon-Gluon vertex, $G_2(t_1,t_2, \kappa)$ takes on the
following dependence on $\kappa$,
\be
G_2= C(t_1,t_2)e^{A  \log^2 (- \kappa/\mu^2) - B(t_1,t_2)\log (- \kappa/\mu^2) }
\label{eq:G2BDS}
\ee
where $A=-f(\lambda)/16$, $B(t_1,t_2)=(\omega(t_1)+\omega(t_2))/2$,
and 
\be
C(t_1,t_2)  =  \mbox{const} \;  (-t_1/\mu^2)^{-\Half \omega(-\mu^2)}(-t_2/\mu^2)^{-\Half
\omega(-\mu^2)}  
\times %A^2_{div}(t_1) \;  A^2_{div} (t_2)
\exp[f(\lambda) \log^2(t_1t_2/\mu^4)/16] 
\ee
%
%with
%
%\be
%A_{div}(t) = \exp\left\{  -\frac{1}{16 }  f\left(\lambda\right) \log^2(-t/\mu^2) 
% +\left[\frac{1}{8 \epsilon}  f^{(-1)}\left(\lambda\right) + \frac{1}{4 }  g\left(\lambda\right)\right] \log(-t/
%\mu^2)  \right\}
% \label{eq:div}
%\ee
%
As expected, with $t_1,
t_2<0$ fixed, $G_2$ is real in the Euclidean region where $-\infty<
\kappa<0$.  The analytic structure can be specified by a cut-plane,
with a branch cut drawn along the positive axis, from $\kappa=0$ to
$\kappa=+\infty$.  For color-ordering $(12345)$, as indicated in
Fig. \ref{fig:regge5both}, the physical region, where $s_1, s_2, s>0$,
can be reached from the Euclidean region via the procedure outlined
earlier. This corresponds to continuing $\kappa$ to the positive axis,
$0<\kappa<\infty$, via the upper-half plane, i.e., $\kappa\rightarrow
|\kappa| +i\epsilon$ and $\log (- \kappa/\mu^2) \rightarrow \log
(|\kappa|/\mu^2) - i\pi$.  That is, in the physical region, the color-ordered
amplitude is
\bea
M_5&\simeq&  \gamma_1\gamma_2 \; (-s_1)^{\omega_1} (-s_2)^{\omega_2} \; G_2(t_1,t_2,|\kappa|+i\epsilon) 
\nn
&=& \gamma_1\gamma_2 \;  e^{-i\pi\omega_1 -i\pi \omega_2} s_1^{\omega_1} 
s_2^{\omega_2}\;   G_2(t_1,t_2,|\kappa|+i\epsilon)  \label{eq:5pointlinearMR}
\eea
Note that, in addition to the product of two Regge phase factors, $G_2$ is also complex in the physical region.

\subsection{Double-Regge Representation  and Analyticity}

The singularity of $G_2(t_1,t_2,\kappa)$ in $\kappa$ is a reflection
of the singularities in $s_1, s_2, s$, for each color-ordered
amplitude. For the ordering $(1,2,3,4,5)$, $M_5$ has has right-hand branch cuts in $s$, $s_1$, and $s_2$.  From (\ref{eq:5pointlinearMR}), its  discontinuity in $s$ comes entirely from the vertex $G_2$. On the other hand,  the discontinuity in $s_1$ is more involved,  receiving  contributions from both the Regge factor, $(-s_1)^{\omega_1}$ and the vertex $G_2$.   The same applies for the discontinuity in $s_2$.    Before  exploring  the unusual singularity structure of the BDS
vertex, Eq. (\ref{eq:G2BDS}), it is instructive to contrast it with
that for the flat-space string theory. 

The 5-gluon super string MHV amplitude can be written 
\bea
M_{5,string}&=& \frac{A_{5, string}}{A_{tree}} \\
&=& \int_0^1 dx \int_0^1 dy x^{- \alpha' t_1} y^{ - \alpha' t_2} (1-x)^{ -\alpha' s_1} (1-y)^{- \alpha' s_2} (1 - xy)^{ - 
\alpha' (s - s_1 - s_2)} K(x,y) \nonumber
\label{eq:A5String}
\eea
where $ K =\alpha'^2\{ t_1 t_2/xy + \half [t_1 s_1 + s_2 t_2 - s (t_1
+ t_2) -s_1 s_2 + 2 i \epsilon_{\mu\nu\sigma\lambda} k^\mu_1 k^\nu_2
k^\sigma_3 k^\lambda_4]/(1 - xy) \}$.  By holding all invariants
Euclidean, one again can verify that $M_{5,string}\rightarrow 1$ when
$\alpha'\rightarrow 0$.  Using the technique of Vertex Operator,
${\cal V}$, introduced by Brower, Polchinski, Strassler and
Tan,~\cite{Brower:2006ea} or more directly from (\ref{eq:A5String}),
one finds for the double Regge~
limit~\cite{Detar:1971dj,Detar:1972nd,Brower:1974yv}
\be
G_{2,string}(t_1,t_2,\kappa) \sim \kappa^{-1}(-t_1- t_2 + \cdots)  \int_0^\infty dx_1 \int_0^\infty dx_2 x_1^{-\omega_1} x_2^{-\omega_2 }  
e^{-x_1-x_2 + (1 /\alpha'\kappa)x_1x_2}\label{eq:G2String2}
\ee
Here we have used the same notation  $\omega_1= \alpha_1-1$ and $\omega_2=\alpha_2-1$ appropriate for a 
linear Regge trajectory with intercept 1.  Note that in $G_{2,string}$ both integrals converge for $\kappa<0$, thus 
defining an analytic function which is real for $\kappa<0$. The function for $\kappa>0$ must be defined by analytic 
continuation, and one finds  a branch cut from $\kappa=0$ to $\kappa=+\infty$. This cut-plane analytic  structure is 
shared by both the flat-space string theory and the BDS vertex,  (\ref{eq:G2BDS}). However, they differ significantly 
in the nature of singularity in $\kappa$.

It can be shown that (\ref{eq:G2String2}) can be expressed as
\be
G_{2,string}(t_1, t_2,\kappa)= (- \kappa)^{-\omega_2}V_1(t_1,t_2, \kappa) + (- \kappa)^{-\omega_1} V_2(t_1,t_2, 
\kappa) \label{eq:G2String}
\ee
where $V_1$ and $V_2$  each admits a power series expansion in $\kappa$. One finds that  $V_1$ and $V_2$  
are {\bf single-valued} for $-\infty<\kappa< \infty$, thus remaining {\bf real} for $\kappa>0$.  In fact,  $V_1$ and 
$V_2$ are {\bf entire functions} of  $\kappa$.
 It follows that a double-Regge expansion leads to a decomposition for the five-point amplitude as a sum of two 
terms, 
\be
M_{5,string} \sim \gamma(t_1)\left[
  (-s)^{\omega_2} (-s_1)^{\omega_1-\omega_2} V_1(t_1,t_2,\kappa)  +  (-s)^{\omega_1} (-s_2)^{\omega_2-
\omega_1} V_2(t_1,t_2,\kappa) \right] \gamma(t_2) \;.  \label{eq:V1V2}
\ee
 The first term has right-hand cuts  in  invariants $s$ and $s_1$ and has no singularity in $s_2$, since $V_1$ is analytic  in $\kappa $. Similarly, the second term has  
singularities in $s$ and $s_2$, but not in $s_1$.

As one moves from the Euclidean region into the physical region where $s_1,s_2, 
s>0$ and $t_1,t_2<0$, the amplitude becomes  complex-valued. Since $V_1$ and $V_2$ are entire in $\kappa$, it 
follows that there is a separation in the singularity structure in  invariants $s_1$ and $s_2$, with
\bea
{\rm Disc_{s_1}} \; M_{5,string} (s,s_1,s_2,t_1,t_2)&\simeq &  2 i \gamma_1\gamma_2 \sin{\pi(\omega_2-
\omega_1)} (-s)^{\omega_2} s_1^{\omega_1-\omega_2} V_1(t_1,t_2,\kappa)  ,\nonumber\\
{\rm Disc_{s_2}}\; M_{5,string}(s,s_1,s_2,t_1,t_2)&\simeq& 2 i  \gamma_1\gamma_2 \sin{\pi(\omega_1-\omega_2)}  
(-s)^{\omega_1} s_2^{\omega_2-\omega_1} V_2(t_1,t_2,\kappa) . \nn \label{singlediscString}
\eea
 In contrast, both terms in (\ref{eq:V1V2}) contribute to the discontinuity in s,
\bea
&&{\rm Disc_{s}} \; M_{5,string} \\
&&\simeq  -2i  \left( \sin{\pi\omega_2} s^{\omega_2} (-s_1)^{\omega_1-\omega_2} 
V_1(t_1,t_2,\kappa)  
+  \sin{\pi\omega_1} s^{\omega_1} (-s_2)^{\omega_2-\omega_1} V_2(t_1,t_2,\kappa) \right )  .  \nonumber 
\label{singlediscString2}
\eea

It is well-known that, given its discontinuity, an analytic function can be re-constructed through a dispersion relation. 
Similarly, given ${\rm Disc_{s_1}} M_{5} $,  ${\rm Disc_{s_2}}  M_{5} $, and ${\rm Disc_{s}} M_{5} $, it is in principle 
possible to reconstruct the full amplitude through repeated dispersion integrals. It is therefore useful to refer to  
(\ref{eq:V1V2}) as a ``dispersive decomposition'' for the 2-to-3 amplitude in the Regge limit.

Let us now return to the BDS vertex, Eq. (\ref{eq:G2BDS}).  Given  $G_2(t_1,t_2,\kappa)$, as a complex number, it is always  possible  to express
this vertex as a sum of two terms
\be
G_2(t_1,t_2,\kappa) =(- \kappa)^{-\omega_2}  C_1(t_1,t_2, \kappa)+  (- \kappa)^{-\omega_1} C_2(t_1,t_2, \kappa)   
\label{eq:G2Lipatov}
\ee
and
\be
M_{5,BDS} \sim \gamma(t_1)\left[
  (-s)^{\omega_2} (-s_1)^{\omega_1-\omega_2} C_1(t_1,t_2,\kappa)  +  (-s)^{\omega_1} (-s_2)^{\omega_2-
\omega_1} C_2(t_1,t_2,\kappa) \right] \gamma(t_2) \;.  \label{eq:C1C2}
\ee
By demanding that $C_2$ and $C_1$ be {\it real in the physical region} where $\kappa= |\kappa| + i\epsilon$, these two coefficients can  be expressed  in terms of the magnitude, $|G_2|$, and its phase, $\phi$, 
\be
C_1(t_1,t_2,\kappa) = \kappa^{\omega_2} |G_2| \frac{\sin(\pi \omega_1-\phi)}{\sin\pi(\omega_2-\omega_1)}\;\; , \quad \quad 
C_2(t_1,t_2,\kappa) = \kappa^{\omega_1} |G_2| \frac{\sin(\pi \omega_2-\phi)}{\sin\pi(\omega_1-\omega_2)}
\ee

However, it is important to examine the analytic structure of  $C_1(t_1,t_2,\kappa)$ and $C_2(t_1,t_2,\kappa)$ in  $
\kappa$. The key difference between $G_{2,string}(t_1, t_2,\kappa)$, the BDS vertex   $G_2$, (\ref{eq:G2BDS}), is the presence of the $\log^2(-\kappa/\mu^2)$ factor in the exponent, leading to $M_5 \sim (-\kappa)^{A\log(-\kappa/\mu^2)}$ at $\kappa= 0$.    
 It is easy to verify that $C_1(t_1,t_2,\kappa)$ and $C_2(t_1,t_2,\kappa)$  contain branch points both at $\kappa=0$ and
$\kappa=\infty$.    It follows that, unlike the corresponding functions $V_1(t_1,t_2,\kappa)$ and $V_2(t_1,t_2,\kappa)
$ for flat-space string theory,  $C_1(t_1,t_2,\kappa)$ and $C_2(t_1,t_2,\kappa)$ are
not single-valued over the positive $\kappa$-axis. Since both $C_1$ and $C_2$ contain a branch point at $
\kappa=0$, both terms will contribute to discontinuities in $s_1$, $s_2$ and $s$. That is,  with $C_1$ and $C_2$ replacing $V_1$ and $V_2$, (\ref{singlediscString}) no longer holds for $M_{5,BDS}$. In particular, one cannot associate 
${\rm Disc_{s_1}}\; M_{5,BDS}$ with  that from the $C_1$ and ${\rm Disc_{s_1}}\; M_{5,BDS}$  with that from $C_2$, as is the case for the flat-space string theory.  

\subsection{Analyticity and Unitarity}

To appreciate the importance of the discussion above, it is
useful to briefly review the constraints coming from enforcing
analyticity and unitarity. It is well-known that the total imaginary
part of an amplitude in the physical region can be interpreted as the
sum of discontinuities. On the other hand, through unitarity, each
discontinuity can be expressed as a product of other amplitudes. These
general relations, with appropriate qualifications, can also be applied
to color-ordered amplitudes.
Consider the color-ordered 4-point amplitude, $A(s,t)$, with color ordering (1234) in the physical region where 
$s>0$ and $t<0$. From 2-to-2 unitarity, the s-channel discontinuity can be expressed as a sum, with contribution 
from allowed planar multi-particle intermediate states
in the s-channel. That is, each contribution can be associated with an allowed re-scattering process.

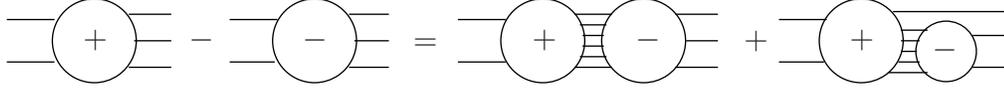
\begin{figure}[bthp]

\begin{center}
%\fcolorbox{white}{white}{
  \begin{picture}(409,60) (0,0)
    \SetWidth{0.5}
%    \SetColor{Black}
    %
\Line(44,26)(26,26)
  \Line(44,10)(26,10)
      \CArc(59,18)(15.81,305,665)
  \Text(55,14.5)[lb]{{\Black{$+$}}}
      \Line(72,28)(88,28)
    \Line(74,18)(88,18)
    \Line(72,8)(88,8)
       \Text(95,14.5)[lb]{{\Black{$-$}}}
         \Line(128,26)(110,26)
    \Line(127,10)(110,10)
     \CArc(142,18)(15.81,305,665)  
        \Text(138,14.5)[lb]{{\Black{$-$}}}
     \Line(155,28)(170,28)
    \Line(157,18)(170,18)
    \Line(155,8)(170,8)
      \Text(180,14.5)[lb]{{\Black{$=$}}}
        \Line(214,26)(196,26)
    \Line(214,10)(196,10)
 \CArc(228,18)(15.81,305,665)
         \Text(225,14.5)[lb]{{{$+$}}}
           \Line(240,28)(254,28)
            \Line(242,24)(252,24)
    \Line(243,20)(251,20)
    \Line(243,16)(251,16)
    \Line(242,12)(252,12)
    \Line(240,8)(254,8)
         \CArc(266,18)(15.81,305,665)
         \Text(264,14.5)[lb]{{\Black{$-$}}}
  \Line(278,28)(294,28)
  \Line(282,18)(294,18)
 \Line(278,8)(294,8)
     \Text(305,14)[lb]{{\Black{$+$}}}
        \Line(317,26)(335,26)
  \Line(317,10)(334,10)
      \CArc(348,18)(15.81,305,665)
     \Text(345,14.5)[lb]{{\Black{$+$}}}
         \Line(363,22)(373,22)
    \Line(363,18)(370,18)
    \Line(363,14)(369,14)
     \Line(362,10)(370,10)
     \Line(358,6)(373,6)
         \Line(360,29)(402,29)
         \CArc(380,14)(11.4,285,645)
            \Text(376.25,10.5)[lb]{{\Black{$-$}}}
    \Line(390,20)(402,20)
    \Line(390,8)(402,8)

  \end{picture}
% }
\end{center}
\caption{2-to-3 unitarity having two types of discontinuities.}
\label{fig:2to3unitarity}
\end{figure}

For 2-to-n  amplitudes, $n>2$,   there are many different re-scattering processes allowed, leading to discontinuities 
in various different invariants.  For instance, for a 2-to-3 process, $a+b\rightarrow c_1+c_2+c_3$, the unitarity 
condition in the physical region can be represented schematically by Fig. \ref{fig:2to3unitarity}.
Each term on the right can again be associated with a physically
realizable re-scattering process.  There are now two types of
discontinuities.  The first term represents the discontinuity in the
total energy invariant, $s_{ab}=-(p_a+p_b)^2$. The second is a sum of
three separate terms, each represents the discontinuity in one of
three sub-energy invariants, $s_{ij}=-(p_{c_i}+p_{c_j})^2$. Although
these discontinuities co-exist in the physical region, each can be
extracted by taking the imaginary part of the amplitude by first
holding all other invariants in the Euclidean region and then
returning back to the physical region by analytic continuation.  As illustrated by the flat-space string amplitude
$A_{5,string}$ for color-ordering $(12345)$, in a
double-Regge expansion, (\ref{eq:V1V2}), we have discontinuities in
invariants $s_1,s_2$ and $s$, given by (\ref{singlediscString}) and
(\ref{singlediscString2}) respectively.

Clearly, for $n>4$, there can be simultaneous discontinuities in
several invariants in the physical region.  However, one must
distinguish between compatible and overlapping invariants.  Since each
discontinuity can be identified with an allowed ``re-scattering''
process, a simultaneous discontinuity can exist in the physical region
only for compatible invariants, e.g., simultaneous in $s$ and one of
the sub-energy variable. For our color-ordered amplitude, (\ref{eq:A5String}), the allowed pairs are $(s,s_1)$ and $
(s,s_2)$. For flat-space string theory, from either (\ref{singlediscString}) or (\ref{singlediscString2}), the associated double-discontinuities are given by 
\bea
{\rm Disc_s\;Disc_{s_1}} \; M_{5,string} (s,s_1,s_2,t_1,t_2)&\sim & \sin {\pi \omega_2 }\sin{\pi(\omega_1-\omega_2)} 
s^{\omega_2} s_1^{\omega_1-\omega_2} V_1(t_1,t_2,\kappa)  ,\nonumber\\
{\rm Disc_s\;Disc_{s_2}}\; M_{5,string}(s,s_1,s_2,t_1,t_2)&\sim& \sin {\pi \omega_1 }\sin{\pi(\omega_2-\omega_1)}  
s^{\omega_1} s_2^{\omega_2-\omega_1} V_2(t_1,t_2,\kappa) . \nn  \label{doublediscString}
\eea

 In contrast, there cannot be simultaneous discontinuities for overlapping invariants, e.g., the pair $(s_1,s_2)$.  The 
double-discontinuities in this pair of overlapping invariants must vanish since it would not correspond to an allowed 
re-scattering  process. 
For flat-space string theory, it follows from (\ref{singlediscString})  
\be
{\rm Disc_{s_1} Disc_{s_2}} \; M_{5,string}(s,s_1,s_2,t_1,t_2) =0 \; ,  \label{Steinmann}
\ee
which  is often referred to as the Steinmann relation
~\cite{Stapp:1971hh,Steinmann,Detar:1972nd,Tan:1972kr,Cahill:1973px}.  
The Steinmann relation has played an important role historically in establishing inclusive distributions as discontinuities and in understanding the Mueller-Regge hypothesis \cite{Stapp:1971hh,Detar:1972nd,Tan:1972kr}. The 
traditional proof for the Steinmann relation relies on having a mass gap,  e.g., \cite{Stapp:1971hh}. 
For a theory with a mass gap, double discontinuities in overlapping singularities are associated with higher order 
Landau singularities, not normal thresholds, and they vanish in the physical 
regions~\cite{Stapp:1971hh,Tan:1972kr}.

However, for 5-point BDS amplitude, $M_{5,BDS}$, the singularities of  $C_1$ and $C_2$ at $\kappa=0$ will contribute to discontinuities in $s_1$ and $s_2$. Expressing $M_{5,BDS}$  as a function of independent variables $(s,s_1,s_2, t_1, t_2)$,  in place of $\kappa$, it follows from  (\ref{eq:G2BDS}) that $M_{5,BDS}$  will contain  in the exponent a term of the form  $\sim \log(-s_1)\log(-s_2)$. Such a term  clearly leads to
\bea
 {\rm Disc_{s_1}Disc_{s_2}}\; e^{\left[ 2 A \log(-s_1)\log(-s_2) +\cdots\cdots\right]} &  \ne & 0 \; , \label{eq:Steinmann-BDS}
\eea
\ignore{
\bea
{\rm Disc_{s_1}Disc_{s_2}}\; M_{5,BDS}(s,s_1,s_2,t_1,t_2)&  \ne & 0 \; , \label{eq:Steinmann-BDS}
\eea
}
where this double discontinuity is taken with $s, t_1, t_2$ fixed. The remainder  in the exponent includes quadratic terms,  $- 2A [ \log(-s_1)\log(-s)  +  \log(-s_2)\log(-s)]$, as well as terms   linear in $\log (-s)$ , $\log (- s_1)$ and $\log(-s_2)$. That is,  the BDS 5-point amplitude does not share the simple separability
property exhibited by (\ref{singlediscString}), (\ref{doublediscString}) and (\ref{Steinmann}) for flat space string  theory.

It is tempting to identify the  difficulty noted above  as  a 
manifestation of massless theory where  Laudau singularities  coalesce. 
This issue has also been commented on in  a recently  updated version of \cite{Bartels:2008ce}, where it 
has been argued  that failure of the 
Steinmann relation would  lead to  ``gluon instability''~\footnote{More recently \cite{Lipatov:2009nt}, it has been emphasized  that n-point BDS amplitudes violate Steinmann rules for $n\geq	 6$. However, it has been argued that this violation is due to a different mechanism, unrelated to the issue of the presence of massless particles. Here we are concerned with 5-point BDS amplitude.}.  Since one could treat a massless YM theory as the zero mass 
limit of a non-abelian Higgs model where the gauge bosons are massive  \cite{Grisaru:1973vw,Grisaru:1974cf}, and since 
the Steinmann rule is expected to hold before taking the massless limit, the difficulty of the singularity at 
$\kappa=0$ noted by us  should be traceable to how the limit is approached~\footnote{We would like to thank Prof. Lipatov for sharing his insight on this issue with us. In such an approach, the vertex $G_2$, will likely have further singularity in $\kappa$, e.g.,  at $\kappa_1<0$, in addition to the branch point at $\kappa_0=0$. Steinmann relation holds only when the discontinuity is taken across $\kappa_0$, but not across $\kappa_1$.  In the massless limit, as we pointed out above, these singularities coincide, preventing one from verifying the Steinmann relation by analytic continuity in $\kappa$.}.  We will return to comment on this issue 
further in Sec. \ref{sec:discussion}.

It is  worth noting that this analytic representation for 5-point gluon amplitudes,  (\ref{eq:G2Lipatov}), has been studied previously.  It was emphasized in  \cite{Fadin:1993wh} that, to one-loop,  both $C_1$ and $C_2$ are real  in the physical region,  as expected.
However, our emphasis here is on the analytic structure  in $\kappa$. In particular, we stress that both $C_1$ and $C_2$ are singular at $\kappa=0$, leading to (\ref{eq:Steinmann-BDS}).  
 It has been suggested in  \cite{Bartels:2008ce}, (v.4),   that the vanishing of overlapping discontinuity in $(s_1, s_2)$, ${\rm Disc_{s_1}Disc_{s_2}}\; M_{5,BDS}(s,s_1,s_2,t_1,t_2)=0$,  can be maintained by not identifying singularities of $C_1$ and $C_2$ at $\kappa=0$ with singularities at $s_1=0$ and $s_2=0$.  That is, $Disc_{s_1}Disc_{s_2}\; M_{5,BDS}(s,s_1,s_2,t_1,t_2)$ should be taken at fixed $\kappa\neq 0$.  This suggested procedure differs from that used here, which is more conventional, and is one normally used for analyzing singularity structure for flat-space string amplitudes.
Clearly,  this is important issue which deserves further investigation.  

 We stress here, although (\ref{eq:Steinmann-BDS})  follows from the  singularity at $\kappa=0$, our observation on  the Steinmann rule violation  can  be attributed  directly to  the  term $\sim \log(-s_1)\log(-s_2)$, coming from 1-loop contribution to $\log M_5$.  When expressed in terms of the $\kappa$ variable, this leads to the $\log^2(-\kappa/\mu^2)$ term in the exponent in (\ref{eq:G2BDS}).   Therefore, at least to 1-loop, one can not avoid identifying singularity at $\kappa=0$  with singularities in $s_1$ and $s_2$ respectively~\footnote{It is well-known that one-loop gluon amplitudes can be re-constructed via a cut-unitarity procedure for box-diagram. The general procedure of cut-unitarity should in principle work also for all orders. It has long been recognized that the analytic properties of multiparticle amplitudes in a massless theory are complicated, even in the so-called ``multi-Regge kinematic" (MRK) region. However, it has been demonstrated that, in the MRK region, by confining to the ``next-to-leading logarithmic approximation", (NLLA), certain analytic simplification can be achieved, which can be cast in the form of a set of bootstrap conditions and prescription for taking discontinuities has been given, consistent with the Steinmann relations~\cite{Fadin:2006bj,Bogdan:2006af}.  It remains a challenge on demonstrating how these could be realized beyond NLLA and on how they apply to the BDS ansatz which is supposed to be exact to all order. Nevertheless, it has been stressed by some that ``there is no doubt in the fulfillment of the Steinmann relation in QCD and in supersymmetric gauge models". Our concern here relates to how to properly interpret a simultaneous discontinuity in overlapping invariants bordering  the physical region and the overall singularity structure for 5-point BDS amplitude, as exhibited by Eq. (\ref{eq:Steinmann-BDS}).}.  A practical difficulty of this observation manifests itself for the 5- and higher-point amplitudes in the helicity-pole limit \cite{Detar:1971gn,Detar:1971dj,Detar:1972nd}. Unlike the case of flat-space  string theory, a planar amplitude no longer takes on a simple form in this  limit. Since this limit is structurally related  to the more general triple-Regge limit, a topic we discuss in Sec. \ref{sec:inclusive},   we turn to this  analysis for flat-space string theory next.

\subsection{Helicity Pole Limit for 5-Point Amplitudes}

We now discuss   the helicity pole limit for five-point amplitude, which provides a simpler 
illustration of the Steinmann rules. This  also serves as a prelude to the discussion for physical region 
discontinuities in ``crossed-invaraints'' for $n\geq 6$ carried out in Sec. \ref{sec:inclusive}, but can be omitted at a 
first reading.

This limit   corresponds to  $s \rightarrow \infty$ with $s_1$ 
and/or  $s_2$ fixed~\cite{Detar:1971gn,Detar:1971dj,Detar:1972nd}.   
To simplify the analysis further, let us first consider a combined helicity-pole and Regge  
limit where $s_1, s\rightarrow \infty$ but  $s>>s_1$, while  holding $s_2$ fixed, i.e., 
$s>>s_1>>s_2$.  (Of course, one can also treat the opposite limit by interchanging indices 1 and 2.) 
In term of $s_1,s_2, \kappa$, this corresponds to taking the limit, 
\be
s_1\rightarrow \infty, \quad\kappa\rightarrow 0\; 
\ee
with $s_2$ held fixed. This limit can be approached in several ways. For instance, we can first take the single-Regge  limit where $s_1\rightarrow \infty$, with  $s_2$ and $\kappa$ fixed, Fig. \ref{fig:regge5both}a, leading to a 
4-point vertex, $G_1^{[3]}(s_2, t_1, t_2, \kappa)$, with one external Reggeon. From this, one can take the small $
\kappa$ limit.

In flat-space string theory, this 4-point vertex $G_1^{[3]}(s_2, t_1, t_2, \kappa)$  can be easily found
\cite{Detar:1972nd},
\be
G_1^{[3]}(s_2, t_1, t_2, \kappa)\sim  \kappa^{-1} \int_0^1 dx x^{-\omega_2} (1-x)^{-\omega(s_2)} [1-(1-
s_2/\kappa ) x]^{\omega_1-1}
\ee
which is shown kinematically in Fig. \ref{fig:regge5both}. Note that, in this representation, the s-dependence for 
$M_{5,string}$ enters through $\kappa$.
In the small $\kappa$ limit, one finds that the amplitude can again be expressed as a sum of two terms
\be
M_{5,string}
\sim \gamma_1 (-s)^{\omega_1} {\cal A}_{string}(s_2, t_2; t_1) + (-s)^{\omega_2} (-s_1)^{\omega_1-
\omega_2}\Gamma(-\omega_2+1)\Gamma(\omega_2-\omega_1)\label{A5HP}
\ee
where  ${\cal A}_{string}(s_2, t_2;t_1)$ takes
 on the interpretation of Reggeon-particle-to-particle-particle amplitude. This decomposition is a generalization of 
the result obtained earlier, Eq. (\ref{eq:V1V2}), but now valid for $s_2$ held at a finite value.  Note that the second 
term is independent of $s_2$ and all the singularity 
 in $s_2$ is reflected in the first term.   We can also  identify the product $\Gamma(-\omega_2-1)\Gamma(\omega_1-
\omega_2)$ with $\gamma_1 \gamma_2 V_1(t_1,t_2, 0)$.

  The amplitude ${\cal A}_{string}$ has a series of poles in $s_2$ when $\alpha(s_2)=\omega(s_2)+1$ takes on positive integers,
\be
{\cal A}_{string}(s_2, t_2; t_1)=  \int_0^\infty dx x^{-\omega_2+\omega_1 -1} (1-x)^{-\alpha(s_2)+1}=
\frac{\Gamma(\omega_1-\omega_2)\Gamma(-\alpha(s_2)+2)}{\Gamma(2-\alpha(s_2)-\omega_2+\omega_1)}
\label{A5Rppp}
\ee
Recall $\alpha_1=\alpha(t_1)$ and $\alpha_2=\alpha(t_2)$. When $\alpha_1=2$, ($\omega_1=1$), it reduces to an on-shell 4-point 
amplitude, without tachyon.  The 
Steinmann relation again holds trivially
\bea
{\rm Disc_s\; Disc_{s_1} }\; M_{5,string}&\sim & s^{\omega_2} s_1^{\omega_1-\omega_2} V_1(t_1,t_2,0)\neq 0 
\nonumber\\
{\rm Disc_s\; Disc_{s_2}}\; M_{5,string}&\sim&  s^{\omega_1} {\rm Disc}_{s_2} {\cal A}_{string}(s_2, t_2; t_1) \neq 0 
\nonumber\\
{\rm Disc_{s_1} Disc_{s_2} } \; M_{5,string} & =  & 0 
\eea
In the Regge limit where $s_2$ is large, all the poles in $s_2$ collapse into a right-hand cut, with
\be
{\cal A}_{string}(s_2, t_2;t_1)\sim (-s_2)^{\omega_2-\omega_1}
\ee
which again leads to Eq. (\ref{eq:V1V2}). However, the  helicity-pole limit is more general, and it holds for $s_2$ 
finite.

For completeness, we record here that, for the flat-space string theory, the helicity-pole limit for a 5-point function 
where $s\rightarrow - \infty$ with $s_1$ and $s_2$ both fixed is
\be
M_{5,string}
\sim \gamma_1 (-s)^{\omega_1} {\cal A}_{string}(s_2, t_2; t_1) +{\cal A}_{string}(s_1, t_1;t_2) (-s)^{\omega_2} 
\gamma_2  \label{eq:HPA5}
\ee
In this representation, the Steinmann condition is manifest.

\newpage
%\section{On-Shell Analytic Continuation for $n\geq 6$ in the 
%Linear Multi-Regge Limit}

\section{Analytic Continuation and Factorization Constraints}
\label{sec:FactorizationConstraints}

Factorization for Reggeon (just as for particle) diagrams places strong recursive contraints on higher point functions. The first time Regge factorization plays a crucial role in gluon (or open string) scattering is at the level of the 6-point function. Here the identical double Regge particle vertex defined in the 5-point function enters again. How the multi-Regge limit factorizes is in fact a subtle interplay of the issues of analytical continuation of the individual planar amplitudes and the projection onto trajectories of definite signature.

Here we examine  this by comparing the 5-point and 6-point multi-Regge
amplitudes. Fortunately this analysis is independent of whether
or not the Steinmann relations hold.

\subsection{Signatured Regge-Regge Particle Vertex}

To determine the correct phase for
each the individual permutations of the planar amplitudes,
$A_n(\pi(1),\pi(2),\cdots, \pi(5))$, contributing to the physics
region for $-k_1 , -k_5, \rightarrow
k_2, k_3, k_{4}$ we begin  in
the Euclidean region as described in Sec.~\ref{sec:analytic} and analytically
continue in {\bf independent} invariants to the physical region~\footnote{For convenience, we now switch in this 
section to an all-outgoing momentum convention.}.

For general n-point amplitude in  the Euclidean region,
 a natural choice for the  $3n-10$ independent invariants
is  the BCP set, $s_1,
s_2, \cdots$, $t_1, t_2, \cdots$, $\kappa_{12}, \kappa_{23}, \cdots$ as
illustrated in Fig.  \ref{fig:multi_gluon.a}, where $\kappa_{i,i+1}=s_is_{i+1}/\Sigma_{i,i+1}$  with
$\Sigma_{i,i+1} = - (k_{i+1} + k_{i+2} + k_{i+3})^2$. 
As an example consider the amplitude the planar permutation, $A_n(1,2,\cdots,n)$,
whose multi-Regge limit is
\be 
M_n= A_n/A_{n,tree} \simeq \gamma_1(t_1) (-s_1)^{\omega_1}
G_2(t_1,,t_2,\kappa_{12})\cdots\cdots(-s_{n-3})^{\omega_{n-3}}\gamma_{n-3}(t_{n-1})\;
. \label{eq:multiRegge-n2} 
\ee 
 This color-ordered amplitude in  the
Euclidean Regge limit  ($s_i \rightarrow - \infty$ at
fixed $t_i < 0,\; \kappa_{i,i+1} < 0$) must be purely real,
since it can be demonstrated that  all BDS invariants remain
negative, $t^{[r]}_i<0$, away from all singularities.  In (\ref{eq:multiRegge-n2}), we have set the mass scale $1/\alpha' = 1$ and $\mu^2 =1$ for
flat space string and BDS amplitudes respectively. An illustration
in the deep Euclidean region  is given for $n=5$ 
and $n=6$ in Fig.~\ref{fig:star} and Fig.~\ref{fig:hex} respectively. 
In the multi-Regge limit we must take  $k^\pm \rightarrow  \infty$ for right and left movers respectively while staying 
in the Euclidean
region as depicted,  for example for the 5-point function, in Fig.~\ref{fig:5Regge.a}.  As mentioned earlier, this can clearly be generalized to any n. 

The continuation to the
physical region can be done in light-cone variables by a suitable
generalization of the procedure described in Sec.~\ref{sec:sig_double}
for the 5-point amplitude. Focus first on  the natural color-ordering $(1,2,3,4\cdots, n)$.  Observing that $s_i \simeq k^+_{i+2}
k^-_{i+3}$ and $\kappa_{i,i+1} \simeq (k^\perp_{i+2})^2$, we may begin
by continuing the longitudinal components so that $s_i \rightarrow s_i
>0 $ in a large semi-circle in the UHP holding $\kappa $'s essentially
fixed. Then each $\kappa$ is continued into its UHP. For $n=5$, as seen earlier,  one gets
\be
M_5(1,2,3,4,5)\simeq \gamma_1 s_1^{\omega_1} e^{ -i \pi \omega_1} 
G_2(t_1,t_2,\kappa_{12} + i\epsilon)s_{2}^{\omega_{2}} e^{ -i\pi \omega_{1}}\gamma_{2}\;
. \label{eq:multiRegge-phase}  
\ee

\begin{figure}[bthp]
\begin{center}
\includegraphics[width = 0.4\textwidth]{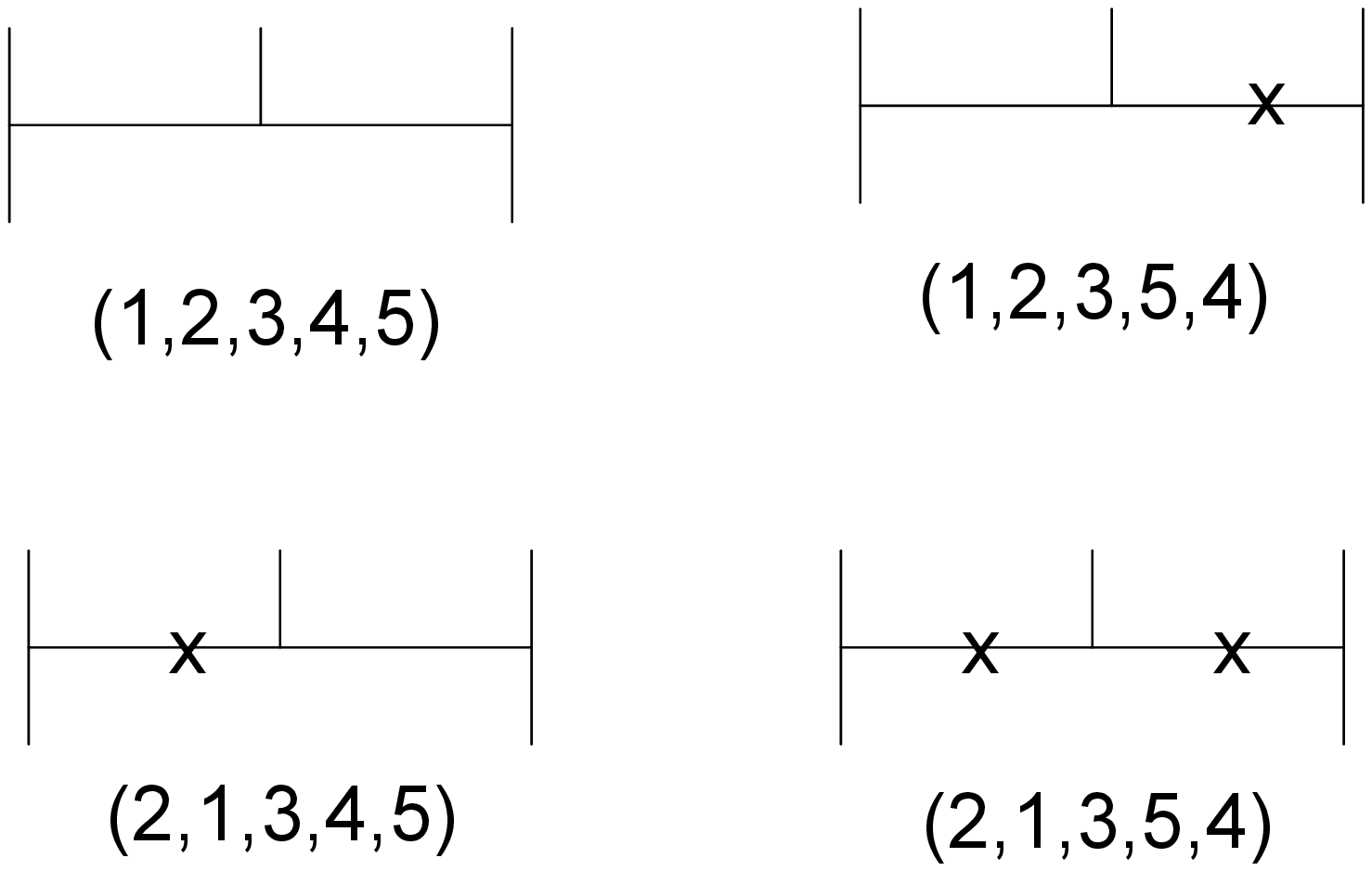}
\end{center}
\caption{Four cyclic orderings characterized by twistings.}
\label{fig:twist5.a}
\end{figure}

Next, we need to treat the continuation for other inequivalent orderings. 
We will illustrate the procedure first for the 5-point functions here, which
in fact suffices to define the general rule.  For 5-point amplitudes, there are 8 inequivalent color orderings, but anticyclic reversal reduces the independent set to four, which can be characterized by two indices $(\eta_1,\eta_2)$,  as depicted in Fig. \ref{fig:twist5.a}, where $\eta_i=-1$ is indicated by a ``cross".  The double-Regge limit
for each planar diagram in the physical region has powers $( - \eta_i s_i)^{\omega_i}$ with
$\eta_i = \pm 1$, $i=1,2$.  which contribute a complex phase $e^{
  - i \pi \omega_i} $ for $\eta_i = 1$ (right-hand cuts) and real
factors for $\eta_i = -1$ (left-hand cuts).  That is, for $\eta_i=-1$, there is no need for continuation.

One important new feature is the fact that, by {\bf on-shell} analytic
continuation, the two Reggeon vertices become complex in the physical
region, but can take on two possible values
\be
G_2(1,2;\pm) = G_2(t_1,t_2; |\kappa_{12}| \pm  i \epsilon)
\ee
For $n=5$, except for $\eta_1=\eta_2=- 1$, $\kappa_{12}$ is continued to the
positive real axis via the UHP so that the two-Reggeon-vertex takes on
$G_2(+)$. However, when $\eta_1 = \eta_2 = -1$, $\kappa_{12} = s_1
s_2/\Sigma_{12}$ is continued via the LHP. This is because both $s_1$
and $s_2$ have left hand cuts and no continuation is required in $s_1$
and $s_2$.  However, $\Sigma_{12}$ is continued into the UHP to its
final positive value, leading to $G_2(-)$.  The general factorization
pattern in the linear multi-Regge limit will invariably involve the
discontinuity $\Delta G= G_2(+)-G_2(-)$. When particles are on-shell, one
finds $\Delta G$ vanishes and the factorization pattern simplifies.  

To summarize, in the physical region,
\be
M_{\eta_1\eta_2}\sim s_1^{\omega_1} \;  s_2^{\omega_2} \; (-\eta_1)^{\omega_1}(-\eta_2)^{\omega_2}    \; \gamma_1 \gamma_2 \;
G_2(1,2;\epsilon(\eta_1,\eta_2))    \; , \label{eq:doublereggeM}
\ee
or, equivalently for the planar amplitudes,
\be
A_{\eta_1\eta_2}\sim s_1^{\alpha_1} \;  s_2^{\alpha_2} \;  (-\eta_1)^{\alpha_1}(-\eta_2)^{\alpha_2}     \;\Gamma(t_1) \Gamma(t_2) \; (\gamma_1 \kappa_{12} \gamma_2)^{-1}
G_2(1,2;\epsilon(\eta_1,\eta_2))     \;  , \label{eq:doubleregge}
\ee
where  $\epsilon(+,+)=\epsilon(-,+)=\epsilon(+,-)= +$ and $\epsilon(-,-)=-$, i.e.,
\be
\epsilon(\eta,\eta') = \frac {1+\eta+\eta'- \eta\eta'}{2}\; .  
\ee
Note that, with our convention for the propagator $\Gamma(t)$, (\ref{eq:BDSPropagator}), a factor of $(\gamma_1\gamma_2)^{-1}$ has to be inserted~\footnote{For flat-space string theory, recall that $\Gamma(t)=\Gamma(1-\alpha(t))$ and $\gamma(t)=1$.}.  Dependence on  polarizations has been suppressed and will be made explicit in the next section~\footnote{In going from $M_{\eta_1\eta_2}$ to $A_{\eta_1\eta_2}$,  a factor of $\eta_1\eta_2$ has been supplied, which will  be accounted for in the next section when treating the color-trace. }.

As emphasized earlier, multi-Regge factorization is generally expected only for ``signatured" amplitudes \cite{Weis:1972ir}.  For the 5-point amplitudes, we can introduce ``signatured" amplitudes, 
\be
\widetilde A_{\sigma_1\sigma_2}=2\; \Omega_{\sigma_1\eta_1}\Omega_{\sigma_2\eta_2} \; A_{\eta_1,\eta_2}\;,
\ee
Expressing these signatured amplitudes as
\bea
\widetilde A_{\sigma_1\sigma_2}&=&  s^{\alpha(1)}_1 s^{\alpha(2)}_2 \; \widetilde  \Pi_{\sigma_1}(1) \;  \widetilde G_{\sigma_1\sigma_2}(1,2) \; \widetilde \Pi_{\sigma_2}(2)   \;.  \label{eq:signafactorization.5}
\eea
this  allow us to define ``signatured" vertex $\widetilde G_{\sigma_1\sigma_2}$,
\bea
\widetilde G_{\sigma_1\sigma_2}(1,2)&= &(\gamma_1\kappa_{12}\gamma_2)^{-1} \left[ G_{2}(1,2; +) -\Delta G \; \sigma_1\sigma_2\xi_1(\sigma_1)^{-1}\xi_2(\sigma_2)^{-1}\right] \; .  \label{eq:signatured2vertex}
\eea
Here,   $\widetilde \Pi_{\sigma_i}(i)= \xi_{\sigma_i}(t_i)\Gamma(t_i)$  is a signatured propagator, generalizing Eq. (\ref{eq:pi}) for each exchange.  Conversely, given signatured amplitudes  $\widetilde A_{\sigma_1\sigma_2}$, one can recover the planar amplitudes, $A_{\eta_1,\eta_2}$,  by inversion,
\be
A_{\eta_1,\eta_2}=(1/2)\; \Omega_{\eta_1\sigma_1}\Omega_{\eta_2\sigma_2} \widetilde A_{\sigma_1\sigma_2} \;. 
\ee

\subsection{Analyticity Constraint on the 6-Point Function}
\label{subsect:linearMR}

 We turn next to 6-point functions  and continue to focus
on the linear multi-Regge limits, as illustrated by
Fig. \ref{fig:multiregge6}b.  In spite of the troubling observation on the analytic structure of the vertex $G_2$ in  $
\kappa$, we can proceed to discuss the question of  continuation into the physical region.   For the multi-Regge limit,  there are now 8 inequivalent color configurations, which can be characterized by three indices, $\eta_1,\eta_2,\eta_3$, $\eta_i=\pm 1$.  We will focus here on the {\bf on-shell}
continuation from Euclidean to the physical region for each color configuration.  In particular, we show how this allows multi-Regge factorization for signatured amplitudes, i.e., generalizing (\ref{eq:signafactorization.5}) to $n=6$,
\bea
\widetilde A_{\sigma_1\sigma_2\sigma_3}
&=&  s^{\alpha(t_1)}_1 s^{\alpha(t_2)}_2 s^{\alpha(t_3)}_3\; \widetilde  \Pi_{\sigma_1}(t_1) \;  \widetilde G_{\sigma_1\sigma_2}(t_1,t_2;\kappa_{12}) \; \widetilde \Pi_{\sigma_2}(t_2) \;  \widetilde G_{\sigma_2\sigma_3}(t_2,t_3;\kappa_{23}) \; \widetilde \Pi_{\sigma_3}(t_3)  \;.  \nn \label{eq:signafactorization.6}
\eea
Again,  color-trace as well as polarization factors will be ignored for now.

One of the key 
differences between the
case $n\leq 5$ vs. $n\geq 6$ is the fact that BDS invariants are no
longer independent.  In Ref. \cite{Brower:2008nm}, we have stressed that, in the Regge limit, constraints among BDS invariants can be 
understood in terms of constraints on cross ratios. Indeed,  one obtains multi-Regge behavior in the Euclidean 
regions only if these constraints are imposed. 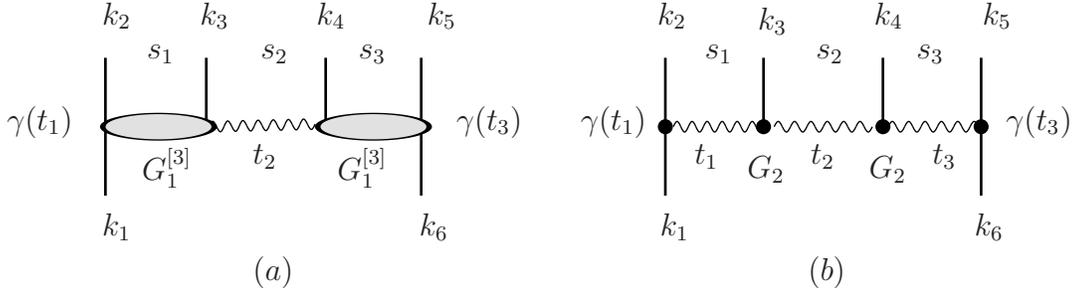
\begin{figure}[h!]
\begin{center}
%\fcolorbox{white}{white}{
  \begin{picture}(408,98) (2,-40)
    \SetWidth{1.0}
%    \SetColor{Black}
    \Line(39,-5)(39,19)
    \Line(77,22)(77,47)
    \Text(75,58)[lb]{{\Black{$k_3$}}}
   \Line(39,47)(39,22)
  \Line(158,-5)(158,19)
    \Text(95,5)[lb]{{\Black{$t_2$}}}
   \Line(122,22)(122,47)
    \Text(38,58)[lb]{{\Black{$k_2$}}}
    \Text(55,45)[lb]{{\Black{$s_1$}}}
    \Text(98,45)[lb]{{\Black{$s_2$}}}
    \Text(119,58)[lb]{{\Black{$k_4$}}}
    \Text(135,45)[lb]{{\Black{$s_3$}}}
    \Text(161,58)[lb]{{\Black{$k_5$}}}
   \Line(158,22)(158,47)
    \Text(38,-22)[lb]{{\Black{$k_1$}}}
    \Text(158,-22)[lb]{{\Black{$k_6$}}}
    \SetWidth{0.5}
    \Photon(78,21)(118,22){2}{7}
 \GOval(59,21)(5,21)(0){0.882}
\GOval(140,21)(5,21)(0){0.882}
 
    \Text(127,-2)[lb]{{\Black{$G_1^{[3]}$}}}
    \Text(53,-2)[lb]{{\Black{$G_1^{[3]}$}}}
    \Text(2,17)[lb]{{\Black{$\gamma(t_1)$}}}
    \Text(172,17)[lb]{{\Black{$\gamma(t_3)$}}}
    \SetWidth{1.0}
    \Line(250,-5)(250,21)
   \Line(287,21)(287,47)
    \SetWidth{0.5}
    \Vertex(287,21){2.83}
  \Photon(287,21)(253,21){2}{6}
    \Text(286,56)[lb]{{\Black{$k_3$}}}
    \SetWidth{1.0}
    \Line(250,47)(250,21)
    \SetWidth{0.5}
    \Vertex(250,21){2.83}
    \SetWidth{1.0}
  \Line(369,-5)(369,21)
  \Text(380,17)[lb]{{\Black{$\gamma(t_3)$}}}
  \Text(262,5)[lb]{{\Black{$t_1$}}}
    \Text(282,0)[lb]{{\Black{$G_2$}}}
  \Text(306,5)[lb]{{\Black{$t_2$}}}
    \SetWidth{0.5}
 \Photon(291,21)(328,21){2}{6}
    \Vertex(332,21){2.83}
 \Photon(334,21)(371,21){2}{6}
  \Vertex(369,21){2.83}
    \SetWidth{1.0}
    \Line(332,21)(332,47)
    \Text(248,58)[lb]{{\Black{$k_2$}}}
    \Text(266,45)[lb]{{\Black{$s_1$}}}
    \Text(308,45)[lb]{{\Black{$s_2$}}}
    \Text(330,58)[lb]{{\Black{$k_4$}}}
    \Text(346,45)[lb]{{\Black{$s_3$}}}
    \Text(328,0)[lb]{{\Black{$G_2$}}}
  \Text(352,5)[lb]{{\Black{$t_3$}}}
    \Text(371,58)[lb]{{\Black{$k_5$}}}
  \Line(369,21)(369,47)
    \Text(249,-22)[lb]{{\Black{$k_1$}}}
    \Text(368,-22)[lb]{{\Black{$k_6$}}}
  \Text(219,17)[lb]{{\Black{$\gamma(t_1)$}}}
    
      \Text(95,-40)[lb]{{\Black{$(a)$}}}
    \Text(305,-40)[lb]{{\Black{$(b)$}}}
    
  \end{picture}
%}
\end{center}
\caption{Regge limits for 6-point amplitude are completely
determined by factorization from the 5-point amplitude. On the left, (a), single-Regge limit 
with vertices $G_1^{[3]}(s_1,t_2,t_1,\kappa_{12})$ and $G_1^{[3]}
(s_3,t_2,t_3,\kappa_{23})$. On the right, (b), linear triple-Regge limit with internal vertices  
$G_2(t_1,t_2,\kappa_{12})$ and  $G_2(t_2,t_3,\kappa_{23})$. }
\label{fig:multiregge6}
\end{figure}
To be on-shell, one can work with independent  invariants, in terms of which these constraints are 
automatically satisfied. To be precise, we have used the set of $3n-10$ independent BCP invariants, 
$s_1, s_2, \cdots$, $t_1, t_2, \cdots$, $\kappa_{12}, \kappa_{23}, \cdots$, and have shown, in the {\it Euclidean multi-Regge region where $s_i, t_j, \kappa_{i,i+1}$ are all negative}, that
\be
M_n\simeq \gamma_1 (-s_1)^{\omega_1} G_2(t_1,,t_2,\kappa_{12})\cdots\cdots(-s_{n-3})^{\omega_{n-3}}
\gamma_{n-3}\; . \label{eq:multiRegge-n}
\ee
Note that  $\kappa_{i,i+1}=s_is_{i+1}/\Sigma_{i}<0$, $\Sigma_i= - (k_{i-2}+k_{i-3}+k_{i-4})^2.$  To illustrate, let us focus on the case $n=6$ and the natural color ordering $(1,2,3,4,5,6)$. As is well-known, there are now three independent cross ratios,  
\be
u_1= \frac{t_1^{[2]} t_6^{[4]}}{t_1^{[3]}t_6^{[3]}}=\frac{t_1s_3}{t_2\Sigma_2}\; ,\;\quad 
u_2= \frac{t_2^{[2]} t_{1}^{[4]}}{t_2^{[3]}t_1^{[3]}}=\frac{t_3s_1}
{t_2\Sigma_1}\; , \;\quad 
 u_3= \frac{t_3^{[2]} t_2^{[4]}}{t_3^{[3]}t_2^{[3]}}=\frac{s_2s}{\Sigma_1\Sigma_2}\;.
  \label{crossr}
\ee
and, in the linear multi-Regge limit, nonlinear constraints among BDS variables together with the {\bf on-shell 
conditions} lead to
\be
u_1\rightarrow 0\;,\quad u_2\rightarrow  0\;,\quad u_3\rightarrow 1
\ee
Only when these conditions are enforced, multi-Regge behavior, Eq. (\ref{eq:multiRegge-n}), follows.  This analysis leading to factorization in the Euclidean region applies  to all color orderings, and it has been generalized  to $n > 6$ in Ref. \cite{Brower:2008nm}.

Let us now turn to the continuation to the physical region. Staying
on-shell, we need to specify the paths of continuation for $\kappa_{12}$
and $\kappa_{23}$ separately. For the multi-Regge limit, generalizing the analysis for $n=5$,   there are 16 color configurations involving  $t_1$, $t_2$ and $t_3$ as BDS invariants. Of these, 8 are inequivalent (due to anti-cyclic symmetry), which can be characterized by three indices, $\eta_1,\eta_2,\eta_3$,  $\eta_i=\pm 1$, e.g., the color ordering $(1,2,3,4,5,6)$ corresponds to $\eta_1=\eta_2=\eta_3=+1$.  Since $\kappa_{12}$
and $\kappa_{23}$  depend on $(s_1,s_2, \Sigma_1)$
and $(s_2,s_3,\Sigma_2)$ respectively, the physical region again
corresponds to $\kappa_{12}\rightarrow |\kappa_{12}| +i\epsilon$ and
$\kappa_{23}\rightarrow |\kappa_{23}| +i\epsilon$. That is, in the physical
region where $s_i>0$, the amplitude with color-order $(123456)$,
Fig. \ref{fig:multiregge6}b, is
\bea
A_{+++} &\simeq&  (-s_1)^{\alpha_1} (-s_2)^{\alpha_2} (-s_3)^{\alpha_3}\nn
&\times&  \Gamma(1)\Gamma(2)\Gamma(3) (\gamma_1\kappa_{12}\gamma_2)^{-1}(\gamma_2\kappa_{23}\gamma_2)^{-1}  \; G_2(1,2;+) G_2(2,3;+)  \;
   \label{eq:6pointlinearMR}
\eea
where  $G_2(i,i+1;\pm)=G_2(t_i,t_{i+i}; |\kappa_{i,i+1}|\pm i\epsilon) $.

It is straightforward to generalize this continuation procedure to other color configurations, leading to apparent factorization for all color orderings in the physical region. However, for the BDS amplitudes,   a non-factorizable result  was obtained in Ref. \cite{Bartels:2008ce} for the color configurations corresponding to $(-,+,-)$. Prompted by this discrepancy, we have examined more closely the corresponding issues in flat-space string theory. Indeed, it has been demonstrated for flat-space string theory that a straight forward generalization of the above procedure would not  lead to a ``faithful" representation of the analytic structure for some of the planar amplitudes. To be precise, in the case of $n=6$, for configurations  $(-,+,-)$ and $(-,-,-)$, holding $u_3=1$ would not account   correctly the analytic structure of the original dual amplitudes. For instance, for the $(-+-)$, a planar amplitude contains right-hand cuts in $s_{34}=-(k_3+k_4)^2$ and $s_{25}=-(k_2+k_5)^2$. Holding $u_3 \simeq 1$ in the course of continuation would not allow both $s_{34}$ and $s_{25}$ to reach the physical region simultaneously via UHP.   

An effective procedure for keeping track of the analytic structure in the multi-Regge limit is to retain the $u_3$ dependence so that,
\bea
M_{\eta_1\eta_2\eta_3}&\simeq& s_1^{\omega_1} s_2^{\omega_2} s_3^{\omega_3}\;  (-\eta_1)^{\omega_1}(-\eta_2)^{\omega_2}(-\eta_3)^{\omega_3}\;   
 \gamma_1\gamma_3\;  B_{\eta_1\eta_2\eta_3}(1,2,3,\Phi_{\eta_1\eta_2\eta_3})\;,  \label{eq:M3}
\eea
where  we have  re-written $\Phi$ for $u_3$.  For all planar configurations other than these two exceptions listed above, $\Phi=1$ is consistent with the process of continuation. For configurations $(-,+,-)$ and $(-,-,-)$, $\Phi$ takes on $e^{-2\pi i}$ and $e^{2\pi i}$ respectively. Due to a branch-cut singularity at $\Phi=0$, holding $\Phi=1$ for these two configurations in the course of continuation to the physical region would lead to incorrect results. 

Following the analysis of   \cite{Weis:1972ir}, one can show  for flat-space string theory  that $B_{\eta_1\eta_2\eta_3}$ can be expressed as a sum of two terms,
\bea
B_{\eta_1\eta_2\eta_3}(1,2,3,\Phi_{\eta_1\eta_2\eta_3})
&= & G(1,2;\epsilon(\eta_1,\eta_2)) \;  G(2,3;\epsilon(\eta_2,\eta_3))  +\Delta B_{\eta_1\eta_2\eta_3}(1,2,3,\Phi_{\eta_1\eta_2\eta_3})\; , \nn  \label{eq:B6}
\eea
where
\bea
\Delta B_{\eta_1\eta_2\eta_3}
&=& \left[ (\Phi_{\eta_1\eta_2\eta_3})^{\alpha_1}-1\right] \frac{e^{i\epsilon(\eta_1,\eta_2)\pi \alpha_1}}{\sin \pi \alpha_1} \frac{ e^{i\epsilon(\eta_2,\eta_3)\pi \alpha_2} }{\sin \pi \alpha_2} \frac{\Delta G(1,2) }{2i}\frac{ \Delta G(2,3)}{2i}\; .  \label{eq:B6b}
\eea
 For color configurations $(-,\pm,-)$,  we have $\Phi = e^{\mp 2\pi i}$ and   $\Delta B$ non-zero, thus breaking naive factorization. It is also clear that this fact depends on  $G(t,t';\kappa)$ having a branch-cut for $\kappa>0$ and $\Delta G\neq 0$.   
 
 If one were to set $\Phi_{\eta_1\eta_2\eta_3}=1$, independent of $\eta_i$, (\ref{eq:B6}) would become  a product  of two vertices, thus leading  to ``naive factorization",
\bea
A_{\eta_1\eta_2\eta_3} &\simeq&  (-\eta_1)^{\alpha_1}(-\eta_2)^{\alpha_2}(-\eta_3)^{\alpha_3}\; 
s_1^{\alpha_1} s_2^{\alpha_2} s_3^{\alpha_3}\\
&\times&  \Gamma(1)\; \Gamma(2) \; \Gamma(3)\;  (\gamma_1\kappa_{12}\gamma_2)^{-1}(\gamma_2\kappa_{23}\gamma_3)^{-1}  \; G_2(1,2;\epsilon(\eta_1,\eta_2)) G_2(2,3;\epsilon(\eta_2,\eta_3))  \; \;. \nonumber  \label{eq:6pointlinearMR.a}
\eea
Alas, analyticity, i.e., causality, dictates that we must retain the correct  $\Phi_{\eta_1\eta_2\eta_3}$ dependence. This invalidates the naive factorization for $A_{\eta_1\eta_2\eta_3}$. Instead, it leads to  factorization for signatured amplitudes, which we turn to next.

\subsection{Signature Factorization and  the BDS Amplitudes}

Let us next examine  the consequence of signature factorization, (\ref{eq:signafactorization.6}). It is easy to transform  signature factorization   back to $A_{\eta_1\eta_2\eta_3}$, 
\be
A_{\eta_1\eta_2\eta_3}= 2^{-3/2} \Omega_{\eta_1\sigma_1}\Omega_{\eta_2\sigma_2} \Omega_{\eta_3\sigma_3} \widetilde A_{\sigma_1\sigma_2\sigma_3}
\ee
This leads to a two-term recursion relation,
\be
B_{\eta_1,\eta_2,\eta_3}(1,2,3)= B_{\eta_1,\eta_2} (1,2) J_1(2,3)+B_{\eta_1,(\eta_3\eta_2)}(1,2) \; e^{i\epsilon(\eta_2,\eta_3)\pi \alpha_2} \;  J_2(2,3)\;  ,  \label{eq:2termrecursion}
\ee
i.e.,  relating  $B_{\eta_1,\eta_2,\eta_3}$ to $B_{\eta_1,\pm \eta_2}$, where $B_{\eta_1,\eta_2} (1,2)=G(1,2;\epsilon(\eta_1,\eta_2))$.  The coefficient functions $J_1(2,3)$ and $J_2(2,3)$ are real in the physical region, related to the  Reggeon-Reggeon vertex by
\bea
G(2,3;+)&=& J_1(2,3)+e^{i \pi\alpha_2} J_2(2,3) \;.   \label{eq:J-representation}
\eea
More directly, one can express $J_1$ and $J_2$  in terms of $G$ and $\Delta G$, e.g.,
\bea
 J_2(i,i+1) &=&   
 \frac{ \Delta G(i,i+1) }{2i\sin \pi \alpha_{i+1}}  \; .  
\eea 

It is easy to   check  that (\ref{eq:2termrecursion}) directly reproduces  the  non-factorization,  (for  color configurations  $(-,\pm,-)$), emphasized  earlier.   Using the fact that, in the physical region, the phase $\Phi_{\eta_1\eta_2\eta_3}$ can be expressed as
$e^{ - i (1-\eta_1)\eta_2 (1-\eta_3)\pi/2}$, one finds that (\ref{eq:B6})  directly leads to the the recursion relation, (\ref{eq:2termrecursion}). Alternatively, one can check that  (\ref{eq:B6b}) directly leads to  signature factorization,  (\ref{eq:signafactorization.6}),  for the flat-space string theory.

 This two-term recursion relation can also be generalized to higher point functions.  Assuming  factorization for higher point signatured amplitudes, it follows that  $B_{\eta_1,\eta_2,\cdots,\eta_{n-1},\eta_n}$  can be expressed as a linear combination of  $B_{\eta_1,\eta_2,\cdots,\eta_{n-1}}$ and $B_{\eta_1,\eta_2,\cdots,-\eta_{n-1}}$,
\be
B_{\eta_1,\eta_2,\cdots,\eta_{n-1},\eta_n}= B_{\eta_1,\eta_2,\cdots,\eta_{n-1}} J_1(n-1,n)+B_{\eta_1,\eta_2,\cdots,-\eta_{n-1}} \; e^{i\epsilon(\eta_{n-1},\eta_n)\pi \alpha_{n-1}} \;  J_2(n-1,n)\;  .  \label{eq:2termrecursion-n}
\ee
It has been shown inductively \cite{Weis:1972ir} that indeed signature factorization holds for $n>6$ for flat-space string theory.

Let us turn next to an examination of 6-point BDS amplitudes. As emphasized earlier, we adopt the procedure of dropping $O(\epsilon)$ for 
\be
\log{M_6}=\log\frac{A_6}{A_{6,tree}}=I_6^{(1)}(\epsilon)+F_6^{(1)}(0)
\ee
in taking the multi-Regge limit.  In this case,  BDS amplitudes for $n\geq 6$ reduce to simple combinations of products of logarithms and dilogarithmic functions. As pointed in Ref.
 \cite{Brower:2008nm}, these dilog functions do not contribute in the Euclidean multi-Regge limit and Regge factorization can be achieved. This relies on the observation that all cross ratios either vanish or approaching 1 in this limit. 

However, as pointed above, analyticity consideration forces one to relax the constraint on the cross ratios in the course of continuation back to the physical region. In the case of $n=6$, there are only three such cross ratios, and the one which requires special attention is the variable $\Phi$, or $u_3$ in (\ref{crossr}). Since its nontrivial dependence enters explicitly, continuation into the physical region can be carried out unambiguously. 
In \cite{Bartels:2008ce}, one finds for $A_{-+-}$ that, in the course of continuation where $\Phi: 1\rightarrow e^{-2\pi i}$, 
$\log{M_6}$ picks up  an extra piece
\bea
\Delta \log{M_6}(-,+,-)&=&\frac{f(\lambda)}{8}\left( \frac{1}{\epsilon}
+\log \frac{\mu^2 s_2}{(-t_1)(-t_3)}      \right) \log \Phi +  \frac{f(\lambda)}{4}\pi i \log  \left( \frac{\Phi}{1-\Phi} \right)  \nn
&=&\frac{f(\lambda)}{4}\pi i\left(-\frac{1}{\epsilon}
+\log \left(\frac{(-t_1)(-t_3)}{\mu^2 s_2}\left[\frac{\Phi}{1-\Phi}\right]\right)\right)\label{lipatovextra}\;.
\eea
Here the term $\log ( {\Phi}/{1-\Phi})  $  comes from analytic continuation of the dilog $Li_2(z)$, $z=1-\Phi$,  onto its second sheet at $z\simeq 0$. With this addition to $M_6(-,+,-)$ due to analytic continuation, it breaks the naive factorization, (\ref{eq:6pointlinearMR.a}). We note that the term $\log \Phi$ is equally important in arriving at a finite result, and  this term can not be expressed as a function of cross ratios. A similar analysis can also be carried out for the $M_6(-,-,-)$, and naive factorization again breaks down.
\ignore{ It has been suggested a term could be added so that naive factorization could be restored.}

Let us next examine how BDS fares with respect to MR factorization expected for signatured amplitudes. Either from (\ref{eq:2termrecursion}) or directly from (\ref{eq:B6}), one can show that the condition for factorization for signatured amplitudes can  be expressed as
\be
e^{\Delta log M_6({-,\pm , -})}= 1\pm(  2i)\frac{e^{\pm i(\phi_{1,2}+\phi_{2,3})} \sin\phi_{1,2}\sin\phi_{2,3}}{\sin\pi \alpha_2}  \label{eq:key}
\ee
where angles $\phi_{i,i+1}$ are defined by
\be
G_2(i,i+1;+)=|G_2| \; e^{i\phi_{i,i+1}}\; .
\ee
It can be checked  the modification to BDS amplitude  due to continuation in $\Phi$ through $Li_2(1-\Phi)$, (\ref{lipatovextra}), does not satisfy the condition above.  It follows that  signature factorization, (\ref{eq:signafactorization.6}), also fails. 

\newpage
\section{Factorization for Multi-Regge Limits}
\label{sec:5plus}

Factorization is an iterative property. Propagators and vertices encountered in lower point functions must be present in higher point functions, along with new higher vertices in non-polynomial expansion such as those present in the multi-Regge (or Gribov) effective field theory diagrams.  Here we focus on factorization for the general linear multi-Regge for 2 to n-2 amplitudes. As we noted in Sec. when we neglect the color trace the 4-point amplitude has 
two degenerate trajectories of opposite signature and in the 5-point function these two trajetories couple to a 2 by 2 Reggeon-Reggeon vertex. Here we include the color trace and demonstrate the form of multi-Regge factorization for the flat space string. For the BDS amplitudes factorization appears to fail at the 6-point level because due to unconventional analyticity properties. It seems likely that this is another aspect of the difficulties noted in Sec. with the Steinmann relation for the BDS amplitudes.

\begin{figure}[ht]
\begin{center}
%\fcolorbox{white}{white}{
  \begin{picture}(290,288) (75,-120)
    \SetWidth{0.5}
%    \SetColor{Black}

    \Text(116,-101)[lb]{{\Black{$k_1$}}}
    \Text(233,-5)[lb]{{\Black{$k_5$}}}
    \Text(155,-5)[lb]{{\Black{$k_3$}}}
    \Text(116,-5)[lb]{{\Black{$k_2$}}}
        \Text(312,-101)[lb]{{\Black{$k_8$}}}
            \Text(188,-5)[lb]{{\Black{$k_4$}}}
    \Text(267,-5)[lb]{{\Black{$k_6$}}}
    \Text(312,-5)[lb]{{\Black{$k_7$}}}
    
       \ArrowLine(122,-15)(122,-48)
   \ArrowLine(122,-48)(155,-48)
    \ArrowLine(155,-48)(155,-15)
  \ArrowLine(116,-88)(116,-15)
     \ArrowLine(273,-15)(273,-48)
  \ArrowLine(279,-48)(279,-15)
 \ArrowLine(312,-54)(312,-88)
    \ArrowLine(318,-88)(318,-15)
    \ArrowLine(312,-15)(312,-48)
   \ArrowLine(233,-48)(233,-15)
     \ArrowLine(239,-15)(239,-48)
   \ArrowLine(122,-54)(122,-88)
         \ArrowLine(200,-48)(200,-15)
    \ArrowLine(194,-15)(194,-48)
    \Line(222,-54)(216,-48)
    \Line(216,-54)(222,-48)
 \ArrowLine(222,-48)(233,-48)
       \ArrowLine(251,-54)(222,-54)
\ArrowLine(177,-54)(216,-54)
   \ArrowLine(195,-48)(177,-48)
    \ArrowLine(216,-48)(200,-48)
  \Line(257,-48)(251,-54)
\Line(251,-48)(257,-54)
\ArrowLine(312,-48)(279,-48)
  \ArrowLine(257,-54)(312,-54)
 \ArrowLine(273,-48)(257,-48)
 \ArrowLine(239,-48)(251,-48)
   \ArrowLine(171,-54)(122,-54)
 \Line(177,-48)(171,-54)
\Line(171,-48)(177,-54)
 \ArrowLine(161,-15)(161,-48)
   \ArrowLine(161,-48)(171,-48)

    \ArrowLine(312,105)(279,105)
    \ArrowLine(312,72)(312,105)
    \ArrowLine(312,111)(312,144)
 \ArrowLine(240,111)(312,111)
    \ArrowLine(318,144)(318,72)
  \ArrowLine(122,144)(122,111)
    \ArrowLine(122,111)(155,111)
    \ArrowLine(155,111)(155,144)
 \ArrowLine(161,144)(161,111)
  \ArrowLine(234,111)(234,144)
   \ArrowLine(240,144)(240,111)
    \ArrowLine(194,72)(194,105)
    \ArrowLine(200,105)(200,72)
  \ArrowLine(273,72)(273,105)
    \ArrowLine(279,105)(279,72)
    \ArrowLine(194,105)(122,105)
   \ArrowLine(273,105)(200,105)
    \ArrowLine(161,111)(234,111)
    \ArrowLine(122,105)(122,72)
    \ArrowLine(116,72)(116,144)
    \Text(116,57)[lb]{{\Black{$k_1$}}}
    \Text(194,57)[lb]{{\Black{$k_4$}}}
    \Text(273,57)[lb]{{\Black{$k_6$}}}
    \Text(312,57)[lb]{{\Black{$k_7$}}}
    \Text(312,153)[lb]{{\Black{$k_8$}}}
    \Text(233,153)[lb]{{\Black{$k_5$}}}
    \Text(155,153)[lb]{{\Black{$k_3$}}}
    \Text(116,153)[lb]{{\Black{$k_2$}}}
    \Text(222,171)[lb]{{\Black{$\tau_5 = 1$}}}
    \Text(332,105)[lb]{{\Black{$\tau_7 = -1$}}}
    \Text(261,40)[lb]{{\Black{$\tau_6 = -1$}}}
    \Text(180,40)[lb]{{\Black{$\tau_4 = -1$}}}
    \Text(142,171)[lb]{{\Black{$\tau_3 =1$}}}
    \Text(75,105)[lb]{{\Black{$\tau_2 = 1$}}}
    \Text(155,-71)[lb]{{\Black{$\eta_2 = -1$}}}
    \Text(200,-71)[lb]{{\Black{$\eta_3 = -1$}}}
    \Text(245,-71)[lb]{{\Black{$\eta_4 = -1$}}}
       \Text(215,20)[lb]{{\Black{(a)}}}
      \Text(215,-120)[lb]{{\Black{(b)}}}
  \end{picture}
%}
\end{center}
\caption{Example of planar diagram $A(12358764)$  contributing
to the multi-Regge limit for $- k_1 - k_8 \rightarrow k_2 + k_3 + k_4 + k_5 + k_6 + k_7$
with twisted vertices  $\tau_4 = \tau_6 = \tau_7 = -1$ in the top diagram, (a), 
implying  twisted links $\eta_2 = \tau_3 \tau_4= -1$ , $\eta_3 = \tau_4 \tau_5= -1$ 
 and $\eta_4 = \tau_5 \tau_6= -1$ in the
bottom diagram, (b).}
\label{fig:planar}
\end{figure}
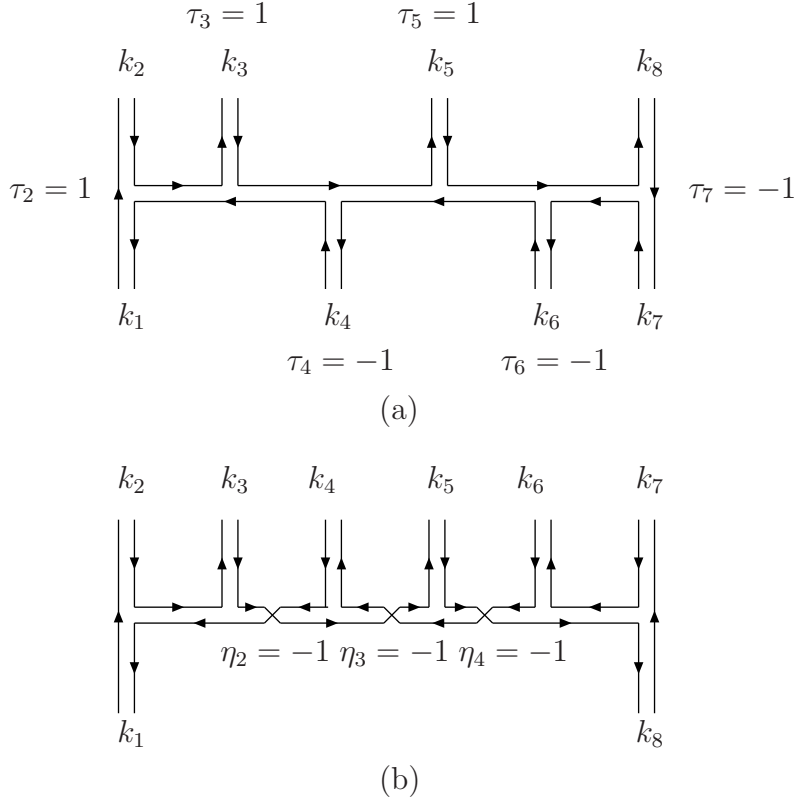

The pattern which emerges can be most easily seen by considering the
general case of the n-gluon amplitude in the linear multi-Regge
limit. The full set of planar amplitudes contributing to the linear
Multi-Regge limit for $1+ n \rightarrow 3 + 4 +\cdots n-1$
scattering are $2^{n-2}$ permutation found by ``flipping'' any of the
$n-2$ final particles to the opposite side of the trace, $Tr[12\cdots
n]$, as illustrated in Fig.~\ref{fig:planar}. 
Here we introduce the notation: $Tr[ijkl\cdots] =
Tr[T^{a_i}T^{a_j}T^{a_k}T^{a_l}\cdots]$.  To count these configurations, let $\tau_i = \pm 1$ for $i = 2,\cdots
n-1$ for each outgoing line, (see Figure \ref{fig:planar}a), $\tau_i =
-1$ indicates the ith gluon has been ``flipped''. There are $2^{n-3}$
choices with Regge powers $(\mp s_i)^{\alpha(t_i)} \equiv (-\eta_i
s_i)^{\alpha(t_i)}$. The case of a real  Regge power, $s_i^\alpha(t_i)$ (
$\eta_i = -1$) requires a twist of one of the two adjacent lines so
$\eta_i = \tau_{i+1} \tau_{i+2}$, (see Figure \ref{fig:planar}b). In addition we note that the n-gluon
planar amplitudes obey exact cyclic and anti-cyclic conditions,
\be
A_n(2,\cdots, n,1) = A_n(1,2,\cdots, n)  \quad , \quad 
A_n(1,2,\cdots, n) = (-1)^n A_n(n,\cdots, 2,1)
\ee
respectively.  As we will verify shortly, this
implies an extra factor of $\tau_2 \tau_3 \cdots \tau_{n-1}$.

\begin{figure}
\begin{center}
%\fcolorbox{white}{white}{
  \begin{picture}(256,131) (170,-80)
    \SetWidth{0.5}
%    \SetColor{Black}
    \Vertex(322,3){1.41}
    \Vertex(338,4){1.41}
    \Vertex(353,4){1.41}
    \Vertex(342,-33){1.41}
    \Vertex(354,-33){1.41}
    \Text(312,26)[lb]{{\Black{$s_3$}}}
    \SetWidth{1.0}
    \LongArrow(176,-33)(176,-30)
    \SetWidth{0.8}
    \Photon(176,-70)(176,-33){1.5}{5.5}
    \SetWidth{1.0}
    \LongArrow(289,13)(289,16)
    \SetWidth{0.8}
    \Photon(289,-25)(289,12){1.5}{5.5}
    \Photon(281,-26)(233,-25){1.5}{7.5}
    \SetWidth{1.0}
    \LongArrow(280,-25)(284,-26)
    \SetWidth{0.8}
    \Photon(338,-26)(290,-25){1.5}{7.5}
    \SetWidth{1.0}
    \LongArrow(337,-25)(341,-26)
    \SetWidth{0.8}
    \Photon(394,-26)(346,-25){1.5}{7.5}
    \SetWidth{1.0}
    \LongArrow(394,-25)(398,-26)
    \LongArrow(400,15)(400,18)
    \SetWidth{0.8}
    \Photon(401,-23)(401,14){1.5}{5.5}
    \Text(283,32)[lb]{{\Black{$k_4$}}}
    \Text(395,35)[lb]{{\Black{$k_{n-1}$}}}
    \Text(172,-80)[lb]{{\Black{$-k_1$}}}
    \Text(397,-80)[lb]{{\Black{$-k_n$}}}
    \Text(316,-37)[lb]{{\Black{$q_3$}}}
    \Text(370,-39)[lb]{{\Black{$q_{n-3}$}}}
    \SetWidth{1.0}
    \LongArrow(402,-30)(402,-27)
    \SetWidth{0.8}
    \Photon(402,-68)(402,-31){1.5}{5.5}
    \Text(225,-41)[lb]{{\Black{$\kappa_{12}$}}}
    \Text(283,-38)[lb]{{\Black{$\kappa_{23}$}}}
    \SetWidth{1.0}
    \LongArrow(174,13)(174,16)
    \SetWidth{0.8}
    \Photon(175,-25)(175,12){1.5}{5.5}
    \Photon(223,-26)(175,-25){1.5}{7.5}
    \SetWidth{1.0}
    \LongArrow(222,-25)(226,-26)
    \LongArrow(229,13)(229,16)
    \SetWidth{0.8}
    \Photon(230,-26)(230,11){1.5}{5.5}
    \Text(171,27)[lb]{{\Black{$k_2$}}}
    \Text(228,28)[lb]{{\Black{$k_3$}}}
    \Text(254,-36)[lb]{{\Black{$q_2$}}}
    \Text(195,-36)[lb]{{\Black{$q_1$}}}
    \Text(195,27)[lb]{{\Black{$s_1$}}}
    \Text(255,25)[lb]{{\Black{$s_2$}}}
  \end{picture}
%}
\end{center}
\caption{Multiperipheral limit for the 2 to n-2 gluon scattering amplitude
in the tree approximation.}
\label{fig:multi_gluon}
\end{figure}
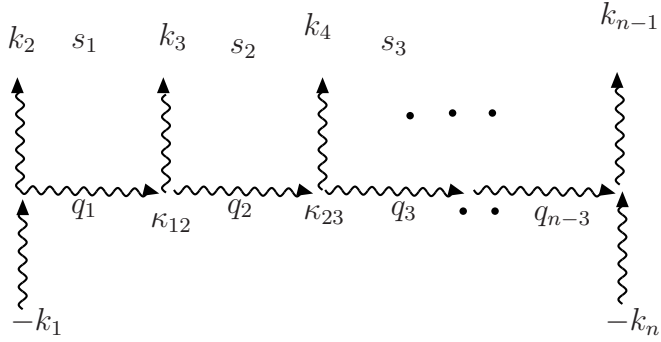

In the multi-Regge limit, the full amplitude (at least for MHV helicities) can be written as a sum over contributions from $2^{n-2}$ distinct color permutations, $\pi(\tau_2,\tau_3,\cdots)$. Each contribution  is  a product of three factors,
\be
 {\cal A}_n(1,2,  \cdots, n) = \sum_{\pi([\tau])} Tr[T^{a_{\pi(1)}}  T^{a_{\pi(2)}}
\cdots T^{a_{\pi(n)}} ] A^{tree}_{n,[\tau]}(k_{1}, \epsilon_{1},  \cdots, k_{n},\epsilon_{n}) \; 
M_{n,[\eta]}(k_{1}, \cdots, k_{n}) 
\ee
the color trace for each cyclic order, the tree diagram and the reduced amplitude $M_n$ with no dependence on the polarization or color labels. 
Thus it is useful to start with the tree approximation.

\subsection{Factorization of the MHV Tree Diagram}
\label{subsec:alln}

The factorization of $A_{n,tree}$ can be understood both from perturbation theory and open string theory.  The Regge limit of the planar Born term, as depicted in Fig.~\ref{fig:multi_gluon},
\be
A^{tree}_{n, [\tau]}(k_1, \epsilon_1,\cdots , k_n,\epsilon_n) 
=   g^n \; s\; \tau_2 \; \epsilon_1\cdot \epsilon_2 \frac{1}{t_1} (\tau_3 \epsilon_3\cdot \gamma_3) \frac{1}{t_2} 
\cdots \frac{1}{t_{n-3}}\tau_{n-1} \epsilon_{n-1}\cdot \epsilon_{n} 
\ee
where $\gamma_i(q_{i-2},q_{i-1}) \simeq - q^\perp_{i-2} -
q^\perp_{i-1} + \cdots$ is a reduced  vertex, related to the
three gluon effective vertex for the peripheral multi-gluon high energy limit,
\be
\Gamma^\mu_{\nu,\nu'}(q,q') =  \frac{ k^\nu_1 k^{\nu'}_n}{s} \gamma^\mu(q,q')
\ee
which was first discussed in ~\cite{BL,Kuraev:1977fs}  for their treatment of perturbative Pomeron in QCD. (For an elementary treatment, see~\cite{FR} and references therein.) The longitudinal components
are fixed by the on shell gauge condition, $k^\mu_i \gamma^\mu_i(q_{i-2},q_{i-1}) =0$, and a gauge choice,
$\gamma^\mu_i \rightarrow \gamma^\mu_i + k^\mu_i$.   We also note that, to leading
order, using
$ s \simeq  s_1\kappa_{12}^{-1}  s_2  \kappa_{23}^{-1}  s_3 \cdots
 \kappa_{n-4,n-3}^{-1} s_{n-3}$, 
\be
A^{tree}_{n, [\tau]}(1,2,\cdots n)
\sim \tau_2 \;  \frac{s_1}{t_1} \; \tau_3  \kappa_{12}^{-1} \;
\frac{s_2}{t_2} \; \tau_4 \kappa_{23}^{-1} \; \frac{s_3}{t_3} \cdots\cdots
  \tau_{n-3}\kappa_{n-4,n-3}^{-1} \; \frac{s_{n-3}}{t_{n-3}} \; \tau_{n-1} 
\ee
where we have dropped $g^n$  and the polarizations
factors to simplify the expression. (For a more explicit treatment, see \cite{DelDuca:1995zy}.)

From the open string perspective (with co-ordinate parameter $w =\tau + i\sigma$ , $\sigma \in [0,\pi]$) the signs ($\tau_i = \pm 1$) are world sheet charge conjugations implemented by applying the twist operator, $\Omega = (-1)^N$, to permute the gluon vertex:
\be
\epsilon_i(k) \cdot \dd_\tau X(\sigma,\tau) e^{i k_i  X(\sigma,\tau) }\rightarrow \epsilon_i(k) \cdot \dd_\tau X(\pi- 
\sigma,\tau) e^{i k_i  X(\pi- \sigma,\tau) }
\ee
It is well known that in the zero slope limit $\alpha' \rightarrow 0$ that both the bosonic and super string reproduces planar trees for $N_c = \infty$ Yang Mills theory. Indeed this is the original inspiration for the Parke-Taylor and MHV developments. 

Next consider the factorization of the color trace for the $2^{n-2}$
permutations enumerated by $\tau_i$. Using the procedure identical to
the established analysis of Chan-Paton~\cite{Paton:1969je} factors in
open string theory, the traces can be factored in the t-channel using
completeness for $U(N_c)$, $\sum_a T^a_{ij}T^a_{lm}=2
\delta_{im}\delta_{jl}$, 
where we have adopted the  normalization: $Tr[T^aT^b]=2\delta_{ab}$. 
Beginning with no   color twists
($\tau_i = 1$) the trace is factorized as
\be
Tr[12 \cdots n] =2^{-(n-3)}  Tr[a_1 a_2 c_1] Tr[c_1 a_3  c_2] Tr[c_2 a_4 c_3] \cdots 
Tr[c_{n-3} a_{n-1} a_n]
\label{eq:colorLinearRegge} 
\ee
The permutation corresponding to flipping the  $i$th outgoing
gluon ($\tau_i = -1$) in the color trace corresponds to complex
conjugation: $Tr[c_{i-1} a_i c_{i-2}] = Tr^*[c_{i-2} a_i
c_{i-1}]$. Thus the triple trace $Tr[abc] = d^{abc} + i f^{abc}$
at each vertex is replace by a factor,
$$T^{abc}_\tau = d^{abc} + i \tau f^{abc} \;. $$ 
This combined with anti-cyclic
symmetry of the n-gluon amplitudes ${\cal A}_n(1,2,\cdots, n)= (-1)^n
{\cal A}_n(n,\cdots 2,1)$ implies that there are always an even number
of D vertices in each monomial in accord with our explicit  $n = 4$
and $n = 5$ forms.

Applying this to the Regge limit of the Born
approximation alone, the sum over $\tau$'s remove all D-terms,
\bea 
{\cal A}_{n,tree} &\sim& 2^{-(n-3)} (g)^{n-2} \sum_{\tau_i} \tau_2
T_{\tau_2}^{a_1a_2c_1} (s_1/t_1)\kappa_{12}^{-1} \tau_3 T_{\tau_3}^{c_1a_3c_2} (s_2/t_2)\kappa_{23}^{-1}
\cdots (s_{n-3}/t_{n-3}) \nn
 &=&2\; ( i g)^{n-2}f^{a_1a_2c_1}
f^{c_1a_3c_2} \cdots f^{c_{n-3}a_{n-1}a_{n}}(s_1/t_1)\kappa_{12}^{-1} (s_2/t_2)\kappa_{23}^{-1} \cdots
(s_{n-3}/t_{n-3}) \nn
\eea
where for simplicity we have ignored the polarization factors.  Without this extra factor, $\tau_2 \cdots \tau_{n-1}$, 
the Born term would not agree with perturbation theory at the tree level. 

\subsection{Full Multi-Regge Factorization}

To proceed to the full Regge limit we now must consider the
factorization of the reduced amplitude. As discussed in Sec.
for both super string theory and the BDS amplitudes, each of the $2^{n-2}$ individual linear Regge amplitudes takes the  form of multi-Regge dependence,
\bea
A_n(\tau_i)/A^{tree}_{n,[\tau]} &\simeq& (-\eta_1)^{\alpha_1-1} (-\eta_2)^{\alpha_2-1}\cdots  (-\eta_{n-3})^{\alpha_{n-3}-1} \; s_1^{\alpha_1-1} s_2^{\alpha_2-1}\cdots  s_{n-3}^{\alpha_{n-3}-1}\nn
&\times&  \gamma(t_1)\gamma(t_{n-3}) B_n(\eta_1,\eta_2,\cdots, t_1,\kappa_{12}, \cdots)\;.
\label{eq:linearRegge}
\eea
Putting the color traces (\ref{eq:colorLinearRegge}) together with polarizations and the Regge amplitudes (\ref{eq:linearRegge}) 
leads to a separation between color-/helicity-factors and planar amplitudes
\bea
{\cal A}_{n Regge} &\simeq& \sum_{\tau} 
\widetilde t^{a_1a_2c_1}_{\tau_2}  \widetilde T^{c_1 a_3 c_2}_{\tau_3} \widetilde T^{c_2 a_4 c_3}_{\tau_4} \cdots\cdots \widetilde t^{c_{n-3}a_{n-1}a_n}_{\tau_{n-1}}   A_n(\eta_1,\eta_2,\cdots)  \label{eq:calAn}
\eea
where $\eta_j=\tau_{j+1}\tau_{j+2}$, and $\{A_n(\eta_1,\cdots)\}$   are planar amplitudes discussed in Sec. \ref{sec:FactorizationConstraints},
\bea
A_n(\eta_1,\eta_2,\cdots) &=&(-\eta_1)^{\alpha_1} (-\eta_2)^{\alpha_2}\cdots  (-\eta_{n-3})^{\alpha_{n-3}} \; s_1^{\alpha_1} s_2^{\alpha_2}\cdots  s_{n-3}^{\alpha_{n-3}}\\
&\times &  (-t_1 \kappa_{12})^{-1}(-t_2 \kappa_{23})^{-1}\cdots \gamma(t_1)\gamma(t_{n-3}) B_n(\eta_1,\eta_2,\cdots, t_1,\kappa_{12}, \cdots) \nonumber
\label{eq:linearRegge2}
\eea
In arriving at (\ref{eq:calAn}), we have supplied a factor of $\eta_1\eta_2\cdots\eta_{n-3}$, which was left out in going from  $M_n$ to $A_n$ for $n=5,6$ in the previous section. We have also made use of the fact that  $\eta_1\eta_2\cdots\eta_{n-3}=\tau_2\tau_{n-1}$, thus removing factors $\tau_2$ and $\tau_{n-1}$  at the ends of the MR chain  coming from the tree. 
The vertices now include color labels and polarization vectors, 
\be
\widetilde T^{cac'}_{{\tau_j}} = ( g\; \epsilon_j
\cdot \gamma_j/2) \; (\tau_{j} d^{cac'} + i f^{cac'})
\ee
and at the ends reduce to
\bea
\widetilde t^{abc}_{\tau_2}(t_1)& =& ( g \; \epsilon_1 \cdot
\epsilon_2/\sqrt 2)  ( d^{abc} + i \tau_2 f^{abc})\;, \nn
 \widetilde t^{c'a'b'}_{\tau_{n-1}}(t_{n-3}) &=& ( g \; \epsilon_{n-1} \cdot
\epsilon_n/\sqrt 2)  ( d^{a'b'c'} + i \tau_{n-1} f^{a'b'c'})\;.   \label{eq:ends}
\eea
 because one of the
Reggeons is replaced by an on-shell gluon.  
Lastly, by inserting a factor $(\eta_j+\tau_{j+1}\tau_{j+2})/2$ for each $\eta_j$, we arrive at
\bea
{\cal A}_{n Regge} &\simeq& \sum_{\eta}\sum_{\tau}\; \widetilde t^{a_1a_2c_1}_{\tau_2} \; \widetilde T^{c_1 a_3 c_2}_{\tau_3}  \cdots\cdots \widetilde t^{c_{n-3}a_{n-1}a_n}_{\tau_{n-1}} \nn
&\times&\left[(\eta_1+\tau_{2}\tau_{3})/2\right] \left[(\eta_2+\tau_{3}\tau_{4})/2\right]\cdots\cdots A_n(\eta_1,\eta_2,\cdots) \label{eq:calAnb}
\eea
where $\tau$ and $\eta$ are now independent sums. 

So far, our discussion has been general, applicable to both BDS and flat-space string theory, (other that the replacement for the propagator $(-1/t)$ factor by $\Gamma(1-\alpha(t))$.) Let us next turn to the assumption  of factorization in signature space.   Following the discussion in Sec . \ref{sec:FactorizationConstraints}, the reduced amplitude factorizes in signature space. More directly, for the planar amplitudes $A_n$, we have
\be
\widetilde A_n(\sigma_1,\cdots)= \left[s_1^{\alpha_1}s_2^{\alpha_2}\cdots \right]  \widetilde  \Pi_{\sigma_1} \;  \widetilde G_{\sigma_1\sigma_2} \; \widetilde \Pi_{\sigma_2} \;\cdots\cdots   \widetilde G_{\sigma_{n-4}\sigma_{n-3}}\;  \widetilde \Pi_{\sigma_{n-3}} \;,
\ee 
where $ \widetilde \Pi_{\sigma_{j}} $ stands for $ \widetilde \Pi_{\sigma_{j}}(j) $, $ \widetilde G_{\sigma_j\sigma_{j+1}} $ for $ \widetilde G_{\sigma_j\sigma_{j+1}}(j,j+1) $, and $A_n$ is given by an inverse transform,
\be
A_n(\eta_1,\cdots)= 2^{-(n-3)/2}\sum_{\sigma}\;\Omega_{\eta_1,\sigma_1}\Omega_{\eta_2,\sigma_2}\cdots  \widetilde A_n(\sigma_1,\sigma_2,\cdots)\;.  \label{eq:An}
\ee
Substituting this into (\ref{eq:calAnb}), the $\eta$ sum can be carried out, leading to
\bea
{\cal A}_{n Regge} &\simeq&  \sum_{\sigma}\sum_{\tau}\; \Pi_j   \left[ \left(\frac{1+\sigma_j}{2}\right)\tau_{j+1}\tau_{j+2}+ \left(\frac{1-\sigma_j}{2}\right)\right] 
\nn
&\times&  \widetilde t^{a_1a_2c_1}_{\tau_2} \; \widetilde T^{c_1 a_3 c_2}_{\tau_3}  \cdots\cdots\widetilde T^{c_{n-4} a_{n-2} c_{n-1}}_{\tau_{n-2}} \widetilde t^{c_{n-3}a_{n-1}a_n}_{\tau_{n-1}} \;
\widetilde A_n(\sigma_1,\cdots)   \;.    \label{eq:calAnc}
\eea
In more explicit
form as a matrix product, first we define $V^{abc}_{\tau \sigma, \tau'\sigma'}\equiv  \widetilde G_{\sigma,\sigma'}  \widetilde T^{abc}_{\tau}  \delta
_{\tau,\tau'}$,    $\gamma^{abc}_{\tau \sigma}\equiv \widetilde t^{abc}_{\tau}  $  and then 
we represent the Reggeon propagator
 in $\tau$ and $\sigma$ 
\be
\Delta_{\tau\sigma,\tau'\sigma'}(s,t) = \frac{ [(1-\sigma)+ (1+\sigma )\tau \tau']}{2} \Gamma(t) \xi_\sigma (t) ( s)^{\alpha(t)}
\ee

Multi-Regge factorization,
\bea
{\cal A}_{n,Regge}& \simeq&  s_1^{\alpha_1} s_2^{\alpha_2}\cdots  s_{n-3}^{\alpha_{n-3}} \; \gamma^{a_1, a_2}(t_1) \Delta(s_1, t_1) V^{a_3}(t_1, \kappa_{12}, t_2) \nn
&\times& \Delta(s_2,t_2) 
V^{a_4}(t_1, \kappa_{23}, t_3) \cdots \gamma^{a_{n-1}a_n}(t_{n-3})  \; ,
\eea
now takes on  the form of  a product of
$4$ by $4$ propagator matrices, with $\tau = \pm 1,\sigma=\pm 1$, and
Reggeon-Reggeon particle vertices that are $4 N^2_c$ by $4 N^2_c$ matrices,
if we include colors. (Helicity labels have been suppressed.)

A more convenient  form is to diagonalize the Reggeon propagator. This is
achieved by performing an $SU(2)$ rotation, $U = \exp[ - i \pi
\sigma_2/2]$ by $45^\circ$ from the $\tau$ (``twist'') basis to the $\chi$
(``color'') basis
\be
U_{\chi,\tau} =  (1/\sqrt{2}) \left(\begin{array}{ll}
1 & -1 \\
1 & 1\\
\end{array}\right)_{\chi,\tau}
\ee
so the Reggeon propagator becomes (in the vector space $(\chi=+1,\chi=-1)$)
\be
\Delta_{\chi\sigma,\chi'\sigma'}=    \left(\begin{array}{ll}
 (1+\sigma)  & 0 \\
0 &(1-\sigma) \\
\end{array}\right)_{\chi,\chi'} \xi_\sigma \Gamma(t)  s^{\alpha(t)}\;  \delta_{\sigma,\sigma'} \; .
\ee
More explicitly, we can express the propagator as a diagonal matrix in $\chi,\sigma$,  with diagonal elements:
\be
\Delta_{\chi\sigma}(s,t) =\Delta_\sigma(s,t)\delta_{\sigma,\chi}
= 2 \xi_\sigma  \Gamma(t)  s^{\alpha(t)}\delta_{\sigma,\chi}
\label{eq:Reggeprop}
\ee
 Factorization in fact only 
involves a pair of  degenerate trajectories ( $\sigma=\pm 1$), 
or 2 trajectories in conventional parlance.

This rotation also separates the D- and  F- color factors. Replacing $\chi$ by $\sigma$,  
\be
\gamma_\sigma^{abc}(t_1)  =   U_{\tau, \sigma} \gamma_\tau^{abc}(t) =( g \; \epsilon_1 \cdot
\epsilon_2)  \; \gamma_\sigma^{abc} =( g \; \epsilon_1 \cdot
\epsilon_2 ) \;  
 \left(\begin{array}{l}
d^{abc}\\
-i f^{abc} \\
\end{array}\right)_\sigma  
\ee
so that $\sigma = -1$ is the F-term, consistent with the vertex, (\ref{eq:gamma}), introduced earlier. We also must 
transform the 2-Reggeon vertex to color space, and, again  replacing $\chi$ by $\sigma$, 
\be
V^{abc}_{\sigma_1\sigma_2}(t_1, \kappa_{12},t_2) =  ( g\; \epsilon_3
\cdot \gamma_3/2) C^{abc}_{\sigma_1\sigma_2}
 \; \widetilde G_{\sigma_1 \sigma_2}(t_1, t_2;\kappa_{12})
\label{eq:RRvertex}
\ee
where
\be
C^{abc}_{\sigma_1\sigma_2}= 
\left(\begin{array}{ll}
i f^{abc} & d^{abc} \\
d^{abc} & if^{abc} \\
\end{array}\right)_{\sigma_1,\sigma_2}
\;.
\ee
Finally, we have
\bea
{\cal A}_{n,Regge} &\simeq& s_1^{\alpha_1} s_2^{\alpha_2}\cdots  s_{n-3}^{\alpha_{n-3}}\; \gamma_{\sigma_1}^{a_1 a_2 c_1}(1) \Delta_{\sigma_1}(1) V^{c_1a_3c_2}_{\sigma_1\sigma_2}(1,2)     \nn
&\times& 
\Delta_{\sigma_2}(2) V^{c_2a_4c_3}_{\sigma_2\sigma_3}(2,3) \cdots\cdots\Delta_{\sigma_{n-3}}(n-3) \gamma_{\sigma_{n-3}}^{c_{n-3}a_{n-1}a_n}({n-3})  \; .   \label{eq:fullsignature}
\eea
Now the restriction to even number of D-vertices 
is explicit. Starting with an F-vertex (for example), the D-vertices
are ``kink'' operators flipping the sign of $\sigma$ so that kink/anti-kink
pairs guarantees this condition. We also note that
the 5-point function is special with only one vertex. In general
the kink/anti-kink pairs can be separated.

\subsection{Illustration: Signature Representation for 4-Gluon and 5-Gluon Amplitudes}

To understand this
somewhat formal construct, let us consider the special cases for $n
=4$ and $5$ which are especially simple.  

For the 4-point
function, one immediately obtains
\bea
A_{4,Regge} &=& \sum_{c_1,\sigma_1}  \gamma_{\sigma_1}^{a_1a_2c_1}(1) \Delta_{\sigma_1}(1) \gamma_{\sigma_{1}}^{c_1a_3a_4}(1) \;  s^{\alpha(t)} \\
&=&g^2\; \left(\epsilon_1 \cdot \epsilon_2\; \epsilon_3 \cdot \epsilon_4\right)  \left[ (- f^{a_1a_2c}f^{ca_3a_4}) \xi_-(t) +( d^{a_1a_2c} d^{ca_3a_4}) \xi_+(t) \right] \Gamma(t) s^{\alpha(t)} \;.\nonumber   \label{eq:fullsignature4}
\eea
Let us now see how this agrees with a more direct analysis.  For $n=4$, there are $4!=24$ color permutations for planar amplitudes in Eq~ (\ref{colorord}). Taking into account of cyclic  symmetry reduces this  to $6$
independent contributions, and they can be enumerated as three pairs of planar amplitudes with singularities in the s-t, u-t and s-u
Mandelstam invariants. Only  the s-t and u-t amplitudes
contribute to the Regge exchange in the t-channel,
\bea
{\cal A}_4(k_i,a_i)&=& Tr[1234] A_4(1234)+ Tr[2134] A_4(2134) \nonumber\\
&+& Tr[1243] A_4(1243) + Tr[2143] A_4(2143) + \mbox{(s-u) terms}
\label{4point}
\eea
Note that, with  $[T^a,T^b]= i 
f_{abc} T^c$ and $\{T^a,T^b\}= d_{abc} T^c$, we obtain
\be
Tr[T^{a_i}T^{a_j}T^{a_k}T^{a_l}] = (1/2) (i f^{ija} + d^{ija} )(i f^{akl} + d^{akl} )
\ee
Combining this with the condition of invariance of the planar 4-gluon amplitude
under anti-cyclic permutations: $A_{st}=A_{++}=A_4(1234)=A_4(2143)=A_{--}$ and
$A_{ut}=A_{-+}=A_4(2134)=A_4(1243)=A_{+-}$, we have
\be
{\cal A}_4(k_i,a_i) =  - f^{ija}f^{akl} (A_{++} - A_{-+}) + d^{ija} d^{akl} (A_{++} + A_{-+})    + \mbox{(s-u) terms}
\ee
In the Regge limit  $s\rightarrow +\infty$, $t<0$ fixed,
the amplitude (\ref{4point}) factorizes with two degenerate trajectories of opposite
signature (or opposite charge conjugation)
\be
{\cal A}_4(k_i,a_i)/A_{4,tree} \simeq  \sum_{\sigma  = \pm 1} \gamma^{a_1 a_2 c}_\sigma (t)  \left( e^{-i\pi \alpha(t)}  
+ \sigma  \right) s^{\alpha(t)-1} \gamma^{ca_3 a_4}_\sigma (t)
\label{signature4}
\ee
where
\be
\gamma^{abc}_\sigma (t) =\gamma(t) \left\lbrace
\begin{array}{ll}
d^{abc} \;, & \sigma = +1\\
-i f^{abc} \;, & \sigma = -1\\
\end{array}
\right .
\label{eq:gamma}
\ee

When compared with Eq. (\ref{eq:fullsignature4}), the only modifications are factors from the tree specifying the  polarizations. 

Note that the gluon exchange corresponds to odd-signature exchange, with F-coupling, as expected.  The F-coupling trajectory, which contains
the gluon pole at $t=0$, has odd signature. We shall occasionally
refer to this as the color ``octet-trajectory''. The D-coupling
trajectory, containing both a color ``singlet'' and an octet
component, has even signature and does not have a pole at $t=0$, and its contribution vanishes at the tree level.
Nevertheless, at one-loop and beyond, the even-signature persists in the
BDS and super string amplitudes. Indeed having contributions from leading
trajectory with both signatures is also characteristic of type-II
oriented open strings with the ends attached to D-branes.

Turning next to $n=5$. From (\ref{eq:fullsignature}), the 5-point function is
\be
A_{5,Regge} = \gamma^{a_1a_2c_1}_{\sigma_1}(1)\Delta_{\sigma_1}(1)V^{c_1a_3c_2}_{\sigma_1,\sigma_2}(1,2) 
\Delta_{\sigma_2}(2) \gamma^{c_2a_4a_5}_{\sigma_2}(2)  \;s_1^{\alpha_1} s_2^{\alpha_2}
\ee
Substituting in various expressions, we obtain
\bea
A_{5,Regge} &=&  g^3\left(\epsilon_1 \cdot \epsilon_2\right) \; \left(\epsilon_3\cdot \gamma_3/2\right)\left(\epsilon_4 \cdot \epsilon_5\right) \left[\gamma^{a_1a_2c_1}_{\sigma_1}
 C^{c_1a_3c_2}_{\sigma_1\sigma_2} \gamma^{c_2a_4a_5}_{\sigma_2}
  \right]   \nn
&\times&  \; \widetilde G_{\sigma_1 \sigma_2}(t_1, t_2;\kappa_{12}) \left( \xi_{\sigma_1}
(t_1)\Gamma(t_1)s_1^{\alpha(t_1)}\right)  \left( \xi_{\sigma_2}
(t_2)\Gamma(t_2) s_2^{\alpha(t_2)}\right)  \label{eq:fullsignature5}
\eea

To clarify this result, let us again return to a more direct analysis for factorization in the double Regge limit of the
5-point function illustrated in Fig.  \ref{fig:regge5both}.  There are now $5!$ color configurations. To contribute to the
double-Regge limit, the planar amplitudes must have Regge singularities
in $t_1=-(k_1+k_2)^2$ and $t_2=-(k_4+k_5)^2$, so 
that $(1,2)$ and $(4,5)$ are adjacent.  The sum over permutations of $Tr[12345]
A(12345)$, with  $(1,2)$ and $(4,5)$ lines adjacent,  yields  8
terms. The color traces can be factored on the Regge exchanges as
$Tr[12345] =(1/4) Tr[12c_1]Tr[c_13c_2] Tr[c_245]$. Altogether, this  leads to 8
combinations of F- and D-terms.

The 8 permutations of $A(12345)$ are analytically continued from the
Euclidean to the physical scattering region independently to give the
factorized form. 
For $n=5$, because amplitudes are odd  under anti-cyclic permutations,
$A(12345)  = - A(54321)$, this further  reduces  to 4 independent contributions
in the Regge limit as illustrated in Fig.~\ref{fig:twist5.a}.
The total contribution in the physical region is 
\bea
{\cal A}_5 &\simeq & [Tr(12345)-Tr(54321)] A_{++}  + [Tr(12354)-Tr(45321)] A_{+-}  \nonumber\\
&+& [Tr(21345)-Tr(54312)]  A_{-+} +[Tr(21354)-Tr(45312)] A_{--} \label{T5}
\eea
where, in the double-Regge limit, $A_{\eta_1\eta_2}$ is given by Eq. (\ref{eq:doubleregge}).
Again, dependence on  polarizations has been suppressed.

Eq. (\ref{T5})  can directly be expressed in  ``signatured" representation,
 in terms of the 2-gluon Regge vertex
(\ref{eq:gamma}) and a new double Regge vertex,
\be
A_{5,Regge} =  \gamma^{a_1a_2c_1}_{\sigma_1}(t_1)(e^{-i\pi\alpha(t_1)} +\sigma_1) 
V^{c_1a_3c_2}_{\sigma_1, \sigma_2} (e^{-i\pi\alpha(t_2)} + \sigma_2)
 \gamma^{c_2a_4a_5}_{\sigma_2}(t_2) 
\label{eq:5Regge}
\ee
where
\be
V^{abc}_{\sigma_1, \sigma_2} = 
\left\lbrace
\begin{array}{ll}
i f^{abc} \;    \widetilde G_{\sigma_1\sigma_2}(1,2)  \; \;, & \quad \sigma_{1} = \sigma_2\\
\; d^{abc} \;   \widetilde G_{\sigma_1\sigma_2}(1,2)  \;,   &\quad  \sigma_{1} = -\sigma_2\\
\end{array} \right . 
\label{Vvertex1}
\ee
with  $ \widetilde G_{\sigma_1\sigma_2}(1,2)$  given by Eq. (\ref{eq:signatured2vertex}). Again, when compared with Eq. (\ref{eq:fullsignature5}), the only modifications are factors from the tree specifying the  helicity and color configurations.

\newpage

\section{Discontinuity in ``Missing Mass'' $M^2$ for $n\geq 6$}

\label{sec:inclusive}
In Sec. \ref{sec:constraints}, we have discussed some aspects of analyticity constraints on the 5-point function in 
various Regge limits. In particular, we point out that, in the double-Regge limit, the BDS amplitude does not 
satisfies the Steinmann rules. Nevertheless, we have demonstrated in Secs. \ref{sec:FactorizationConstraints} and  \ref{sec:5plus} that this deficiency does 
not affect the discussion of signature factorization property in the physical region  for n-point amplitudes  in the linear multi-Regge limit.  In this section, we return to a closer examination of analyticity and unitarity constraints for higher point 
amplitudes. 

For $n \geq 6$, there now exists threshold singularities in the physical region in ``crossed invariants'' involving both 
initial and final momenta. Consider the amplitude for a 3-to-3 process, 
\be
a+b+x'\rightarrow a'+b'+x.
\ee
The 3-to-3 amplitude has discontinuity in  $M^2=-(p_a+p_b-p_x)^2$, the invariant in the so-called ``missing mass'' 
channel. Just as the 2-to-2 unitarity in the forward limit of $t=0$ leads to a total cross section, the discontinuity in 
$M^2$ in the forward limit where $p_a=p_{a'}$, $p_b=p_{b'}$ and $p_x=p_{x'}$  leads to the inclusive cross section 
for the process
\be
a+b\rightarrow x + {\rm anything},
\ee
 \be
 \frac {d\sigma}{dp_x} \sim (1/s) {\rm Disc}_{M^2} T_6(a,b,x'\rightarrow a',b',x)  \label{eq:inclusive}
 \ee
 where $M^2>0$.  This is a generalized Optical  Theorem. If multi-Regge applies to the 6-point
 function, taking the discontinuity in $M^2$ leads to a non-trivial
 prediction for the inclusive cross section.
 
We focus in this section on such discontinuities in the physical regions. 
 Surprisingly, we find the absence of Regge contribution
 in the ``triple-Regge'' limit of the 6-point function. More
 generally,  BDS amplitudes do not
 lead to a well-defined ``Reggeon-particle'' 4-point amplitude with
 the expected $M^2$-discontinuity, based on flat-space string
 expectation. The absence of such  flat-space behavior 
 also holds for  $ n > 6$. 
 
\begin{figure}[bthp]
\begin{center}
\includegraphics[width = 0.8\textwidth]{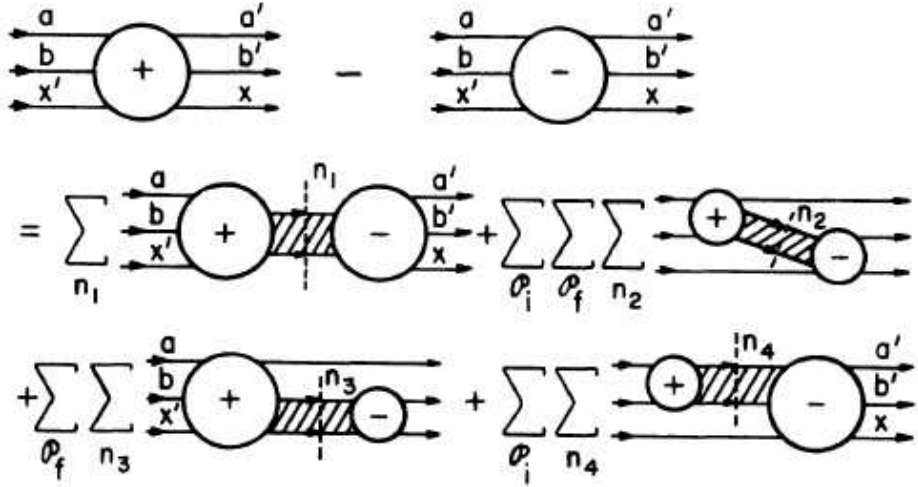}
\end{center}
\caption{3-to-3 unitarity in the physical region. The sum over $n_1$  for the  first term on the right represents all allowed intermediate states in the $abx'$ channel. ${\cal P}_i$ and ${\cal P}_f$ represent sums over different initial- and final-state combinations for various intermediate states, labelled by sums $n_2$, $n_3$ and $n_4$. }
\label{fig:3to3unitarity}
\end{figure}

We end this introduction by illustrating   the relevant 3-to-3 unitarity condition  which can be represented 
schematically   by Fig. \ref{fig:3to3unitarity}.
There are now four types of terms on the right hand side of this
unitarity relation.~\cite{Stapp:1971hh,Tan:1972kr,Cahill:1973px,Detar:1971dj} Each
term can be associated with a physically realizable
re-scattering process, and  each is a discontinuity in an
appropriate invariant. For the third and the fourth terms, they
represent discontinuities in initial or final sub-energy invariants, a
generalization of that discussed earlier for the 2-to-3 unitarity. The
second term is new; it represents discontinuities in ``crossed
invariants'' involving both initial and final momenta, e.g., the missing-mass invariant, $M^2=
-(p_a+p_b-p_x)^2$ introduced above.  We examine  in this section  properties of the  discontinuity in $M^2$ in 
various  Regge  limits. 

Another important reason for studying the helicity-pole/triple-Regge limit is the fact that  functions  ${\rm Li}_2(1-u_i)
$, which enter in the  n-point amplitudes for $n \geq 6$, now become even more important. In all the Regge limits that we have 
studied  in \cite{Brower:2008nm}, the cross ratios, 
$u_1,u_2,u_3$, (\ref{crossr}),  remain finite, moreover taking on  values of either 0 or 1, in the linear multi-Regge 
limit, and as a consequence the terms 
${\rm Li}_2(1-u_i)$ in the BDS ansatz for $n\geq 6$ gave only constants, and thus did not influence the limit. Any 
other 
well-behaved function $f(u_1,u_2,u_3)$
of the cross ratios $u_i$  at 0 and 1 that one could in principle add to the BDS ansatz while still 
respecting dual 
conformal invariance  \cite{Drummond:2007aua,Drummond:2007au,Drummond:2007bm} would have 
become also irrelevant. It is therefore crucial for the consistency of the (corrected) BDS ansatz to examine limits 
where this does not 
happen, and the $u_i$'s are infinite. Notice that from the point of view of BDS invariants, $t_i^{[r]}$'s, the simplest 
limit 
possible would involve only two variables (one independent and one dependent) going to infinity. This will turn out 
to be the helicity-pole/triple-Regge limit, and in this limit indeed we find that  two of the
 $u_i$'s  will become large.

\subsection{Triple-Regge Limit and Expectations from Flat Space String Theory}

\begin{figure}[bthp]
\begin{center}
%\fcolorbox{white}{white}{
  \begin{picture}(427,150) (0,-25)
    \SetWidth{0.5}
%    \SetColor{Black}
    \Text(395,5)[lb]{{\Black{$b$}}}
    \Text(395,36)[lb]{{\Black{$x$}}}
    \Text(396,84)[lb]{{\Black{$b'$}}}
    \Text(142,93)[lb]{{\Black{$2$}}}
    
       \Text(70,-25)[lb]{{\Black{$(a)$}}}
        \Text(220,-25)[lb]{{\Black{$(b)$}}}
         \Text(345,-25)[lb]{{\Black{$(c)$}}}
    
    \SetWidth{0.5}
    \GOval(66,44)(8,24)(0){0.882}
    \Photon(91,45)(114,46){3.5}{2}
    \ArrowLine(118,7)(112,44)
    \ArrowLine(113,45)(119,83)
    \ArrowLine(51,51)(43,83)
    \ArrowLine(83,51)(89,82)
    \ArrowLine(66,52)(66,83)
    \ArrowLine(72,52)(74,82)
    \ArrowLine(77,52)(81,82)
    \ArrowLine(61,52)(59,82)
    \ArrowLine(57,52)(52,82)
    \Line(134,95)(134,0)
    \Line(194,59)(194,60)
    \ArrowLine(254,67)(270,91)
    \ArrowLine(269,45)(254,67)
    \ArrowLine(269,0)(255,20)
    \ArrowLine(255,20)(268,37)
    \Photon(254,66)(228,56){3.5}{3}
    \GOval(212,52)(5,19)(0){0.882}
    \GOval(213,28)(4,19)(0){0.882}
    \Photon(254,19)(225,22){3.5}{3}
    \ArrowLine(201,56)(187,88)
    \ArrowLine(187,1)(201,24)
    \Line(202,32)(202,48)
    \Line(209,32)(209,47)
    \Line(216,32)(216,47)
    \Line(224,32)(224,48)
  \DashLine(186,40)(239,40){10}
    \ArrowLine(322,44)(307,90)
    \ArrowLine(308,0)(322,47)
    \ArrowLine(374,67)(388,90)
    \LongArrow(150,40)(172,40)
    \LongArrow(283,40)(305,40)
    \Text(122,12)[lb]{{\Black{$b$}}}
    \Text(122,81)[lb]{{\Black{$x$}}}
    \Text(314,91)[lb]{{\Black{$a'$}}}
    \Photon(321,45)(352,46){3.5}{3}
    \ArrowLine(389,46)(374,67)
    \ArrowLine(375,22)(388,42)
    \Photon(351,45)(375,68){3.5}{3}
    \Photon(351,46)(375,22){3.5}{3}
    \ArrowLine(389,0)(375,23)
    \Text(319,2)[lb]{{\Black{$a$}}}
    \Text(388,59)[lb]{{\Black{$x'$}}}
    \ArrowLine(43,8)(51,38)
    \Text(49,8)[lb]{{\Black{$a$}}}
    \Line(30,95)(30,0)
  \end{picture}
%}
\end{center}
\caption{Triple-Regge behavior of inclusive cross section  as $M^2$-discontinuity. }
\label{fig:TRRegion}
\end{figure}
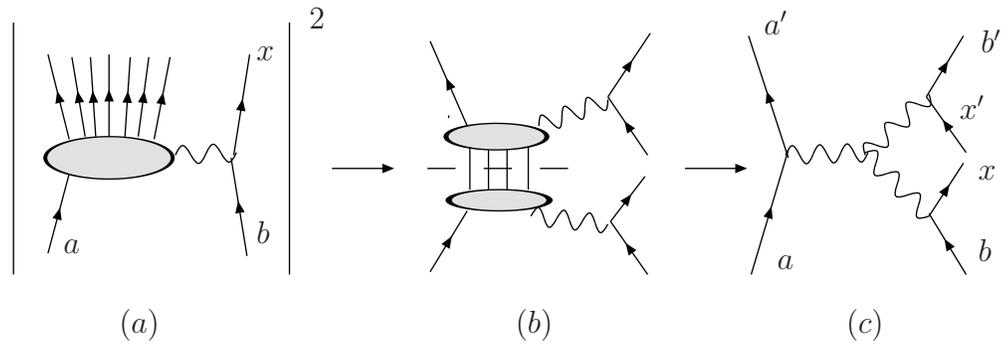

Let us first consider the ``triple-Regge'' limit. Kinematically, an inclusive cross section, Eq. (\ref{eq:inclusive}), can 
be treated as   a 2-to-2 cross section, with $M^2$ the mass-squared  for one of the two final particles, as illustrated 
schematically by the  Fig. \ref{fig:TRRegion}a. As such, it is a function of  three independent invariants, two being the energy 
and momentum-transfer invariants, $s$ and  $t$,  and the third being $M^2$. The triple-Regge limit corresponds to having
\be
s/M^2\rightarrow \infty,\quad M^2\rightarrow \infty
\ee
with $t$ fixed. The  standard Regge behavior first  leads to a factor $|s/M^2|^{2\alpha(t)}$, with $M^2$ serving as a 
scale.   There will be a second Regge factor, $\left(M^2\right)^{\alpha(0)}$, which accounts for  the increasing 
multiplicity of final states as $M^2$ grows. (Fig. \ref{fig:TRRegion}b.) As the $M^2$-discontinuity of a 6-point 
amplitude, the inclusive cross section  thus  takes on a triple-Regge form~\cite{Detar:1971gn,Detar:1971dj}
\be
d\sigma \sim (1/s) {\rm Disc}_{M^2} A_6  \sim (1/s) G(t) (M^2)^{\alpha_0} |s /M^2|^{2\alpha_2}\; ,
\label{inclusiveforward}
\ee
which can be represented schematically by Fig. \ref{fig:TRRegion}c.

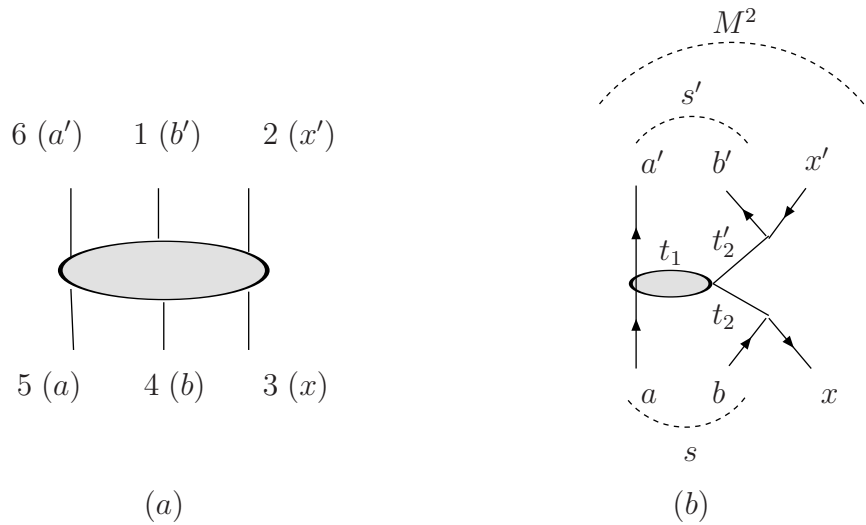
\begin{figure}
\begin{center}
%\fcolorbox{white}{white}{
\begin{picture}(331,170) (95,-90)

%  \begin{picture}(331,160) (95,-67)
    \SetWidth{0.5}
 %   \SetColor{Black}
 
    \Text(140,-90)[lb]{{\Black{$(a)$}}}
    
       \Text(340,-90)[lb]{{\Black{$(b)$}}}

    \ArrowLine(375,-13)(391,-32)
    \Text(355,42)[lb]{{\Black{$b'$}}}
    \Text(390,43)[lb]{{\Black{$x'$}}}
    \Text(355,-45)[lb]{{\Black{$b$}}}
    \Text(396,-45)[lb]{{\Black{$x$}}}
    \Text(328,-45)[lb]{{\Black{$a$}}}
       \Text(335,7)[lb]{{\Black{$t_1$}}}
         \Text(355,10)[lb]{{\Black{$t'_2$}}}
             \Text(355,-17)[lb]{{\Black{$t_2$}}}
 \DashCArc(361,26)(65,40,140){2}
   \GOval(338,0)(5,15)(0){0.882}
  \Text(355,95)[lb]{{\Black{$M^2$}}}
    \Text(343,68)[lb]{{\Black{$s'$}}}
    \Text(344,-67)[lb]{{\Black{$s$}}}
   \ArrowLine(325,-32)(325,1)
    \ArrowLine(325,1)(325,37)
    \DashCArc(346,36.15)(26.92,38.74,141.26){2}
    \DashCArc(343.6,-25.96)(28.15,-140.13,-37.27){2}
    \Text(328,42)[lb]{{\Black{$a'$}}}
      \ArrowLine(389,36)(375,17)
    \ArrowLine(360,-31)(374,-13)
         \Line(354,0)(375,18)
              \Line(354,0)(375,-12)
    \ArrowLine(374,18)(359,35)
      \Text(90,50)[lb]{{\Black{$6\; (a')$}}}
    \Text(92,-45)[lb]{{\Black{$5\; (a)$}}}
    \GOval(147,5)(11,39)(0){0.882}
    \Line(179,12)(179,36)
    \Line(179,-3)(179,-25)
     \Text(185,50)[lb]{{\Black{$2\; (x')$}}}
    \Text(140,-45)[lb]{{\Black{$4\; (b)$}}}
    \Text(185,-45)[lb]{{\Black{$3\; (x)$}}}
   \Line(112,10)(112,36)
    \Line(145,17)(145,36)
   \Line(112,-2)(113,-25)
    \Line(147,-7)(147,-25)
     \Text(136,50)[lb]{{\Black{$1\; (b')$}}}
  \end{picture}
%}
\end{center}
\caption{6-point amplitude  with color-ordering specified by the left figure, (a). Some of the invariants appropriate for 
the triple-Regge limit are shown in (b).}
\label{fig:6pt}
\end{figure}

We next generalize this triple-Regge behavior to the non-forward limit for the 3-to-3 process, $a+b
+x'\rightarrow a' + b' +x$.  We can begin with any color ordering so long as    $(a, b, x)$ are adjacent, i.e.,  an 
amplitude with  singularities in $M^2=-(p_a+p_b-p_x)^2$. 
For definiteness, let us consider first the
color-ordering $(123456)$ identified with $(a,a',b',x',x,b)$, as indicated by
Fig. \ref{fig:6pt}a.    Here, $a,b,x'$ are incoming and $a',b',x$ are outgoing, as indicated by arrows in Figs. 
\ref{fig:TRRegion}b, \ref{fig:TRRegion}c and \ref{fig:6pt}b. With an all-incoming momentum convention, one has  
$k_1=-p_{b'}$, $k_2= p_{x'}$, $k_3= - p_{x},$ $k_4=p_b$, $k_5=p_a$, and $k_6=-p_{a'}$. Amplitudes for
other orderings can then be obtained by appropriate substitutions and
analytic continuations. 

Let us denote adjacent BDS invariants by
\bea
t_4^{[2]}=  s ,  \quad t_6^{[2]}&=&s', \quad 
t_5^{[2]}=  t_1, \quad  
t_3^{[2]}=  t_2,  \quad t_1^{[2]}=  t_2'\; , \quad   t_2^{[2]}=s_{12}\;,  \nonumber\\
 t_3^{[3]}&=&  t_6^{[3]}\, =   M^2,\quad 
t_1^{[3]}= t_4^{[3]}= \Sigma\; , \quad t_2^{[3]}=t_5^{[3]}= \Sigma'\; .  \label{BDSinvariants}
\eea
The triple-Regge limit corresponds to 
\be
s/M^2 \rightarrow \infty,\quad s'/M^2\rightarrow \infty, \quad M^2\rightarrow \infty
\ee
with all other invariants fixed.
Since there are only eight independent variables, there will be a constraint among these nine invariants. As we 
have shown in Ref. \cite{Brower:2008nm},  the constraint simplifies in various Regge limits.
To isolate the singularity in $M^2$, we shall first go to the Euclidean region and then analytically continue $M^2$ to 
the physical region where $M^2>0$.~\footnote{For the physical region, some of the BDS invariants will have to 
be continued to positive values. For appropriate continuation procedure, see Refs. 
\cite{Tan:1972kr,Cahill:1973px,Detar:1971dj}.}   The forward limit has $s=s'$, $t_2=t_2'$, $t_1=0$, etc. 
Away from  the  forward  limit, (\ref{inclusiveforward}) generalizes to 
 \be
{\rm Disc}_{M^2} A_6  \sim   G(t_2,t_2';t_1) (M^2)^{\alpha(t_1) -\alpha(t_2) - \alpha(t_2')} (-s)^{\alpha(t_2)} (-
s')^{\alpha(t_2')}\label{TRstring}
  \ee
More precisely, for $M_6=A_6/A_{6,tree}$, one finds,  for  flat-space string theory, 
 \be
{\rm Disc}_{M^2} M_6  \sim   G(t_2,t_2';t_1) (M^2)^{\omega (t_1) -\omega(t_2) - \omega(t_2')} (-s)^{\omega(t_2)} (-
s')^{\omega(t_2')}\;, \label{TRstring2}
  \ee
where we recall that $\omega(t) = \alpha(t)-1$.

 It is also useful  to first examine a more general limit:  
  \be
 s\rightarrow - \infty\;, \quad s'\rightarrow - \infty 
 \ee
 with $M^2<0 $ fixed before taking the discontinuity in 
$M^2$.  This is historically referred to as the helicity-pole limit ~\cite{Detar:1971gn}.    For flat-space string theory, one finds~\cite{Detar:1972nd}
 \be
M_6  \sim  {\cal A}(M^2, t_2,t_2';t_1) (-s)^{\omega(t_2) } (- s')^{\omega(t_2')}  +   {\cal B}
\label{hplimitgood}
  \ee
where $ {\cal B}$ has no discontinuity in $M^2$.  ${\cal A}(M^2, t_2,t_2';t_1)$  is known as the Reggeon-particle-to-Reggeon-particle amplitude.~\footnote{The corresponding  amplitude for the closed string sector plays an important role in an eikonal sum for multiple ``graviton" exchanges. See, e.g., Refs. 
 \cite{Brower:2007qh,Brower:2007xg} and work by D. Amati, M. Ciafaloni and G. Veneziano 
 \cite{Amati:1987wq,Amati:1987uf}. } In flat-space 
string theory, it takes on  the form  
\be
{\cal A}_{string}(M^2,  t_2, t_2';t_1)\sim \int_0^\infty dx \; x^{-\omega(t_1) + \omega(t_2) +\omega (t_2') -1} (1-x)^{-
\omega(M^2)}\;.
\ee
Note that this is analogous to the  limit taken for the 5-point function, (\ref{A5HP}), where the amplitude is also  expressed as a sum of two pieces, each with unique singularity 
structure. 
 ${\cal A}_{string}(M^2, t_2,t_2';t_1)$ is structurally analogous to ${\cal A}_{string}(s_2,t_2;t_1)$, (\ref{A5Rppp}), the Reggeon-particle-particle-particle 
amplitude discussed earlier for the 5-point function.

We emphasize that the discontinuity of the Reggeon-particle-to-Reggeon-particle amplitude, ${\cal A}_{string}(M^2,  
t_2, t_2',t_1)$, in $M^2$ directly leads to the inclusive cross section in the forward helicity-pole limit, as illustrated in 
Fig. \ref{fig:TRRegion}. This is a generic  feature which should hold in general.  For $M^2$ large, one can easily 
see that
\be
{\cal A}_{string}(M^2,  t_2, t_2',t_1)\sim (-M^2)^{\omega(t_1) - \omega(t_2) -\omega(t_2') } \;, \label{eq;reggeonparticle}
\ee
consistent with Eq. (\ref{TRstring2}).  From the perspective of a dispersion representation in $M^2$, the piece ${\cal B}$ in (\ref{hplimitgood}) represents a subtraction. For completeness, we record here for ${\cal B}$ for scalar tachyon 
amplitude,  which is kinematically simpler. It consists of three terms, (Eq. (4.24) of Ref.  \cite{Detar:1972nd}),
\be
{\cal B} \sim (-s')^{\omega(t_1)} U_1 + (-s)^{\omega(t_1)} U_2+(-s)^{(\omega(t_1)+\omega(t_2)-\omega(t'_2))/2} (-
s')^{(\omega(t)+\omega(t'_2)-\omega(t_2))/2} U_{12}
\label{hplimit}
\ee
with  $U_1$, $U_2$ and $U_{12}$    independent of $M^2$.

\subsection{BDS}
\label{sec:BDS}

We now turn to BDS and see if our flat-space based expectations are satisfied. We first consider the helicity-pole/
triple-Regge limits. As pointed earlier, here we will deal with a situation where the di-logarithm functions begin to 
play an even more  important role. We will demonstrate that, under BDS ansatz, one finds a surprising result where   the  Reggeon-particle-to-Reggeon-particle amplitude, ${\cal A}_{bds}(M^2,  t_2, t_2',t_1)$, vanishes.

Before carrying out this analysis, it is useful to recall that, under dimensional regualrization,  the physical gluon pole does not lie on the Regge trajectory, (see Sec. 3). In order to avoid dealing with such  issues,  we shall avoid approaching singular points, e.g., $t_1=0$ in (\ref{hplimitgood}). In general, in addressing various Regge/helicity pole limits, e.g., leading to (\ref{eq;reggeonparticle}),  we shall keep all fixed variables Euclidean,  e.g., $t_2,t_2', t_1, \Sigma, \Sigma'<0$ in (\ref{BDSinvariants}).   In these regions, no unusual behavior is expected from the tree-amplitudes.

As mentioned earlier, in the notation used by BDS, for $n=6$, with color order $(123456)$,  there are 9 invariants 
($t_i^{[2]}, i=1,...,6$ and $t_i^{[3]}, i=1,2,3$)
and one constraint among them. In our earlier study for  the multi-Regge limits~\cite{Brower:2008nm}, we have seen that it is convenient to 
introduce  $t_1,t_2,t_3, s_1,s_2,s_3$
and $\Sigma_1,\Sigma_2$, (or equivalently $\kappa_1,\kappa_2$), as independent variables, with $s$ as 
dependent variable. 
As we have also shown in  \cite{Brower:2008nm},  the single-Regge limit,  involves taking three $t_i^{[r]}$'s to 
infinity, two independent and one dependent,
e.g. $s_1,\Sigma_1$ and $s$ (or equivalently $s_1,s$ to infinity and $\kappa_1$ fixed).  For the  helicity pole limit, it 
is even simpler since this limit involves only taking two invariants large. For this limit, we have found convenient to 
start with color-ordering indicated by Fig. \ref{fig:6pt}a, and have also introduced a set of invariants, suggested by 
the inclusive cross-section, which are related to the BDS invariants by  (\ref{BDSinvariants}).

An equally useful set of notations  for the BDS invariants has been  introduced  in \cite{Brower:2008nm}  for the poly-Regge limit,  
(which makes full use of the cyclical symmetry of the problem), 
\bea
&&t_6^{[2]}=s_{23},\;\;\;
t_1^{[2]}=t_2,\;\;\;
t_2^{[2]}=s_{12},\;\;\;
t_3^{[2]}=t_1,\;\;\;
t_4^{[2]}=s_{31},\;\;\;
t_5^{[2]}=t_3,\nonumber\\
&&
t_1^{[3]}=s_1,\;\;\;
t_2^{[3]}=s_2,\;\;\;
t_3^{[3]}=s_3.
\eea
These notations have also been  used in \cite{Detar:1972nd,Brower:1974yv}, (see Fig. \ref{fig:polyregge}), and they  
are related to  (\ref{BDSinvariants}),  the $M^2$-discontinuity notation, by the substitions
 $s_3\leftrightarrow M^2$, $s_2\leftrightarrow \Sigma' $, $s_1\leftrightarrow \Sigma$, 
$t_3\leftrightarrow t_1$, $t_2\leftrightarrow t'_2$,
 $t_1\leftrightarrow t_2$,  $s_{23}\leftrightarrow  s'$, $s_{31} \leftrightarrow s$ and $s_{12} \leftrightarrow s_{12}$.

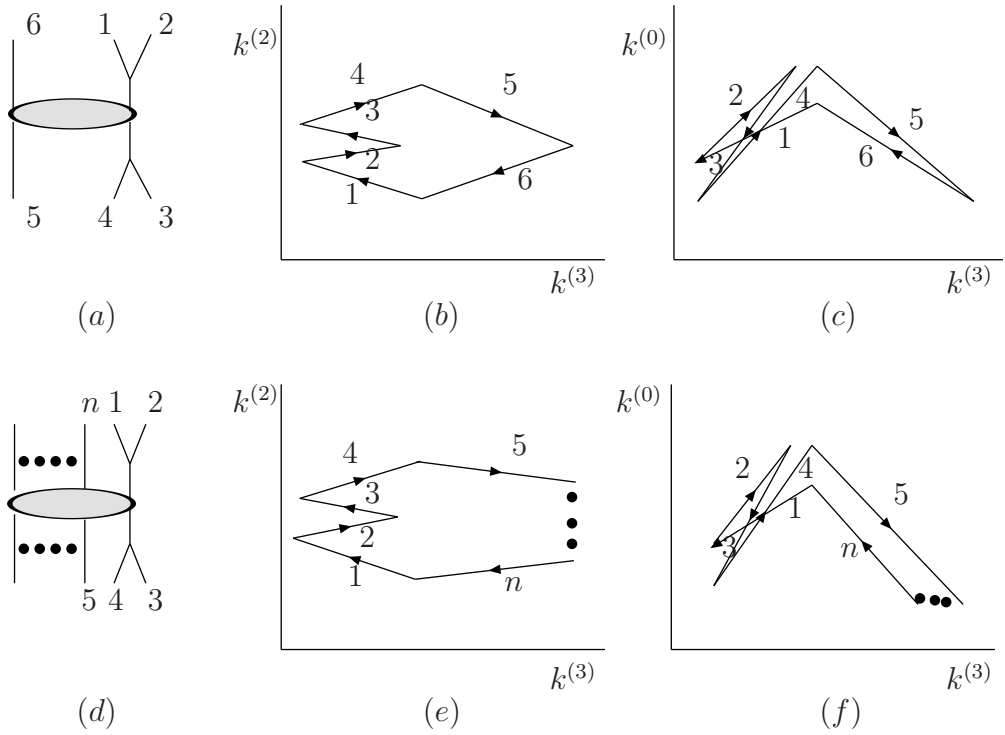
\begin{figure}[bthp]

\begin{center}
%\fcolorbox{white}{white}{
  \begin{picture}(400,300) (0,-10)
    \SetWidth{0.5}
   % \SetColor{Black}
    %
      \Text(59 ,258)[lb]{{\Black{$1$}}}
      \Text(82,258)[lb]{{\Black{$2$}}}
       \Text(82,185)[lb]{{\Black{$3$}}}
        \Text(59,185)[lb]{{\Black{$4$}}}
    \Text(32,185)[lb]{{\Black{$5$}}}
      \Text(32.,258)[lb]{{\Black{$6$}}}
    \GOval(49.35,227.01)(5.64,23.27)(0){0.882}
 \Line(27,255.)(27,228)
    \Line(27,224)(27,195)
    \Line(65,255)(71,240)
 \Line(79,257)(71,240)
     \Line(71,210)(65,195)
 \Line(71,228)(71,240)
  \Line(71,224)(71,210)
    \Line(71,210)(79,195)
     
 \Text(51,145)[lb]{{\Black{$(a)$}}}

 \Line(128,268)(128,172)
    \Line(128 ,172.)(250,172.)
         \ArrowLine(181,238)(238,215)
   \ArrowLine(238,215)(181,195) 
    \ArrowLine(135,223)(181,238)
 \ArrowLine(173,215)(136 ,223)
    \ArrowLine(136,209)(173,215)
\ArrowLine(181,195)(136,209)
   \Text(152.28,193.88)[lb]{{\Black{$1$}}}
   \Text(160,207)[lb]{{\Black{$2$}}}
    \Text(160.04,225.6)[lb]{{\Black{$3$}}}
 \Text(154.4,239)[lb]{{\Black{$4$}}}
    \Text(212.21,235.47)[lb]{{\Black{$5$}}}
   \Text(217.85,199.52)[lb]{{\Black{$6$}}}
   
     \Text(110,250)[lb]{{\Black{$k^{(2)}$}}}
  \Text(230,160)[lb]{{\Black{$k^{(3)}$}}}
   
      \Text(182,145)[lb]{{\Black{$(b)$}}}
      
          \Line(276,172)(400,172)
   \Line(276 ,268)(276 ,172)
   
  \ArrowLine(285,194)(330,245)
    \ArrowLine(322,245)(285,194)
  \ArrowLine(285,209)(322,245)
    \LongArrow (330,231)(285,209)
   \ArrowLine(389,194)(330,231)
    \ArrowLine(330,245)(389,194)
    
       \Text(316,215)[lb]{{\Black{$1$}}}
    \Text(297.51,231.24)[lb]{{\Black{$2$}}}
     \Text(290,205)[lb]{{\Black{$3$}}}
   \Text(323,230)[lb]{{\Black{$4$}}}
    \Text(365.9,222.08)[lb]{{\Black{$5$}}}
    \Text(346.86,207.98)[lb]{{\Black{$6$}}}

     \Text(380,160)[lb]{{\Black{$k^{(3)}$}}}
  \Text(257,250)[lb]{{\Black{$k^{(0)}$}}}

       \Text(332,145)[lb]{{\Black{$(c)$}}}

 \Line(71,80)(71 ,95)
   \Line(71,95)(77, 110)
    \Line(71,95)(65, 110)
  
 \Line(71 ,80)(71,65)
  \Line(71,65)(65,50)
  \Line(71,65)(77,50)
 
  \Line(54,110)(54,85)
       \Line(54,74)(54,50)    
%     \Line(47,110)(47,85)
  %  \Line(47,74)(47,50)
   \Line(27.5,76)(27.5,50)
  \Line(27.5,110)(27.5,85)
    
      \Vertex(49,63){1.99}
    \Vertex(43,63){1.99}
    \Vertex(37,63){1.99}
    \Vertex(31,63){1.99}
      \Vertex(49,96){1.99}
    \Vertex(43,96){1.99}
    \Vertex(37,96){1.99}
    \Vertex(31,96){1.99}
    
   \GOval(49,80)(5.64,23.27)(0){0.882}
    
  \Text(53,115)[lb]{{\Black{$n$}}}
    \Text(63,115)[lb]{{\Black{$1$}}}
    \Text(78,115)[lb]{{\Black{$2$}}}
     \Text(78,40)[lb]{{\Black{$3$}}}
    \Text(63,40)[lb]{{\Black{$4$}}}
    \Text(53,40)[lb]{{\Black{$5$}}}

    \Text(51,-5)[lb]{{\Black{$(d)$}}}
   
      \Line(128,25)(128,125)
  \Line(128,25)(250,25)
   
      \ArrowLine(135,82)(180.5,96)
    \ArrowLine(178.5,51.5)(132.5,67)
    \ArrowLine(132.5,67)(172,75)
  \ArrowLine(172,75)(135,82)
    \Text(151.58,95.88)[lb]{{\Black{$4$}}}
    \Text(160.04,80.37)[lb]{{\Black{$3$}}}
    \Text(157.92,64)[lb]{{\Black{$2$}}}
  \Text(153.69,49)[lb]{{\Black{$1$}}}
  \Text(110,115)[lb]{{\Black{$k^{(2)}$}}}
   \ArrowLine(179.07,95.88)(239,88.13)
 \ArrowLine(238.29,58.52)(178.5,51.5)
    \Vertex(237.59,82.49){1.99}
    \Vertex(237.59,72.62){1.99}
    \Vertex(237.59,64.86){1.99}
    \Text(215.73,98.7)[lb]{{\Black{$5$}}}
   \Text(212.91,47.24)[lb]{{\Black{$n$}}}
          \Text(230,10)[lb]{{\Black{$k^{(3)}$}}}
   
      \Text(182,-5)[lb]{{\Black{$(e)$}}}

         \Line(275,125)(275,25)
         \Line(275,25)(397,25)

  \Vertex(368.72,44.29){1.99}
  \Vertex(374.36,43.58){1.99}
   \Vertex(378.59,42.88){1.99}
     \Text(300,90)[lb]{{\Black{$2$}}}
  \Text(340,60)[lb]{{\Black{$n$}}}
 \Text(360,80)[lb]{{\Black{$5$}}}
   \Text(295,60)[lb]{{\Black{$3$}}}
          \Text(324,90)[lb]{{\Black{$4$}}}
    \Text(320,75)[lb]{{\Black{$1$}}}

       \ArrowLine(368,42)(328,87)       
 \ArrowLine(291,49)(328,102)
    \ArrowLine(320,102)(291,49)
   \ArrowLine(291,65)(320,102)
   \LongArrow(328,87)(291,64)
  \ArrowLine(328,102)(385,42)

  \Text(255,115)[lb]{{\Black{$ k^{(0)}$}}}
     \Text(380, 10)[lb]{{\Black{$k^{(3)}$}}}

          \Text(332,-5)[lb]{{\Black{$(f)$}}}

  \end{picture}
% }
\end{center}

\caption{Configuration of momenta in the helicity pole limit. (a) 6-point amplitude; (b) and (c) momenta 
for 6-point amplitude in the $(k^{(3)},k^{(2)})$ and $(k^{(3)},k^{(0)})$ planes. (d) n-point amplitude; (e) and (f)
momenta for n-point amplitude in the $(k^{(3)},k^{2)})$ and $(k^{(3)},k^{(0)})$ planes.}

\label{extraregge}
\end{figure}

For  the helicity pole limit,  we only have two  $t_i^{[r]}$ invariants becoming  large,
 $t_4^{[2]}=s_{31}=s \rightarrow -\infty$ and $t_6^{[2]}=s_{23} = s' \rightarrow -\infty$. The constraint among
the BDS invariants implies in this limit that we have $s_{31}\simeq s_{23}$ ($s\simeq s'$), this being therefore 
the 
simplest limit one can take on the 6-point BDS ansatz. As mentioned earlier, all fixed variables will be held Euclidean, e.g., $t_1<0$, away from the singular point $t_1=0$.  We can also characterize the limit by saying that
$s/M^2$ is large, with $M^2$ and $s/s'$  fixed, as discussed earlier. (Recall the analogous limit for $n=5$.)
In this limit, two of the three $u_i$ cross ratios go to infinity, so it is in principle a good way to test for
the presence of an additional function $f(u_1,u_2,u_3)$, as it could become important in the limit 
where its arguments are large (the same way as ${\rm Li}_2(1-u_i)$ in the BDS 
6-point amplitude does). For the determination of the $f(u_1,u_2,u_3)$ function via dual Wilson loops, 
it is of interest to exhibit the configuration of momenta (or dual Wilson loop) in the helicity pole 
limit. We therefore present it, together with the generalization to higher n-points, in Fig. \ref{extraregge}.

We obtain in this limit for the BDS 6-point amplitude, (expressing in both the notation appropriate for inclusive 
distribution and  in  the poly-Regge notation for
comparison with the poly-Regge limit),
\bea
M^{BDS}_6&\simeq& (-s)^{ (\omega(t_1)+\omega(t_2)-\omega(t_2'))/2}
(-s')^{(\omega(t_2)+\omega(t_2')-\omega(t_2))/2}U(t_1,t_2,t_2',s_{12},\Sigma,\Sigma')+\cdots\cdots\nn
&\leftrightarrow& (-s_{23})^{(\omega(t_2)+\omega(t_3)-\omega(t_1))/2}
(-s_{31})^{(\omega(t_3)+\omega(t_1)-\omega(t_2))/2}U(t_1,t_2,t_3,s_{12},s_1,s_2)+\cdots  \nonumber\\
\label{limi}
\eea
or, if we substitute $s=s'$,
\bea 
M^{BDS}_6&=& (-s)^{\omega(t_1)}U(t_1,t_2,t_2',s_{12},\Sigma,\Sigma')+\cdots\cdots \nn
&\leftrightarrow& (-s_{23})^{\omega(t_3)}U(t_1,t_2,t_3,s_{12},s_1,s_2)+\cdots\cdots
\label{limiti}
\eea
Notice that (\ref{limi}) only corresponds to the last term in (\ref{hplimit}), and that there is  no $M^2$ 
dependence at all. That is, the Reggeon-particle-to-Reggeon-particle amplitude, ${\cal A}_{bds}(M^2,  t_2, t_2',t)$, 
vanishes.

We emphasize that in order to obtain ${\cal A}(M^2,  t_2, t_2',t)\neq 0$,  the first term in (\ref{hplimitgood}), we would 
need to add in $\log M_6^{BDS}$
a term
\be
\Delta \log M_6^{BDS}\simeq 
-\frac{f}{8} \ln \frac{t_6^{[2]}t_2^{[2]}}{t_4^{[2]}}\ln\frac{t_3^{[2]}t_5^{[2]}}{t_1^{[2]}}+O(1)
=-\frac{f}{8}\ln u_{6,4;2}\ln u_{1,3;5}+O(1)
\ee
where 
\be
u_{i,j;k}\equiv\frac{x^2_{i,k}x^2_{j,k}}{x^2_{i,j}}
\ee
are not cross ratios, and cannot be written in terms of them. Thus such a term would be prohibited by dual 
conformal invariance \cite{Drummond:2007au,Drummond:2007aua,Drummond:2007bm}.

The absence of proper $M^2$ discontinuity raises further concerns on the reliability of BDS ansatz for multi-gluon amplitudes. There are several possibilities. 
It is important to point out, from Fig. \ref{fig:TRRegion}, at 1-loop, the effective diagrams
giving $M^2$ dependence would come from the Passarino-Veltman
reduction to "two mass hard" (2mh) scalar boxes, i.e., scalar boxes
with 2 adjacent external massive (virtual) lines and the other two
external lines massless (on-shell).   But it is known that for MHV 
amplitudes, there are no contributions to any 1-loop n-point functions
from 2mh scalar boxes. In fact, the Passarino-Veltman reduction
obtains only ``two mass easy" (2me) scalar boxes, with non-adjacent external massive
lines, for ${\cal N}=4$ SYM at leading order in $N$.   At higher loops it becomes more involved to show how the $M^2$
dependence vanishes, but at 2-loops, the explicit 6-point calculation
of \cite{Bern:2008ap} finds the BDS ansatz is correct (up to a
function of cross ratios, that cannot generate the $M^2$ dependence,
as we argued), and at 3-loops, the explicit IR divergence formula
agrees with the BDS ansatz \cite{Sterman:2002qn}.  Therefore, it is possible that the absence of proper $M^2$ discontinuities is the property of MHV amplitudes only.

However, our analogous findings in the helicity pole limit on the vanishing of ``Reggeon-particle" amplitudes for both $n=5$ and $n=6$ BDS amplitudes suggest a more serious deficiency. An interesting possibility relates to the fact that, as pointed out earlier, to reconstruct the full amplitudes, it is technically insufficient to keep only  $O(1)$ terms for $\log M_n$ as $\epsilon\rightarrow 0$. That is, from $M_n=e^{\log M_n}$, the $O(1)$ for $M_n$ will receive contributions from terms in $\log M_n$ to all orders in $\epsilon$, due to the presence of $\epsilon^{-2}$ and $\epsilon^{-1}$ terms in $\log M_n$.  In fact, for $n>4$, $O(\epsilon)$ terms  at 1-loop will involve more than box-diagrams.~\footnote{We would like to thank Marcus Spradlin and Anastasia Volovich for emphasizing this fact to us.}  Therefore, it is conceivable that proper $M^2$ discontinuities can be restored in this more general setting. However, as we have also pointed out in the Introduction, we do not consider this more general treatment in our  analysis here.

\subsection{Poly-Regge Limit}

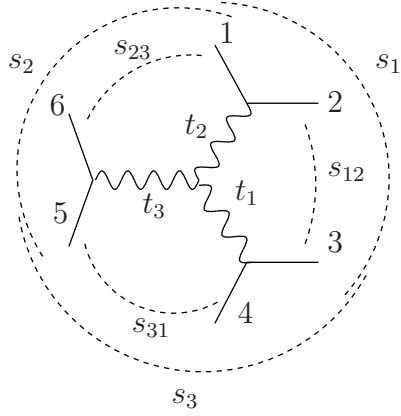
\begin{figure}[bthp]

\begin{center}
%\fcolorbox{white}{white}{
  \begin{picture}(167,165) (28,-17)
    \SetWidth{0.5}
 %   \SetColor{Black}

    \Text(167,110)[lb]{{\Black{$s_1$}}}
      \Text(28,110)[lb]{{\Black{$s_2$}}}
        \Text(90,-17)[lb]{{\Black{$s_3$}}}
       
           \Text(149,69)[lb]{{\Black{$s_{12}$}}}
              \Text(68,115)[lb]{{\Black{$s_{23}$}}}
                 \Text(75,10)[lb]{{\Black{$s_{31}$}}}
                 
                     \Text(114,60)[lb]{{\Black{$t_1$}}}
                      \Text(95,85)[lb]{{\Black{$t_2$}}}
                          \Text(79,55)[lb]{{\Black{$t_3$}}}
                 
        \Text(108,121)[lb]{{\Black{$1$}}}            
    \Text(149,95)[lb]{{\Black{$2$}}}
        \Text(149,42)[lb]{{\Black{$3$}}}
           \Text(115,15)[lb]{{\Black{$4$}}}
               \Text(45,53)[lb]{{\Black{$5$}}}
                 \Text(44,94)[lb]{{\Black{$6$}}}

      \Photon(100,69)(118,98){3.5}{3}
    \Photon(61,69)(100,69){3.5}{4}
    \Photon(100,67)(120,38){3.5}{3}
 \Line(118,98)(145,98)
    \Line(118,38)(145,38)
    \Line(118,98)(106,120)
   \Line(60,69)(51,94)
    \Line(60,69)(51,44)
    \Line(118,38)(106,15)
 
   \DashCArc(90.5,53.32)(34.52,-166.06,-61.44){2}
    \DashCArc(79,67.5)(65.02,-20.25,20.25){2}
      \DashCArc(96.93,72.76)(43.43,84.63,153.71){2}

  \DashCArc(101.45,67)(70.26,-170.32,-28.81){2}
   \DashCArc(105.27,69.68)(66.38,-41.49,87.32){2}
    \DashCArc(92.29,73.14)(60.85,70.8,212.76){2}

  \end{picture}
%}
\end{center}
\caption{Color-ordered amplitude with invariants appropriate for the poly-Regge limit. }
\label{fig:polyregge}
\end{figure}

We now re-visit the poly-Regge limit, studied in  \cite{Brower:2008nm}, and we  re-produce here
the schematic depiction for this limit  in Fig. \ref{fig:polyregge}.
In that limit, in accordance with 
\cite{Detar:1972nd}, 
we have  found that
\be
{A}_6\simeq \prod_{j=1}^3 (-s_j)^{\alpha(t_j)}V(t_i;s_{ij})
\ee
where $V(t_i;s_{ij})$ is also referred to as the triple Regge vertex. The poly-Regge limit is defined as 
$|s_{ij}| \gg| s_i| \rightarrow\infty$, with $t_i$ and $\eta_{ij}=s_{ij}/s_is_j$ kept fixed. 
In Sec. 3 of \cite{Detar:1972nd}, the triple Regge vertex $V$ of flat space string theory
was analyzed in the limit that $\eta_{ij}$ are also large. This is a very interesting case 
to consider since  $u_1=\eta_{23}t_1,u_2=\eta_{31}t_2, u_3=\eta_{12}t_3$.
This  means we  encounter a situation  where all $u_i\gg 1 $.  This is  therefore the best case to study for 
determining the 
relevance of the ${\rm Li}_2(1-u_i)$ terms in the BDS ansatz.

In \cite{Detar:1972nd}, the amplitude of flat space string theory was found to become in the above limit, 
when re-expressed in terms of $s_i$ and $s_{ij}$,~\footnote{This limit allows an interpolation between 
the poly-Regge limit and the helicity pole limit.  The expression quoted here is for bosonic string with  tachyons. The 
corresponding amplitude for superstring is similar and we will not report it here.}
\bea
{A}_{6}
 &\sim& \left(-s_3\right)^{\alpha_3-\alpha_1-\alpha_2}\left(-s_{13}\right)^{\alpha_1}\left(-
s_{23}\right)^{\alpha_2}\;
 \Gamma\left(-\alpha_1\right)
\Gamma\left(-\alpha_2\right) \Gamma\left(\alpha_1+\alpha_2-\alpha_3\right)\\
&+&\; \left(-s_1\right)^{\alpha_1-\alpha_2-\alpha_3}\left(-s_{12}\right)^{\alpha_2}\left(-
s_{13}\right)^{\alpha_3}\; \Gamma\left(-\alpha_2\right) \Gamma\left(-\alpha_3\right)\Gamma\left(\alpha_2+\alpha_3-\alpha_1\right)\nonumber\\
&+&\; \left(-s_2\right)^{\alpha_2-\alpha_1-\alpha_3}\left(-s_{23}\right)^{\alpha_3}\left(-
s_{21}\right)^{\alpha_1}\; 
\Gamma\left(-\alpha_3\right) \Gamma\left(-\alpha_1\right) \Gamma\left(\alpha_3+\alpha_1-\alpha_2\right)\nonumber\\
&+& \frac{1}{2}\;\left(- s_{23}\right)^{(\alpha_3+\alpha_2-\alpha_1)/2} \; \left(-s_{13}\right)^{
(\alpha_3+\alpha_1-\alpha_2)/2} \; \left(-s_{12}\right)^{
(\alpha_1+\alpha_2-\alpha_3)/2}\times \nonumber\\
&& \Gamma\left(-(\alpha_3+\alpha_2-\alpha_1)/2\right)
 \Gamma\left(-(\alpha_3+\alpha_1-\alpha_2)/2\right)
\Gamma\left(-(\alpha_1+\alpha_2-\alpha_3)/2\right) \nonumber
\label{dwresult}
\eea
where $\alpha_i=\alpha(t_i)$, $i=1,2,3$. 
However, from the BDS result, we obtain again just the fourth term,  namely
\bea
\left( A_6\right)_{BDS}&\sim & (-s_{23})^{(\alpha_3+\alpha_2-\alpha_1)/2}(-s_{13})^{
(\alpha_3+\alpha_1-\alpha_2)/2}(-s_{12})^{
(\alpha_1+\alpha_2-\alpha_3)/2}\; \Gamma(t_1,t_2,t_3) +\cdots\cdots\nn
\label{resul}
\eea
and the product of Gamma functions in the last term in (\ref{dwresult}) is  replaced by a new  vertex 
\be
\Gamma(t_1,t_2,t_3)=\exp\left\{\left(\frac{f^{-1}(\lambda)}{8\epsilon}+\frac{g(\lambda)}{4}\right)
(\ln t_1+\ln t_2+\ln t_3)\right\} \label{eq:coefficient}
\ee
Note that, in order to express (\ref{eq:coefficient})  in terms of $\alpha_i$, or equivalently $\omega_i$, one would have to modify factors associated with IR divergent terms.

To restore all  terms  in (\ref{dwresult}), one needs 
\be
\Delta \log M_6^{BDS} \simeq -\frac{f(\lambda)}{8}\log \frac{t_5^{[2]}}{t_3^{[2]}t_1^{[2]}}
\log\frac{t_6^{[2]}t_4^{[2]}}{t_2^{[2]}t_3^{[3]}}+O(1)=
-\frac{f(\lambda)}{8} \log u_{1,5;3} \log \frac{u_{2,4;6}}{x_{3,6}^2}+O(1)
\ee
which  cannot be expressed as functions of cross ratios. This result is analogous to the findings of   Sec. \ref{sec:BDS}, and makes it  clearer that the full flat space string theory result does not appear in ${\cal N}=4$ SYM, but rather only a subleading term in its expansion.

\subsection{Mueller Regge Limit}

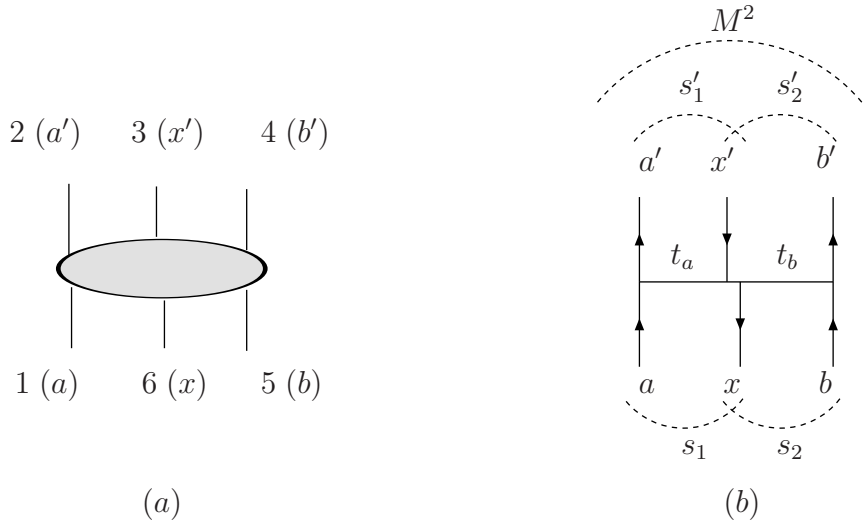
\begin{figure}[bthp]
\begin{center}
%\fcolorbox{white}{white}{
  \begin{picture}(331,172) (95,-67)
    \SetWidth{0.5}
 %   \SetColor{Black}
  %  \ArrowLine(375,-13)(391,-32)
    \Text(355,42)[lb]{{\Black{$x'$}}}
    \Text(395,43)[lb]{{\Black{$b'$}}}
    \Text(360,-42)[lb]{{\Black{$x$}}}
    \Text(396,-42)[lb]{{\Black{$b$}}}
    \Text(328,-42)[lb]{{\Black{$a$}}}
       \Text(340,5)[lb]{{\Black{$t_a$}}}
         \Text(380,5)[lb]{{\Black{$t_b$}}}
  
 \DashCArc(361,26)(65,40,140){2}
   
    \Text(355,95)[lb]{{\Black{$M^2$}}}
    \Text(343,68)[lb]{{\Black{$s_1'$}}}
    \Text(344,-67)[lb]{{\Black{$s_1$}}}
   \Text(380,68)[lb]{{\Black{$s_2'$}}}
    \Text(380,-67)[lb]{{\Black{$s_2$}}}
   \ArrowLine(327,-32)(327,0)
    \ArrowLine(327,0)(327,32)
    \Line(327,0)(400,0)
    \DashCArc(346,36.15)(26.92,38.74,141.26){2}
    \DashCArc(343.6,-27)(28.15,-140.13,-37.27){2}
        \DashCArc(380.15, 36.15)(26.92,38.74,141.26){2}
    \DashCArc(380.15,-27)(28.15,-140.13,-37.27){2}

    \Text(328,42)[lb]{{\Black{$a'$}}}
      \ArrowLine(400,0)(400,32)
    \ArrowLine(400,-32)(400,0)
         \ArrowLine(365,0)(365,-32)
              \ArrowLine(360,32)(360,0)
     \Text(90,50)[lb]{{\Black{$2\; (a')$}}}
    \Text(92,-45)[lb]{{\Black{$1\; (a)$}}}
    \GOval(147,5)(11,39)(0){0.882}
    \Line(179,12)(179,35)
    \Line(179,-3)(179,-26)
 
      \Text(185,50)[lb]{{\Black{$4\; (b')$}}}
    \Text(140,-45)[lb]{{\Black{$6\; (x)$}}}
    \Text(185,-45)[lb]{{\Black{$5\; (b)$}}}
    \Line(112,10)(112,37)
    \Line(145,17)(145,36)
    \Line(113,-2)(113,-25)
    \Line(148,-7)(148,-25)
     \Text(136,50)[lb]{{\Black{$3\; (x')$}}}
     
      \Text(140,-90)[lb]{{\Black{$(a)$}}} 
        \Text(360,-90)[lb]{{\Black{$(b)$}}} 
  \end{picture}
%}
\end{center}
\caption{6-point amplitude with momentum-color-ordering appropriate for Mueller-Regge limit. }
\label{fig:6ptMueller}
\end{figure}

For completeness, we end by a discussion on  the so-called ``Mueller-Regge" limit where kinematically the momenta can be 
arranged in the following suggestive multi-Regge order, Fig. \ref{fig:6ptMueller}.  To allow discontinuity in $M^2$, we need to consider color ordering with $(axb)$ and $(a'x'b')$ adjacent. 
It is convenient to consider first  the color-ordering, $(123456)=(aa'x'b'bx)$, (Fig. \ref{fig:6ptMueller}a);  other color 
orderings can be obtained by appropriate substitutions and continuations. 

 Let us now denote adjacent BDS invariants as, (Fig. \ref{fig:6ptMueller}b), 
\bea
t_5^{[2]}&\equiv&s_2, \quad t_3^{[2]}\equiv s_2',\quad
t_6^{[2]}\equiv s_1, \quad t_2^{[2]}\equiv s_1', \quad 
t_1^{[2]}\equiv  t_a, \quad  t_4^{[2]}\equiv  t_b   \nonumber\\
 t_2^{[3]}&\equiv& M^2,\quad 
t_4^{[3]}  \equiv  \Sigma, \quad t_3^{[3]}=\Sigma'      \label{muellerinv}
\eea
 Note that, for this color-ordering,
\be
s=-(p_a+p_b)^2=-(k_1+k_5)^2, \; s'=-(p_{a'}+p_{b'})^2= -(k_2+k_4)^2, 
\ee
are non-adjacent  invariants. 

We consider $s_1\simeq s'_1\rightarrow - \infty$, and $s_2\sim s_2'\rightarrow -\infty$, with $t_a$, $t_b$, $\cdots$ 
fixed. This is effectively a linear double-Regge limit. One finds that,~\cite{Detar:1971dj}
\be
M_{6,string}\sim  (-s_1)^{\omega(t_a)}(-s_2)^{\omega(t_b)} G^{[4]}_{2,string}(\kappa, \kappa', \cdots)
\ee
where 
\be
\kappa=\frac{s_1s_2}{M^2}, \quad \kappa'=\frac{s_1's_2'}{M^2} \; , \quad {\rm and} \quad \kappa\simeq \kappa'
\ee
are also kept fixed. By continuing $M^2$ back to the physical region, it is clear that taking the discontinuity in 
$M^2$ is the same as taking the discontinuity in $\kappa$ and $\kappa'$. For the  inclusive cross section where $
\kappa=\kappa'$, one finds that 
$G^{[4]}_{2,string}(\kappa, \kappa, \cdots)$ in flat-space string theory can be expressed as an integral 
over the Reggeon-particle-Reggeon vertex introduced earlier, $G_2(t_a,t_b, \kappa)$,~\cite{Detar:1971dj}
\be
G_{2,string}^{[4]}(\kappa, \kappa, \cdots) \sim  \int_0^1 dz z^{-\omega(\Sigma)} (1-z)^{-\omega(\Sigma')} G_{2,string}
(t_a,t_b, z(1-z)\kappa)
\ee
A similar but more involved expression for the non-forward limit can also be obtained.  Note that  $G^{[4]}_{2,string}
$ is real for $\kappa<0$ in the Euclidean region, and has a right-hand cut for $\kappa>0$. With $s_1=s_1', s_2= 
s_2'<0$, one has
\be
{\rm Disc}_{M^2} M_{6,string}\simeq  (-s_1)^{\omega(t_a)}(-s_2)^{\omega(t_b)} {\rm Disc}_{M^2}  G^{[4]}_{2,string}
(\kappa, \kappa, \cdots)
\ee
The discontinuity can then be taken, similar to that for a 5-point function. Unlike the case of 5-point function, however, this discontinuity now directly relates to an observable.

Let us next turn to the BDS amplitude. We can now do the Mueller Regge limit in exactly the same way, using the 
variables in 
(\ref{muellerinv}).
In the limit $s_1\simeq s_1'\rightarrow  - \infty; s_2\simeq s_2'\rightarrow -\infty$,  
$M^2\rightarrow - \infty$, with $\kappa=s_1s_2/M^2$ and $\kappa'=s_1's_2'/M^2$ fixed, we 
obtain from the BDS amplitude
\be
M_6\simeq (\sqrt{s_1s_1'})^{\omega(t_a)}(\sqrt{s_2s_2'})^{\omega(t_b)}G^{[4]}_2(\kappa,\kappa',\cdots)
\ee
where the vertex $G^{[4]}_2(\kappa,\kappa',\cdots)$ is given by 
\bea
\log G^{[4]}_2(\kappa,\kappa',\cdots)&=&\frac{f(\lambda)}{8}\log(- t_a)\log (-t_b)
+\frac{1}{2}\left(\frac{f^{(-1)}(\lambda)}{4\epsilon}+\frac{g(\lambda)}{2}\right)\log(t_at_b)\nonumber\\
&&
-\frac{f(\lambda)}{8}\left\{{\rm Li}_2\left(1-\frac{\kappa' s_1}{\Sigma' s_1'}\right)
+{\rm Li}_2\left(1-\frac{t_at_b}{\Sigma'\Sigma}\right)+{\rm Li}_2\left(1-\frac{\kappa s_1'}{\Sigma s_1}\right)
\right.\nonumber \\
&&\left.
+\frac{1}{2}\left[\log^2\left(\frac{\kappa 's_1}{\Sigma' s_1'}\right)+\log^2\left(\frac{t_at_b}
{\Sigma'\Sigma}\right)+\log^2\left(\frac{\kappa s_1'}{\Sigma s_1}\right)\right]
\right\}\label{vertexG}
\eea

Let us first note that, in the physical region, $\Sigma<0$ and $\Sigma'<0$. With $\kappa>0$ and $\kappa'>0$, those 
four terms in (\ref{vertexG})  involving $\kappa$ or $\kappa'$ will lead to discontinuities in $M^2$. 
 Note that the last two lines of (\ref{vertexG}) go to zero if the arguments of the log's go to 
 infinity. Thus if we have $s_1\simeq s_1'; s_2\simeq s_2'$ and $\kappa\simeq \kappa'$ 
 is taken to be much larger than $\Sigma, \Sigma'$ on top of the Mueller Regge limit, the 
 $\kappa\simeq \kappa '$ dependence completely drops out, and like in the previous 
 cases, there is no $M^2$ dependence in the vertex $G$. Indeed, in flat-space string theory, one moves smoothly 
from the Mueller double-Regge limit to  the triple-Regge/helicity-pole limits, with $\kappa\rightarrow \infty$, $
\kappa'\rightarrow \infty$.
Then, if $\kappa=\kappa '$ (forward limit), and much larger than $\Sigma, \Sigma'$,
\be
M_6\simeq (-s_1)^{\omega (t_a)}(-s_2)^{\omega(t_b)}G^{[4]}_2(\kappa,\kappa',\cdots)
\ee
where 
\bea
\log G ^{[4]}_2
&=&\frac{f(\lambda)}{8}\log t_a\log t_b
+\frac{1}{2}\left(\frac{f^{(-1)}(\lambda)}{4\epsilon}+\frac{g(\lambda)}{2}\right)\log(t_at_b)\\
&&
-\frac{f(\lambda)}{8}\left\{{\rm Li}_2\left(1-\frac{t_at_b}{\Sigma'\Sigma}\right)+\frac{1}{2}
\log^2\left(\frac{t_at_b}{\Sigma'\Sigma}\right)\right\}\nonumber\\
&=&\log[\gamma(t_a)\gamma(t_b)]
+\frac{f(\lambda)}{8}\log t_a\log t_b
-\frac{f(\lambda)}{8}\left\{{\rm Li}_2\left(1-\frac{t_at_b}{\Sigma'\Sigma}\right)+\frac{1}{2}
\log^2\left(\frac{t_at_b}{\Sigma'\Sigma}\right)\right\}\nonumber
\label{vertex}
\eea

Since there is no $M^2$ dependence in (\ref{vertex}), there is no discontinuity, 
${\rm Disc}_{M^2}A_6=0$, same as for the helicity pole
limit or the poly-Regge limit.

For the Mueller double-Regge limit, however, $\kappa$ and $\kappa'$ should be kept fixed at 
finite values. In that case, the exact cancellation between discontinuities in $M^2$ from ${\rm Li}_2$ and the terms 
in the last line in (\ref{vertexG}) no longer holds.  The expression can be further simplified, but will not be provided here.

\newpage

\section{Discussion}
\label{sec:discussion}

A central issue for this paper, as well as others \cite{Bartels:2008ce}, \cite{DelDuca:2008jg}, is the continuation of multi-Regge amplitudes from the Euclidean region to various physical regions which are distinguished by different color orderings.  The problem that must be dealt with is that the order in going on-shell in the continuation and in taking the multi-Regge limit do not commute.  This is discussed in Sec. \ref{sec:analytic} and more explicitly in Sec. \ref{subsect:linearMR}.  In the body of this paper it is emphasized that multi-Regge factorization is expected only for signatured amplitude.  Thus for the 6-point amplitude, 6 of the 8 inequivalent color configurations allow for the straightforward results that $\Phi = 1$ (in (\ref{eq:M3})) is consistent with the naive continuation. There are two other configurations, where $\Phi$ takes  on $e^{2\pi i}$ and $e^{-2\pi i}$ respectively by circling the branch point at $\Phi=0$.  It is these latter two continuations which have been the subject of controversy \cite{Bartels:2008ce}, \cite{DelDuca:2008jg}.  We proceed to discuss the consequences of the two opposing positions.

\subsection{Comments on  analytical continuation in the literature}

In the Euclidean multi-Regge limit, the finite parts of $\log M_6^{BDS}$ contain $\log s$ and dilogs, and 
all cross ratios either vanish or approaching 1 in this limit. Thus in \cite{Brower:2008nm} we have found 
that dilogs don't contribute in the Euclidean region and factorization can be achieved.

However, as mentioned, in \cite{Bartels:2008ce} it was found that analytical continuation 
of the $u_3\equiv \Phi$ cross ratio in a physical region for $n=6$ gluons leads to an  extra term, 
that apparently breaks factorization. 

The conventional  procedure is to drop $O(\epsilon)$ for 
\be
\log{M_6}=\log\frac{A_6}{A_{6,tree}}=I_6^{(1)}(\epsilon)+F_6^{(1)}(0)
\ee
in taking the mulit-Regge limit.  In this case,  BDS amplitudes for $n\geq 6$ reduce to simple combinations of products of logarithms and dilogarithmic functions. As pointed in Ref.
 \cite{Brower:2008nm}, these dilog functions do not contribute in the Euclidean multi-Regge limit and naive Regge factorization can be achieved. This relies on the observation that all cross ratios either vanish or approaching 1 in this limit. 

However, as pointed out above, analyticity consideration forces one to relax the constraint on the cross ratios in the course of continuation back to the physical region. In the case of $n=6$, there only three such cross ratios, and the one which requires special attention is the variable $\Phi$, or $u_3$ in (\ref{crossr}). Since its nontrivial dependence enters in a single dilog term, continuation into the physical region can be carried out explicitly. 
In \cite{Bartels:2008ce}, one finds for $A_{-+-}$ that, in the course of continuation where $\Phi: 1\rightarrow e^{-2\pi i}$, 
$\log{M_6}$ picks up  an extra piece
\be
\Delta \log{M_6(-,+,-)}=\frac{f(\lambda)}{4}\pi i\left(-\frac{1}{\epsilon}
+\log \left(\frac{(-t_1)(-t_3)}{\mu^2 (s_2)}\left[\frac{\Phi}{1-\Phi}\right]\right)\right)\;.
\ee
With this additional term in $M_6(-,+,-)$, it breaks both the naive factorization, (\ref{eq:6pointlinearMR.a}), and that required for factorization in signature space, (\ref{eq:key}).  A similar analysis can also be carried out for the $M_6(-,-,-)$.  \ignore{It has been suggested a term could be added so that naive factorization could be restored, however it appears that such a ``repair" of the BDS 6-point amplitudes involves adding  a term which is not a function of cross-ratios, thus  unsatisfactory.  No satisfactory ``modification" is currently known, so the tentative conclusion is} It follows  that $M(-,\pm,-)$ violates both naive factorization and that required for  signatured factorization, and factorization cannot be regained  simply by adding terms  which are functions of cross-ratios.

In the published version of the paper by Del Duca, et al., \cite{DelDuca:2008jg}, which appeared after the first version of our paper, a different proposal  for the analytical continuation of the multi-Regge (asymptotic) form was presented (Appendix C).
The BDS ansatz is defined in the $\epsilon$-expansion, so it is a priori hard to check what would happen 
to the full amplitude if we keep $\epsilon$ finite, but in \cite{DelDuca:2008jg}, v.5,  this procedure was tested on the 
one-loop amplitude. It was claimed  that  the $\epsilon\rightarrow 0$ and multi-Regge limits don't commute, 
and the extra term found by [3] 
 is not present if we keep $\epsilon$ finite and take it to zero after the multi-Regge limit.
 It should be emphasized that this is not the conventional way of understanding the BDS ansatz. We also note that in the latest update to \cite{Bartels:2008ce}, it has been argued that   the above claim in  \cite{DelDuca:2008jg}, v.5, was invalid due to an arithmetic error. As of this writing, no retraction by the authors of  \cite{DelDuca:2008jg}, v.5, has appeared.

Independent of the discussion of \cite{DelDuca:2008jg}, one might wonder why  it might be reasonable to take $\epsilon\rightarrow 0$ after the Regge limit.
Although conventional wisdom favors the continuation of  \cite{Bartels:2008ce}, there is one example where that choice can be 
reconsidered. From general principles (Mandelstam counting \cite{Mandelstam:1965zz}, etc.) 
in an IR finite renormalizable Yang-Mills theory, the gluon 
Regge trajectory $\alpha (t)$ must satisfy $\alpha(0)=1$. If one first makes a Laurent expansion in $\epsilon
\rightarrow 0$, then for ${\cal N}=4$ SYM we have $\alpha (t)\rightarrow 1+\frac{1}{4\epsilon}f^{(-1)}(\lambda)
-\frac{1}{4}f(\lambda)\log (-t/\mu^2)$. On the other hand, if one does not expand in $\epsilon\rightarrow 0$, but 
keeps $\epsilon\neq 0$ and finite, and takes $t \rightarrow 0$ instead, from general expectations, we should 
have $\alpha (t=0)=1$, contradicting the above relation, for which $\alpha(t\rightarrow 0)\rightarrow \infty$.
A way to solve the contradiction is to introduce a gauge invariant IR cut-off at fixed $\epsilon$, leading to a
gluon mass $m$ (for instance via Higgs mechanism, to guarantee consistency), with the result at small $\lambda=g^2N
\ll \epsilon$  \cite{Grisaru:1973vw,Grisaru:1974cf}, $\alpha(t)\simeq 1-[(t-m^2)/t] (g^2N/(8\pi^2)) \log (-t)$, consistent 
with the above $\alpha(t)$ in the Laurent $\epsilon$ expansion at $m^2=0$. 
Then $\alpha(t=m^2)=1$ and only then we can take $m^2\rightarrow 0$.  Although this does not bear directly on \cite{Bartels:2008ce} vs. \cite{DelDuca:2008jg}, it emphasizes that the order of limits is often essential in drawing physical conclusions.  In this connection, it is also worth noting that the conventional proof for the Steinmann relation relies  on having a mass gap, e.g., \cite{Stapp:1971hh}.  It is therefore also possible difficulties of this and other related  issues could in principle
be resolved by  calculations of the ${\cal O}(\epsilon)$ terms in the exponent of the IR 
divergent BDS amplitudes.  Further analyses along these lines could help in clarifying the role of Steinmann relation  in a conformal theory.

\subsection{Brief Summary}

In this paper we have investigated the issue of analyticity of the ${\cal N}=4$
SYM amplitudes, using the BDS ansatz, in regard to the multi-Regge limits. 
In particular, we have looked at issues of analytical continuation that were not 
addressed in our previous paper \cite{Brower:2008nm}, analyzed the behavior of universal Regge vertices 
appearing in all n-point amplitudes, and some Regge limits directly related to 
unitarity conditions, the helicity pole and triple-Regge limits. 
By way of comparison, flat space string theory amplitudes are generally used 
as a primer for Regge behavior, so we have compared them with the ${\cal N}=4$
SYM amplitudes. The results that we found are unusual.

We have found  that the IR cut-off ${\cal N}=4$ SYM planar amplitudes, as characterized by BDS
 in the Regge limits differ  from those of flat-space
super string theory in several important aspects. Thus, intuition and specific properties of flat-space
string theory can be applied to planar MSYM  with at best a great deal of
caution. For $n > 5$, it has been suggested that  BDS amplitudes should be modified by adding a function of the cross-ratios.
However this cannot eliminate the various mismatches in properties of
flat-space string theory and the various Regge limits of ${\cal N}=4$ SYM.  This can be seen more directly by examining (\ref{lipatovextra}) and (\ref{eq:key}), i.e., terms which must be added do not appear to be expressible as functions of cross ratios. 
Thus, one must conclude that the Regge behavior of conformal  $ {\cal N} =4$  SYM 
theory is rather different from that of flat-space string theory. Perhaps 
this is not unexpected, as flat space string theory has a mass-scale, 
a mass-gap, and linearly rising Regge trajectories with recurrences:
all absent from ${\cal N}=4$ SYM. With this in mind, the different properties of the 
Regge limits of the BDS amplitudes and flat-space string theory  should not be surprising. 
However, since flat space string properties considered here follow both from generic rules of planar unitarity and the existence of on-shell Reggeon vertex operator~\cite{Brower:2006ea} on the worldsheet,
these difference are worth more careful study particularly with respect to
implementation of the IR cut-off in the BDS construction.

{\bf Acknowledgements}:  We would like to thank 
A. Brandhuber, L. Dixon, P. Heslop, S. Ramgoolam, M. Spradlin, G. Travaglini and A. Volovich for discussions.   
RCB and CIT would like to  thank the Galileo Galilei Institute for Theoretical Physics for the hospitality and the INFN 
for partial support during the completion of this work.
RCB's research  is supported in part by the Department of 
Energy under
Contract.~No.~DE-FG02-91ER40676. 
HN's research  has been done with partial support 
from  MEXT's program ``Promotion of Environmental Improvement for 
Independence of Young Researchers" under the Special Coordination 
Funds for Promoting Science and Technology, and also with partial support from MEXT KAKENHI grant nr. 
20740128. HJS's research is supported in part by the DOE under Grant DE-FG02-92ER40706.  CIT's research is 
supported in part by 
the U.~S.~Department of 
Energy under  Grant DE-FG02-91ER40688, TASK A.

{\bf Note Added:} The authors would like to thank Zvi Bern, Lance Dixon,and Lev Lipatov for conversations and correspondence 
and for pointing out the need to revise the discussion on analytic continuations in the original version of this paper. We also note that  a paper by V. Del Duca, C. Duhr, and E. W. N. Glover, \cite{DelDuca:2008jg},  appeared after our paper was posted. Another recent paper by   Bartels, et al., \cite{Bartels:2008sc} has further extended their work in \cite{Bartels:2008ce}.    We would also like to thanks Dr. Del Duca  for bringing to our attention Refs. \cite{DelDuca:1995zy,DelDuca:1993pp} which are relevant to our discussion in Sec. \ref{subsec:alln}.

\bibliographystyle{utphys}
\bibliography{BNST4}

\end{document}